\documentclass[useAMS,usenatbib]{mn2e}

\usepackage{graphicx}
\usepackage{rotating}

\newcommand{\Msun}{$M_{\odot}$}
\newcommand{\acc}{$M_{\odot}$\,yr$^{-1}$}

\newcommand{\kms}{km\,s$^{-1}$}
\newcommand{\vs}{$v \sin i$}
\newcommand{\teff}{$T_{\rm eff}$}
\newcommand{\lgg}{$\log\,{g}$}
\newcommand{\vmic}{$\xi$}

\title[Chemical abundances of HAeBe stars]{Chemical abundances of magnetic and non-magnetic Herbig Ae/Be stars}

\author[Folsom et al.]{C.P. Folsom$^{1}$\thanks{E-mail: cpf@arm.ac.uk}, S. Bagnulo$^{1}$, G.A. Wade$^{2}$,
E. Alecian$^{3}$, J.D. Landstreet$^{1,4}$, 
\newauthor
S.C. Marsden$^{5,6}$, I.A. Waite$^{7}$ \\
$^{1}$Armagh Observatory, College Hill, Armagh Northern Ireland BT61 9DG\\
$^{2}$Department of Physics, Royal Military College of Canada, P.O. Box 17000, Station `Forces', Kingston, Ontario, Canada, K7K 7B4\\
$^{3}$Observatoire de Paris, LESIA, Place Jules Janssen, F-92195 Meudon Cedex, France \\
$^{4}$Physics \& Astronomy Department, The University of Western Ontario, London, Ontario, Canada, N6A 3K7 \\
$^{5}$Australian Astronomical Observatory, PO Box 296, Epping, Sydney, 1710, Australia\\
$^{6}$Centre for Astronomy, School of Engineering and Physical Sciences, James Cook University, Townsville, 4811, Australia\\
$^{7}$Faculty of Sciences, University of Southern Queensland, Toowoomba, 4350, Australia 
}

\begin{document}

\date{Received: 201?; Accepted: 201?}

\pagerange{\pageref{firstpage}--\pageref{lastpage}} \pubyear{201?}

\maketitle

\label{firstpage}

\begin{abstract}
The photospheres of about 10-20\% of main sequence A- and B-type stars 
exhibit a wide range of chemical peculiarities, often associated with 
the presence of a magnetic field.  It is not exactly known at which  
stage of stellar evolution these chemical peculiarities 
develop. To investigate this issue, in this paper we study the 
photospheric compositions of a sample of Herbig Ae and Be 
stars, which are considered to be the pre-main sequence progenitors of A and 
B stars. We have performed a detailed abundance analysis of 20 Herbig 
stars (three of which have confirmed magnetic fields), and one dusty young star. 
We have found that half the stars in our sample show $\lambda$ Boo chemical 
peculiarities to varying degrees, only one star shows weak Ap/Bp 
peculiarities, and all the remaining stars are chemically normal. 
The incidence of $\lambda$ Boo chemical peculiarities we find in Herbig stars 
is much higher than what is seen on the main sequence.  
We argue that a selective accretion model for $\lambda$ Boo star formation 
is a natural explanation for the remarkably large number of $\lambda$ Boo 
stars in our sample.  We also find that the magnetic Herbig stars do 
not exhibit a range of chemical compositions remarkably different from the 
normal stars: one magnetic star displays $\lambda$ Boo chemical peculiarities 
(HD 101412), one displays weak Ap/Bp peculiarities (V380 Ori A), and 
one (HD 190073) is chemically normal. This is completely different 
from what is seen on the main sequence, where all magnetic A and cool 
B stars show Ap/Bp chemical peculiarities, and this is consistent 
with the idea that the magnetic field precedes the formation of the 
chemical peculiarities typical of Ap and Bp stars. 
\end{abstract}

\begin{keywords}
stars: magnetic fields,
stars: abundances,
stars: chemically peculiar,
stars: pre-main-sequence
\end{keywords}

\section{Introduction}
In main sequence A and B stars a wide range of peculiar chemical abundances are observed.  
While much study has been devoted to these peculiar objects, a lot remains to be learnt 
about the physical processes behind these chemical peculiarities.  
An avenue of study that has not yet been properly investigated is to examine the chemistry 
in the pre-main sequence progenitors of main sequence peculiar stars.  

Herbig Ae and Be (HAeBe) stars are pre-main sequence A and B stars, 
and thus evolve into the wide range of different main sequence A and B stars.  
Observationally, Herbig stars are generally identified by optical emission lines, infrared excess, 
and are usually associated with nebulous regions \citep{Vieira2003-HAeBe_ID}.  
Chemically, these stars are thought to possess approximately solar 
abundances \citep{Acke2004-HAeBe_Abun}, similar to most young, nearby main sequence stars.  

A small number of HAeBe stars have recently been found to have strong, globally ordered magnetic fields 
\citep{Donati1997-major, Wade2005-HAeBe_Discovery, Wade2007-HAeBe_survey, Catala2007-HD190073, 
Alecian2008-HD200775, Alecian2008-ClusterHAeBeLetter}.  
The strength and morphology of these magnetic fields are very similar to those seen 
in the magnetic chemically peculiar Ap and Bp stars 
\citep{Alecian2008-HD200775,Folsom2008-hd72106,Alecian2009-v380ori}, 
making these magnetic HAeBe stars strong candidates for the progenitors 
of Ap and Bp stars \citep{Wade2005-HAeBe_Discovery}.  
This discovery in particular raises the question of whether any sign of 
chemical peculiarity can be found on the pre-main sequence.  

In main sequence A and cooler B stars, magnetic fields are always seen together with 
the characteristic chemical peculiarities of Ap and Bp stars \citep{Auriere2007}.  
Hotter magnetic B stars also usually show chemical peculiarities, such as He-weak or He-strong stars. 
Consequently, if magnetic HAeBe stars evolve into Ap and Bp stars, then at some point they must 
develop chemical peculiarities.  
Detecting such peculiarities on the pre-main sequence would provide new constraints 
on the timescales on which, and the conditions under which, these peculiarities evolve.    

A few detections of chemical peculiarities in very young stars near or on the zero-age main sequence 
provide tantalising hints that chemical peculiarities may be common in magnetic Herbig Ae/Be stars.  
Some notable cases of this are V380 Ori \citep{Alecian2009-v380ori}, 
HD 72106 A \citep{Folsom2008-hd72106}, NGC 6611 W601 \citep{Alecian2008-ClusterHAeBeLetter}, 
and perhaps NGC 2244-334 \citep{Bagnulo2004-B_NGC2244-334} which is a very young main sequence star.  
Recent modelling by \citet{Vick2011-model-pms-peculiarities} of chemical diffusion 
in the presence of modest mass loss suggest it is possible for chemical peculiarities 
to form during the pre-main sequence phase.  
Interestingly, \citet{Cowley2010-hd101412-lambdaBoo} found a different form of chemical peculiarities, 
$\lambda$ Bo\"otis peculiarities, in the Herbig Ae star HD 101412.  

$\lambda$ Bo\"otis stars are mostly main sequence A stars, 
with strong underabundances of many metals, particularly iron peak elements.  
Lighter elements, specifically C, N, O, S, and in some cases Na, have normal abundances, 
while iron peak elements are usually depleted by $\sim$1 dex 
\citep[e.g.][]{Venn1990-lambdaBoo-accreation-abun,Andrievsky2002-lambdaBoo-abun, Heiter2002-lambdaBoo-abun}.  
This is in contrast to almost all other chemically peculiar A and B stars, which are characterised 
by strong overabundances of iron peak elements. 
$\lambda$ Boo stars have not been found to have magnetic fields, 
unlike Ap and Bp stars \citep{Bohlender1990-lambdaBoo-non-magnetic}. 
$\lambda$ Boo stars have a distribution of \vs\, values that is the same as for 
normal A and B stars \citep[e.g.][]{Abt1995-Astar-vsini}, unlike Ap or Am stars.  

The cause of the peculiar abundances seen in $\lambda$ Boo stars remains unknown, 
though a number of hypotheses have been suggested.  The most popular hypothesis 
is that the peculiarities are a result of a selective accretion 
process \citep{Venn1990-lambdaBoo-accreation-abun,Waters1992-selective-accreation}.  
In this scenario, gas depleted in heavier elements is accreted preferentially, building up a layer 
of relative underabundance in heavier elements at the surface of the star.  
A proposed mechanism for this suggests that heavier elements are bound into dust grains,  
which are then preferentially driven away from the star by its radiation, 
while gas, which is depleted in heavier elements as a consequence of dust formation, 
is more readily accreted \citep[e.g.][]{Andrievsky2000-lambdaBoo-dust-gas-decoupeling}.
Thus, because of the lack of significant convective mixing 
in the atmospheres of A-type stars, a layer that is underabundant in heavier elements 
is quickly built up at the surface of the star.  

A variation on this hypothesis is that the $\lambda$ Boo peculiarities result from selective 
accretion of gas in the interstellar medium, rather than pre-existing circumstellar material 
\citep{Kamp2002-lambdaBoo-ism-cloud}. 
In this scenario the star passes through a diffuse interstellar cloud, 
in which heavier elements are already preferentially bound into dust grains.  
The star then accretes the metal poor gas, while driving away the dust by radiation pressure.  

Other hypotheses have been put forward to explore the origins of $\lambda$ Boo stars.  
For example it has been suggested that $\lambda$ Boo stars 
are actually spectroscopic binaries \citep{Faraggiana2005-lambdaBoo-binary}, 
thus the apparent under abundances would be due to continuum emission from a secondary, 
but this hypothesis is largely discounted \citep{Stutz2006-lambdaBoo-not-binary}. 
Radiatively driven atomic diffusion, which is the most widely accepted mechanism for forming 
peculiarities in most chemically peculiar A and B stars, 
has been considered for $\lambda$ Boo stars with the addition of 
mass loss \citep{Michaud1986-lambdaBoo-diffusion-massloss}. 
However, any turbulent or rotational mixing impedes the process sufficiently to prevent it from forming
$\lambda$ Boo peculiarities \citep{Charbonneau1993-lambdaBoo-diffusion-massloss}.  
Since $\lambda$ Boo stars rotate with similar \vs\, to normal A stars, 
they potentially can have significant rotational mixing.  
Thus atomic diffusion is generally considered to be insufficient to cause $\lambda$ Boo peculiarities. 

Numerical models of $\lambda$ Boo star atmospheres including the accretion of material 
depleted in heavier elements, together with atomic diffusion, have been made 
by \citet{Turcotte1993-accreation-diffusion} and \citet{Turcotte2002-lambdaBoo-mixing-model}.  
They find that those models can produce $\lambda$ Boo peculiarities quickly ($\sim$0.1 Myr) 
for gas accretion rates of $\sim$$10^{-13}$ \acc, 
but the peculiarities in these models do not last long after accretion has stopped ($\sim$1 Myr).  
The peculiarities in these models appear to persist despite rotational mixing 
\citep{Turcotte1993-accreation-diffusion}, but larger surface convection zones 
require larger accretion rates to produce the same peculiarities 
\citep{Turcotte2002-lambdaBoo-mixing-model}.  

In order to help address the question of $\lambda$ Boo star formation, 
and to investigate when Ap/Bp stars develop chemical peculiarities, 
we have analysed high resolution spectra of 20 Herbig Ae/Be stars, and one dusty young star. 
We have determined atmospheric parameters and performed a detailed abundance analysis for 
these stars, with the goal of detecting chemical peculiarities.  
Three of these stars have confirmed magnetic field detections, 
and thus may be the progenitors of Ap/Bp stars.  The other 18 stars 
have been carefully searched for magnetic fields, but no confirmed magnetic fields have been found.

\section{Observations}

Observations for this study were obtained with the Echelle Spectropolarimetric 
Device for the Observation of Stars (ESPaDOnS) instrument at the 
Canada France Hawaii Telescope (CFHT).  ESPaDOnS is a high-resolution \'echelle spectropolarimeter, 
providing nearly continuous wavelength coverage from 3700 to 10500 \AA\, at a resolution of $R=65000$.  
All observations were obtained in spectropolarimetric mode, which provides circularly polarised 
Stokes $V$ spectra, as well as total intensity Stokes $I$ spectra.  
Data reduction was performed with the Libre-ESpRIT \citep{Donati1997-major} package,
which is optimised for the ESPaDOnS instrument, and performs calibrations and optimal spectrum extraction.

The observations were obtained over several years, between 2004 and 2006 as part of an extended campaign 
investigating the presence of magnetic fields in Herbig Ae and Be stars.  
This study is discussed as a whole by \citet{Alecian2012-big-HAeBe-magnetism}, 
and individual results are reported by \citet{Wade2005-HAeBe_Discovery}, 
\citet{Catala2007-HD190073}, \citet{Alecian2008-HD200775}, 
\citet{Alecian2008-ClusterHAeBeLetter}, \citet{Folsom2008-hd72106}, and \citet{Alecian2009-v380ori}.  
The observations that were analysed for this paper are reported in Table \ref{observations_table}. 
Additional observations from \citet{Alecian2012-big-HAeBe-magnetism} were used 
to check for variability in spectral lines.  
The high resolution and high S/N necessary for the detection of magnetic fields in metallic lines also 
makes these observations very well suited to a detailed abundance analysis, 
thus we have an excellent dataset for our study.  

The stars analysed in this study are only a subset of the full set of  
targets observed by \citet{Alecian2012-big-HAeBe-magnetism}. 
The stars selected for our study were chosen to cover a range of \teff\, and \vs\, values.  
The stars were also selected to generally have only modest amounts of emission in their optical spectra, 
which allows for a more complete and accurate abundance analysis.  Preference was given to stars 
that were not clearly spectroscopic binaries, and to observations with a peak S/N over 200.  
Stars with a confirmed magnetic field and a \teff\, below 15000 K were all included 
\citep[except for HD 72106 A, which was studied in detail by][]{Folsom2008-hd72106}.  

Observations of one star (HD~101412) were obtained from the 
Anglo-Australian Telescope (AAT) with the University College London Echelle Spectrograph (UCLES) 
together with SEMPOL polarimeter. 
This instrument consists of a bench mounted cross-dispersed \'echelle spectrograph (UCLES) 
fibre fed from a Cassegrain mounted polarimeter unit (SEMPOL), 
and is fundamentally a similar instrument to ESPaDOnS.  
Details of the instrument can be found in \citet{Semel1993-sempol-early} and \citet{Donati1997-major,
Donati2003-sempol-monitoring-cool-active}, and the observing run that the observations of 
HD 101412 were obtained in is described by \citet{Marsden2011-hd141943-sempol-hd101412}.  
Data reduction and optimal spectral extraction were performed with a generic version 
of ESpRIT \citep{Donati1997-major}. 

Supplemental archival observations from the FORS1 instrument at the Very Large Telescope (VLT) were used 
for examining Balmer lines in many stars, and are listed in Table \ref{observations_fors1_table}.  
FORS1 is a low resolution spectropolarimeter, which produces an observed spectrum entirely in one order.  
This is useful as it reduces the possibility of normalisation errors across Balmer lines. 
The observations were obtained from the study of \citet{Wade2007-HAeBe_survey}, 
and a detailed description of the observations and data reduction techniques can be found therein.  
While \citet{Wade2007-HAeBe_survey} focused on detecting magnetic fields, 
we only concern ourselves with Balmer line profile shapes.  

\begin{table}
\centering
\caption{Table of observations for which an abundance analysis was performed.  
The Heliocentric Julian date, total integration time, and peak signal-to-noise ratio 
(per 1.8~\kms\, spectral pixel) in Stokes $I$ are presented for each observation.  }
\label{observations_table}
\begin{tabular}{cccc}
\hline \hline \noalign{\smallskip}
Object    & HJD          & Integration & Peak S/N \\
          &              &  Time (s)   & $I$      \\
\noalign{\smallskip} \hline \noalign{\smallskip}
HD 17081  & 2453422.7286 &  480        & 878   \\ 
HD 31293  & 2453423.8512 & 2400        & 471   \\
HD 31648  & 2453423.8840 & 2400        & 334   \\
HD 36112  & 2453421.7718 & 2400        & 239   \\
HD 68695  & 2453423.9585 & 2400        & 112   \\
HD 139614 & 2453422.0790 & 3600        & 222   \\
HD 141569 & 2454167.0581 & 5400        & 877   \\
HD 142666 & 2453424.0544 & 3600        & 288   \\
HD 144432 & 2453423.1169 & 3200        & 312   \\
HD 163296 & 2453607.7370 & 1200        & 538   \\
HD 169142 & 2453606.8006 & 2000        & 394   \\
HD 176386 & 2453607.7587 & 1600        & 479   \\
HD 179218 & 2453608.8447 & 1600        & 532   \\
HD 244604 & 2453607.0733 & 3600        & 276   \\
HD 245185 & 2453421.8207 & 4800        & 146   \\
HD 278937 & 2453422.7664 & 4800        & 167   \\
T Ori     & 2453607.1183 & 3600        & 204   \\
HD 101412 & 2454194.9520 & 7200        & 150   \\
HD 190073 & 2454167.1543 & 2700        & 496   \\
V380 Ori  & 2453609.0940 & 4800        & 256   \\
\noalign{\smallskip} \hline \hline
\end{tabular}
\end{table}

\begin{table}
\centering
\caption{FORS1 observations used for Balmer line fitting, obtained by \citet{Wade2007-HAeBe_survey}. 
The Heliocentric Julian date, total integration time, and peak signal-to-noise ratio 
(per 64~\kms\, spectral pixel) in Stokes $I$ are presented for each observation.  }
\label{observations_fors1_table}
\begin{tabular}{cccc}
\hline \hline \noalign{\smallskip}
Object    & HJD          & Integration & Peak S/N \\
          &              &  Time (s)   & $I$      \\
\noalign{\smallskip} \hline \noalign{\smallskip}
HD 17081  & 2453331.5530 &    12       &  3800    \\
HD 31293  & 2453331.6774 &   140       &  4050    \\
HD 31648  & 2453331.6929 &   319       &  4225    \\
HD 36112  & 2453331.7097 &   320       &  3725    \\
HD 68695  & 2453332.7761 &   640       &  1950    \\
HD 141569 & 2453062.8424 &  1295       &  5650    \\
HD 142666 & 2453063.8535 &  2800       &  2775    \\
HD 144432 & 2453062.8983 &  2100       &  4575    \\
HD 244604 & 2453331.6217 &  1600       &  4150    \\
HD 245185 & 2453331.6527 &  1600       &  3925    \\
HD 278937 & 2453330.6795 &  1860       &  2550    \\
T Ori     & 2453332.6371 &  1440       &  1900    \\
HD 101412 & 2453062.7977 &  1350       &  1575    \\
HD 190073 & 2453330.5153 &   215       &  3700    \\
V380 Ori  & 2453330.7576 &   480       &  1450    \\
\noalign{\smallskip} \hline \hline
\end{tabular}
\end{table}

\section{Fundamental parameters}

Temperature and surface gravity were first measured by fitting Balmer lines, 
and then confirmed and, if possible, refined by enforcing local thermodynamic equilibrium (LTE) 
ionisation and excitation balances in the metallic line spectra 
of the stars during the modelling procedure.  

Balmer line fitting was performed on FORS1 spectra when available, since it is easier to 
normalise broad spectral features when they are recorded entirely in one spectral order.  
When FORS1 spectra were not available, ESPaDOnS spectra were used and careful attention 
was paid to the normalisation of Balmer lines.  
Continuum points well outside the wings on both sides of a Balmer line were carefully chosen, 
a low order polynomial was then fit through these points, 
and the polynomial was checked to ensure it varied only a small amount over the Balmer line.  
In the ESPaDOnS spectra, careful data reduction and 
the absence of instrumental artifacts produced an excellent correspondence between 
spectral orders in regions where the orders overlap.  These overlap regions were used 
to evaluate and minimise errors in normalisation across spectral orders. 
For stars with both FORS1 and ESPaDOnS spectra available 
the normalised Balmer lines were compared to ensure the accuracy of the 
normalisation procedure for ESPaDOnS spectra.  A good agreement between 
the normalised ESPaDOnS and FORS1 Balmer lines was consistently found.  
In all cases potential normalisation errors are included in the quoted uncertainties 
of physical parameters.  

Careful attention was paid to avoid contamination of the Balmer lines by emission.  
In our observations the cores of most Balmer lines were partially infilled by emission.  
H$\alpha$ was often heavily contaminated by emission, and in many stars was avoided entirely.  
Despite this, the wings of H$\gamma$ and H$\delta$ generally remain free from emission.  
The wavelength range of possible contamination in the higher Balmer lines can usually 
be assessed by the more prominent emission in H$\alpha$.  
As a consequence of these complications, the fitting procedure was performed by hand 
and rather conservative uncertainties were adopted.  
Synthetic Balmer lines were computed from ATLAS9 \citep{Kurucz1993-ATLAS9etc} model 
atmospheres using solar chemical abundances.  

Problematically, for many of our stars there is a substantial degeneracy in 
effective temperature and surface gravity when fitting only the wings of Balmer lines.  
As a result there is a large covariance between \teff\, and \lgg,
and our uncertainties often represent long ellipses in the two dimensional parameter space.  
Typically these degenerate regions have a slope of +250 K in \teff\, for +0.1 dex in \lgg, 
though the slope generally becomes steeper towards hotter temperatures.  

When good quality fits to the observations were obtainable, fitting of \teff\, and \lgg\, 
for the metallic spectra was included in our determination.  
The fitting procedure is described in Sect.~\ref{Abundance Analysis}. 
This method relies on the ionisation and excitation balance in the star 
assuming LTE, but a number of potential problems 
can occur.  Lines with any emission infilling must be avoided.  
Poor quality atomic data, undiagnosed line blends and 
non-LTE effects may also be problematic.  
This also requires lines with a wide range of excitation energies and multiple ionisation states.  
Consequently, this technique could not successfully be applied to all stars in our sample.  
In a number of cases the excitation potential could not provide a useful constraint, 
but the ionisation potential was still useful.  In these cases the constraint provided by 
Balmer line fits, combined with the ionisation balance could still provide fairly 
precise \teff\, and \lgg\, values.  

In order to place the stars of our sample on the H-R diagram, 
we needed to determine accurate luminosities.  Photometric Johnson $V$ magnitudes 
and $B-V$ colours were derived from the Hipparcos $V$ magnitudes using the 
conversion from \citet{Hipparcos1997}.  For stars without Hipparcos observations, 
Johnson magnitudes and colours were taken from \citet{Vieira2003-HAeBe_ID}, 
\citet{Herbst1999-haebe-photometry}, and \citet{deWinter2001-HAeBe-phot}.  
These magnitudes and colours are consistent with those used by \citet{Alecian2012-big-HAeBe-magnetism}. 

Distance were derived using Hipparcos parallaxes, when available, 
from the new reduction by \citet{van_Leeuwen2007-Hipparcos_validation}.  
When no parallax was available, the star was associated 
with an OB association or star forming region, and a literature distance to that was taken. 
Associations were taken from \citet{Alecian2012-big-HAeBe-magnetism}, which agree with 
\citet{Vieira2003-HAeBe_ID}.  Uncertainties for these distances were 
taken to be the spatial extent of the associations, which were estimated from
their angular size on the sky, assuming the associations are approximately spherical.
For HD 169142 no association could reliably be 
made, and thus a literature photometric distance from \citet{Sylvester1996-vega-phot-multiwave} 
was used, and generous uncertainties were assumed.  
Literature sources for individual distances, along with the distances 
themselves, can be found in Table, \ref{fund_param_table}.

Using these distances, bolometric luminosities were calculated. 
$E(B-V)$ was calculated using the intrinsic colours of 
\citet{Kenyon1995-colors-bolometric}, and total extinction was calculated 
using an $R_V$ of 5, as suggested for HAeBe stars by \citet{Hernandez2004-Rv-extinction}.  
The bolometric correction of \citet{Balona1994-bolometric-corr} was used. 

\begin{figure*}
\centering
\includegraphics[width=6.0in]{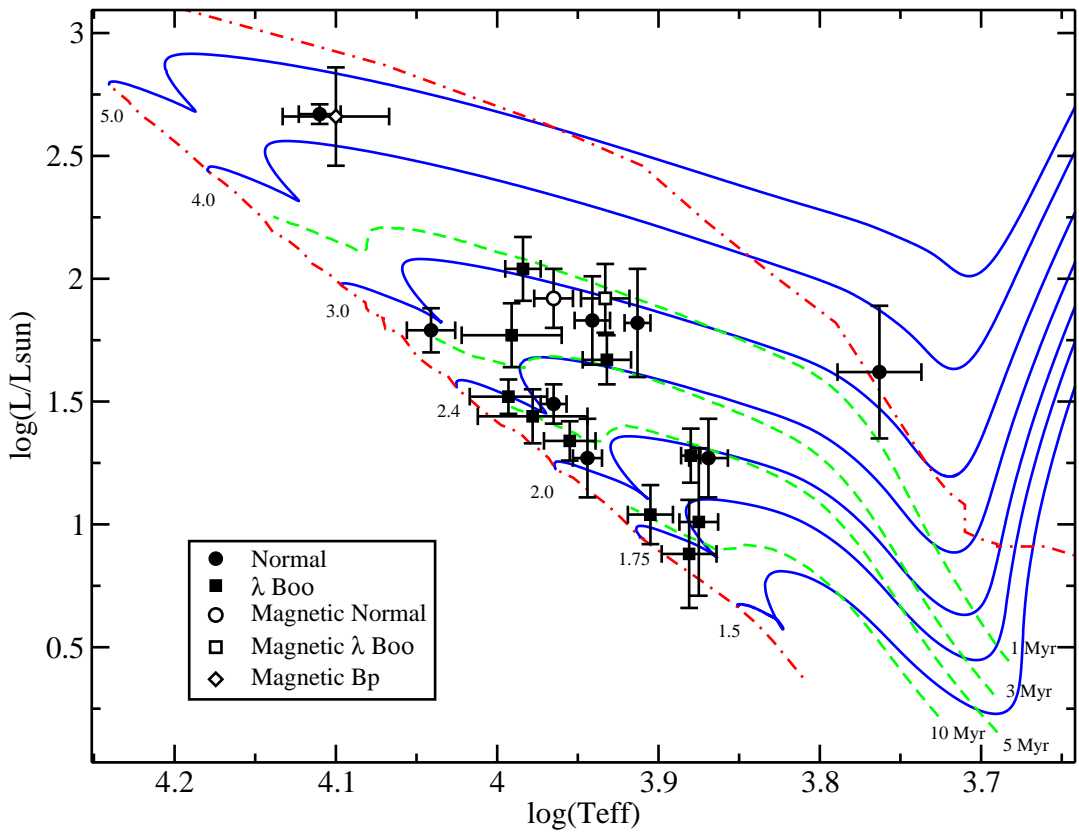}
\caption{H-R diagram for the stars in this study.  
Evolutionary tracks (solid lines) are labelled by mass in solar masses.  
Isochrones (dashed lines) are labelled by age in Myr.  The birth line (right dot-dashed line) 
for an accretion rate of $10^{-5}$ \acc, and the zero-age main sequence line 
(ZAMS, left dot-dashed line) are also shown. Circles are chemically normal stars, 
squares are $\lambda$ Boo stars, the diamond is the possible Bp star V380 Ori A.  
Open symbols are stars with confirmed magnetic field detections. }
\label{h-r_diagram}
\end{figure*}

Using our luminosities and temperatures, we placed the stars on the 
H-R diagram, as shown in Fig. \ref{h-r_diagram}.  Pre-main sequence 
evolutionary tracks and isochrones were calculated with the {\sc cesam} 
\citep{Morel1997-CESAM} evolutionary code, assuming solar metallicities. 
The birth line, the locus of points on the H-R diagram where a star 
becomes optically visible, was taken from \citet{Palla1993-PMS-Evol}, 
for an accretion rate of $10^{-5}$ \acc.  By comparison to these evolutionary tracks 
and isochrones, we determined stellar masses, and ages 
with respect to the birthline.  These values, along with fractional 
pre-main sequence ages ($\tau$) and radii, are presented in Table \ref{fund_param_table}.
The uncertainties on mass and age for a star were based on the range of 
evolutionary tracks and isochrones that intersect the ellipse on the 
H-R diagram described by the uncertainties in the star's luminosity and \teff. 
Note that the choice of a birthline may introduce a further systematic uncertainty into our ages.  
For example, using the birthline of \citet{Behrend2001-evol-birthline-mass-dependent}, 
with a mass dependent accretion rate, would generally increase our ages by $\sim$0.5 Myr 
(but with the actual increase varying with mass).
While we do not present uncertainties on $\tau$, the values should be considered approximate.

\begin{table*}
\centering
\caption{Fundamental parameters derived for the stars in this study.  
$\tau$ is the fractional pre-main sequence age.  
The `Magnetic' column lists which stars have confirmed magnetic field detections (M), 
and which stars we consider to be non-magnetic (N).  
The references for distances are: 
$^a$\citealt{van_Leeuwen2007-Hipparcos_validation} (Hipparcos);
$^b$\citealt{deZeeuw1999-OB-distances};
$^c$\citealt{Brown1994-Ori-OB1-dist-census};
$^d$\citealt{Dolan2001-distance-lambda-ori};
$^e$\citealt{Vieira2003-HAeBe_ID};
$^f$\citealt{Sylvester1996-vega-phot-multiwave}.
The luminosity for HD 190073 was taken from \citet{Catala2007-HD190073} (indicated by *). 
}
\label{fund_param_table}
\begin{tabular}{cccccccccc}
\hline \hline \noalign{\smallskip}
   ID      & \teff~(K)    & \lgg        & distance (pc)& $\log(L/L_\odot)$& $M/M_\odot$    & $R/R_\odot$  & age (Myr)        & $\tau$& Magnetic \\
\noalign{\smallskip} \hline \noalign{\smallskip}
HD 17081   &$12900\pm400$ & $3.8\pm0.2$ & $120\pm 3^a$ & $2.67 \pm 0.04$ & $4.4 \pm0.2 $ & $4.3\pm0.3$ &$0.3 ^{+0.1}_{-0.1}$& $0.48$ & N \\
HD 31293   &$9800 \pm700$ & $3.9\pm0.3$ & $139\pm19^a$ & $1.77 \pm 0.13$ & $2.55\pm0.2 $ & $2.7\pm0.5$ &$2.6 ^{+0.5}_{-0.6}$& $0.65$ & N \\
HD 31648   &$8800 \pm190$ & $4.1\pm0.2$ & $137\pm25^a$ & $1.27 \pm 0.16$ & $2.1 \pm0.25$ & $1.9\pm0.4$ &$6.2 ^{+4.0}_{-2.0}$& $0.84$ & N \\
HD 36112   &$8190 \pm150$ & $4.1\pm0.4$ & $279\pm70^a$ & $1.82 \pm 0.22$ & $2.8 \pm0.5 $ & $4.0\pm1.0$ &$1.5 ^{+1.5}_{-1.0}$& $0.52$ & N \\
HD 68695   &$9000 \pm300$ & $4.3\pm0.3$ & $410\pm36^b$ & $1.34 \pm 0.08$ & $2.2 \pm0.15$ & $1.9\pm0.2$ &$5.2 ^{+3.0}_{-1.0}$& $0.76$ & N \\
HD 139614  &$7600 \pm300$ & $3.9\pm0.3$ & $147\pm37^b$ & $0.88 \pm 0.22$ & $1.7 \pm0.1 $ & $1.6\pm0.4$ &$13.5^{+11 }_{-5.0}$& $0.82$ & N \\
HD 141569  &$9800 \pm500$ & $4.2\pm0.4$ & $116\pm 8^a$ & $1.52 \pm 0.07$ & $2.4 \pm0.2 $ & $2.0\pm0.3$ &$4.0 ^{+1.5}_{-1.0}$& $0.83$ & N \\
HD 142666  &$7500 \pm200$ & $3.9\pm0.3$ & $145\pm18^b$ & $1.28 \pm 0.11$ & $1.95\pm0.15$ & $2.5\pm0.3$ &$6.0 ^{+1.5}_{-1.5}$& $0.63$ & N \\
HD 144432  &$7400 \pm200$ & $3.9\pm0.3$ & $160\pm29^a$ & $1.27 \pm 0.16$ & $1.95\pm0.2 $ & $2.6\pm0.5$ &$5.5 ^{+2.5}_{-1.5}$& $0.58$ & N \\
HD 163296  &$9200 \pm300$ & $4.2\pm0.3$ & $119\pm11^a$ & $1.49 \pm 0.08$ & $2.3 \pm0.1 $ & $2.2\pm0.2$ &$4.1 ^{+1.0}_{-0.2}$& $0.71$ & N \\
HD 169142  &$7500 \pm200$ & $4.3\pm0.2$ & $145\pm50^f$ & $1.01 \pm 0.30$ & $1.7 \pm0.2 $ & $1.9\pm0.7$ &$8.5 ^{+16 }_{-3.0}$& $0.52$ & N \\
HD 176386  &$11000\pm400$ & $4.1\pm0.3$ & $128\pm13^a$ & $1.79 \pm 0.09$ & $2.8 \pm0.2 $ & $2.2\pm0.3$ &$2.7 ^{+1.0}_{-1.0}$& $0.93$ & N \\
HD 179218  &$9640 \pm250$ & $3.9\pm0.2$ & $254\pm38^a$ & $2.04 \pm 0.13$ & $3.1 \pm0.3 $ & $3.7\pm0.6$ &$1.1 ^{+0.7}_{-0.6}$& $0.55$ & N \\
HD 244604  &$8700 \pm220$ & $4.0\pm0.2$ & $380\pm79^c$ & $1.83 \pm 0.18$ & $2.75\pm0.4 $ & $3.6\pm0.8$ &$1.9 ^{+1.1}_{-1.0}$& $0.61$ & N \\
HD 245185  &$9500 \pm750$ & $4.0\pm0.4$ & $450\pm50^d$ & $1.44 \pm 0.11$ & $2.3 \pm0.2 $ & $1.9\pm0.4$ &$5.5 ^{+2.0}_{-2.0}$& $0.90$ & N \\
HD 278937  &$8000 \pm250$ & $4.1\pm0.2$ & $318\pm43^b$ & $1.04 \pm 0.12$ & $1.8 \pm0.1 $ & $1.7\pm0.3$ &$9.5 ^{+5.0}_{-3.0}$& $0.76$ & N \\
T Ori      &$8500 \pm300$ & $4.2\pm0.3$ & $380\pm44^c$ & $1.67 \pm 0.10$ & $2.45\pm0.15$ & $3.1\pm0.4$ &$3.3 ^{+1.2}_{-1.3}$& $0.75$ & N \\
HD 101412  &$8600 \pm300$ & $4.0\pm0.5$ & $600\pm100^e$& $1.92 \pm 0.14$ & $3.0 \pm0.3 $ & $4.2\pm0.8$ &$1.2 ^{+0.8}_{-0.7}$& $0.50$ & M \\
HD 190073  &$9230 \pm260$ & $3.7\pm0.3$ & $>340    ^a$ & $1.92 \pm 0.12$*& $2.9 \pm0.3 $ & $3.6\pm0.5$ &$1.6 ^{+0.7}_{-0.6}$& $0.62$ & M \\
V380 Ori A &$12600\pm1000$& $4.0\pm0.5$ & $398\pm91^c$ & $2.66 \pm 0.20$ & $4.4 \pm0.7 $ & $4.5\pm1.2$ &$0.4 ^{+0.6}_{-0.3}$& $0.55$ & M \\
V380 Ori B &$5800 \pm350$ & $4.1\pm0.3$ & $398\pm91^c$ & $1.62 \pm 0.27$ & $3.3 \pm0.8 $ & $6.4\pm2.2$ &$0.0 ^{+1.5}_{-1.5}$& $0.0 $ & N \\
\noalign{\smallskip} \hline \hline
\end{tabular}
\end{table*}

\section{Abundance analysis}
\label{Abundance Analysis}

The abundance analysis and determination of \vs\, and microturbulence was performed by directly 
fitting synthetic spectra to the observations.  Model spectra were calculated using 
the {\sc Zeeman} spectrum synthesis code \citep{Landstreet1988-Zeeman1,Wade2001-zeeman2_etc}.  
This code performs polarised radiative transfer under the assumption of LTE.  
Optimisations to the code for stars with negligible magnetic fields have been included.  
A Levenberg-Marquardt $\chi^2$ minimisation procedure \citep[e.g.][]{numerical-recipes-Fortran} 
was used to determine best fit parameters.  

The optimisations to the {\sc Zeeman} spectrum synthesis routine exploit symmetries 
in a non-magnetic chemically homogeneous model of a star to dramatically reduce the 
amount of computation required. 
As in the original version of {\sc Zeeman}, the visible disk of the star 
is divided into a number of surface elements.  
In this version it is assumed that the local emergent spectra only vary 
with projected radial distance from the centre of the disk of the star. 
Thus radiative transfer is only performed for a set of surface elements 
with different radial distances, and the results are reused for different angular positions, 
which minimises the number of times radiative transfer must be performed.  
The optimisations also assume the local line absorption (Voigt) 
and anomalous dispersion (Faraday-Voigt) profiles only vary vertically through the atmosphere, 
and not with position on the stellar disk, 
substantially reducing the number of times these profiles must be calculated.  
In this version of the code, limb darkening is still calculated directly, by performing 
radiative transfer at different radial positions on the disk of the star.  
Rotational broadening is also calculated directly by summing local 
emergent spectra that have been Doppler shifted across the disk of the star.  

Detailed comparisons of the optimised code to the original 
version have been performed.  For the same input parameters, with no magnetic field, 
identical spectra are produced.  The original code has been checked against other 
spectrum synthesis codes in detail by \citet{Wade2001-zeeman2_etc}, 
with a very good agreement found.  Similarly, the optimised version of 
{\sc Zeeman} has been checked against {\sc Synth3} \citep{Kochukhov2007-synth3}
and again a good agreement was found. 

The Levenberg-Marquardt $\chi^2$ minimisation procedure \citep{numerical-recipes-Fortran} 
provides a robust and efficient routine for iteratively fitting synthetic spectra to observations.  
Best fit models from this routine have been compared to fits obtained by hand for a wide 
variety of stars, and good agreement between the two methods has been consistently found.  

Input atomic data were taken from the Vienna Atomic Line Database (VALD) \citep{Kupka1999-VALD}, 
using an `extract stellar' request and solar chemical abundances.  
Updated line lists with peculiar abundances were extracted as necessary.  
Model atmospheres were computed using the ATLAS9 code \citep{Kurucz1993-ATLAS9etc} with solar abundances.  
Model atmospheres were calculated on a grid in steps of 250 K in \teff\, and 0.5 in \lgg, 
then interpolated linearly when more precise values were desired. 
Interpolation on this grid of models introduces less then 1\% relative error to the line depths in the resulting spectrum. 

Spectra were initially fit simultaneously for chemical abundances, \vs, microturbulence, 
and radial velocity, using the best fit \teff\, and \lgg\, determined from the Balmer lines.  
Then, if a good constraint could be obtained, the fits were repeated including \teff\, 
and \lgg\, as free parameters.  
If this produced a good fit with well constrained parameters, then these  \teff\, and \lgg\, 
values were used instead of the initial Balmer line fit values.  
If only one of \teff\, and \lgg\, was well constrained from the metallic line fit, 
then the fit was repeated using the Balmer line value for the unconstrained parameter. 
We considered a parameter to be well constrained if the fitting routine reliably converged to the same value 
for different initial conditions, and if that value was consistent with the Balmer line profiles.  
The fits were generally performed on 5 independent regions $\sim$500 \AA\, long. 
The approximate wavelength ranges usually were 4400-4800, 4900-5500, 5500-6000, 6000-6500, 
and 6600-7600 \AA, and occasionally 4150-4280 \AA, 
with significant gaps due to non-photospheric features.  
These windows typically contained several hundred spectral lines 
(the number varying greatly with wavelength and \teff),
however many of these lines are very weak and hence do not provide significant constraints 
on the best fit abundances or atmospheric parameters.  
The exact wavelength ranges varied between stars, due to varying regions of emission 
and different Balmer line widths. 

Microturbulence and \vs\, were determined by $\chi^2$ minimisation, 
simultaneously with the other stellar parameters, using the entire spectral $\sim$500 \AA\, window.  
This method relies on resolved, rotationally broadened spectral lines 
for constraining \vs, and a range of both weak and strong lines 
for constraining microturbulence. 

We verified this method of determining \teff\, and \lgg\, by analysing 
well studied stars, and comparing to literature results.  In this study we have 
analysed $\pi$ Cet, and compared our results with the very precise study 
by \citet{Fossati2009-normal-star}.  
\citet{Fossati2009-normal-star} used photometry and spectral energy distributions, 
as well as Balmer lines, and excitation and ionisation balances when determining their parameters.  
They also used different software tools to perform their modelling, thus their results are truly independent.  
Our results are fully consistent with theirs, as discussed in Sect. \ref{pi-Cet}, 
which demonstrates the accuracy of our methodology.  

For this method to be successful, lines contaminated by emission must be avoided.  
While a line entirely in emission is easily identified, lines with small amounts of 
emission infilling are harder to identify.  When multiple observations were available, 
variability in the emission could be used to identify lines with emission infilling.  
If only one observation was available, special attention was paid to 
lines with low excitation potentials and large oscillator strengths, 
such lines being more likely to contain emission.  
If inconsistent fits were obtained between normal and low excitation potential lines, 
and no other clear explanation for the inconsistency could be found 
(such as an error in \teff\, or the atomic data), 
the low excitation potential lines were considered to likely contain emission 
and were excluded from the final fit. 
Additionally, the shapes of line profiles were examined and, if they could not be 
reproduced by the synthetic spectra, they were considered to be contaminated by 
a non-photospheric component. 
The presence of veiling in our spectra is assessed in detail in Section \ref{Possible Veiling}, 
and we find it is not present at a significant level. 

The use of multiple spectral windows is valuable, as it allows us to take 
the average and standard deviation of abundances determined across the windows.  
The standard deviation in particular provides a reasonable estimate of 
the uncertainty of derived parameters.  Since line-to-line abundances scatter 
is introduced by errors in the atomic data as well as errors in the model atmosphere, 
these effects are included in the standard deviation.  
The standard deviation may underestimate the impact of such errors, 
but provides a robust and easily understandable uncertainty estimate.  
The use of multiple spectral windows also allows us to verify the 
atmospheric parameters derived from metal line fits.  
If the parameters differ substantially from window to window, 
they are likely poorly constrained in some if not all windows, 
in which case we can rely more heavily on the 
parameters from Balmer line fitting.  

For chemical elements with fewer than four lines providing good constraints on 
the abundance in any spectral window (including all cases of elements with 
lines in only one or two spectral windows) 
the uncertainties were estimated by eye rather than using a standard deviation.  
These uncertainties were chosen to include line-to-line scatter, 
uncertainties in the atmospheric parameters, and potential normalisation errors.

\begin{figure*}
\centering
\includegraphics[width=5.4in]{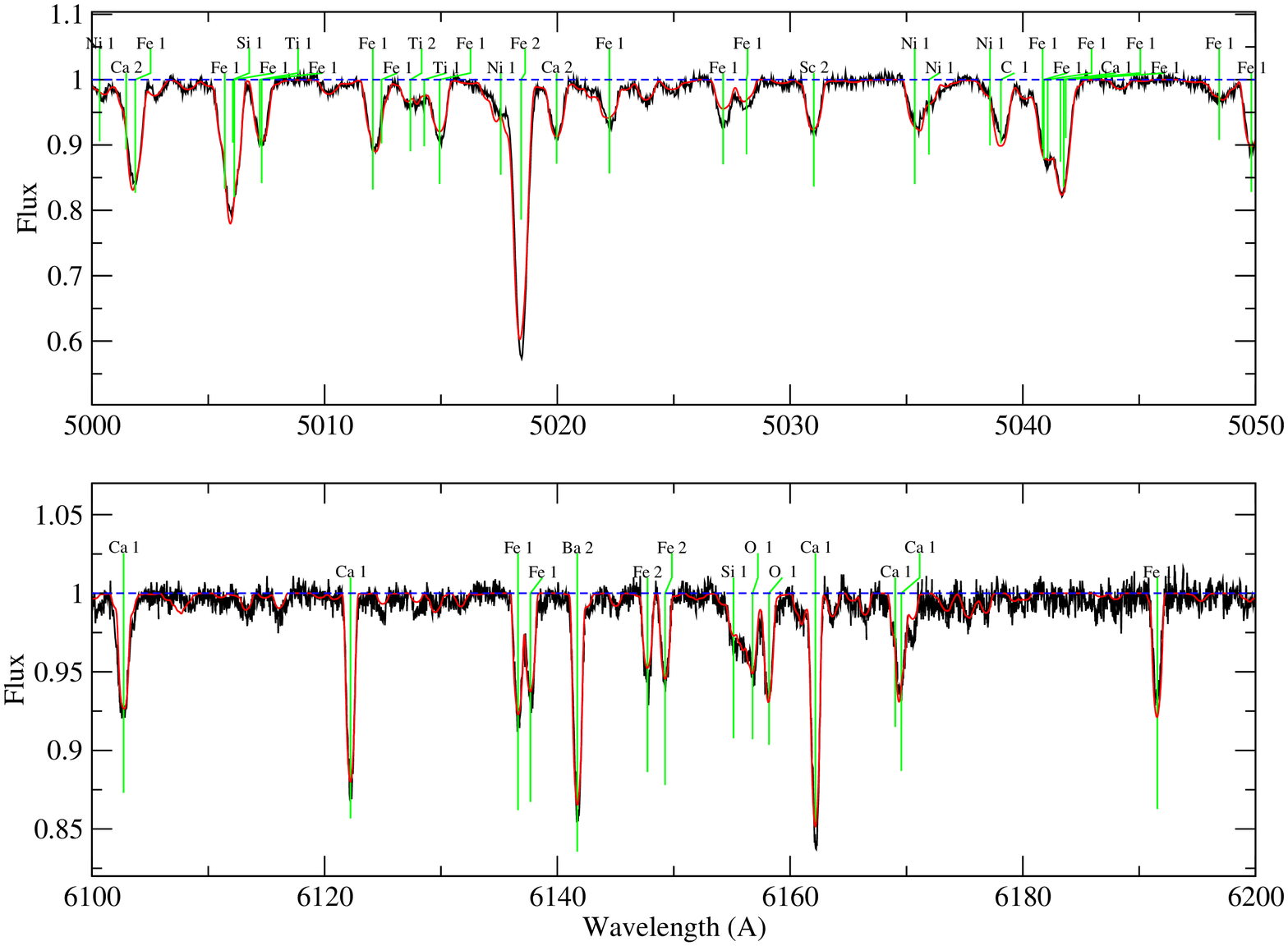}
\caption{Comparison of the observed spectrum (jagged line) to the 
best fit synthetic spectrum (smooth line) for HD 139614.  
Two independent wavelength regions are presented.  
Lines have been labelled by their major contributing species. }
\label{fit-in-text}
\end{figure*}

Examples of typical best fit synthetic spectra compared to the observed spectrum of HD 139614 
in the 5000-5050 \AA\, and 6100-6200 \AA\, windows are presented in Fig. \ref{fit-in-text}.  
Final averaged best fit abundances and atmospheric parameters are presented 
in Table \ref{abun-tab}, together with uncertainties.  Elements with abundances 
based on less than four lines are indicated with an asterisk in the table.  

We find 11 stars with varying degrees of $\lambda$ Boo peculiarity, 
one star (V380 Ori A) that appears to be a weak Ap/Bp star, and 9 stars that are chemically normal 
(although results for V380 Ori B are very uncertain).  
The properties of the individual stars are discussed in detail in the Appendix.  
The final best fit abundances are plotted relative to solar abundances from 
\citet{Grevesse2005-solar_abun} in Figures \ref{abunplots1}, \ref{abunplots2}, \ref{abunplots3},
and \ref{abunplots4}.  
Illustrations of our fits for all of the stars in our study 
can be found in Figures \ref{fit-appendix} to \ref{fit-v380ori}.

\begin{centering}
\begin{table*}
\caption[]{Best fit parameters for the stars in our sample. Chemical abundances are in units of $\log(N_{X}/N_{tot})$. 
Elements marked with an asterisk are based on less then $\sim$4  useful lines 
and have uncertainties estimated by eye.  Microturbulence is given by \vmic. }
\scriptsize{
\begin{tabular}{lccccccccccc}
\hline\hline
              &           HD 17081 &           HD 31293 &           HD 31648 &           HD 36112 &           HD 68695 &          HD 139614 &          HD 141569 &  Solar    \\
              &        ($\pi$ Cet) &           (AB Aur) &                    &                    &                    &                    &                    &           \\
\hline                                                                                                                  
   \teff\,(K) &   $12900 \pm 400 $ &    $9800 \pm 700 $ &    $8800 \pm 190 $ &    $8190 \pm 150 $ &    $9000 \pm 300 $ &   $7600  \pm 300 $ &    $9800 \pm 500 $ &           \\
  \lgg\,(cgs) &     $3.8 \pm 0.2 $ &     $3.9 \pm 0.3 $ &     $4.1 \pm 0.2$  &     $4.1 \pm 0.4$  &     $4.3 \pm 0.3 $ &    $3.9  \pm 0.3 $ &     $4.2 \pm 0.4 $ &           \\
  \vs\,(\kms) &    $20.9 \pm 1.2 $ &       $116 \pm 9 $ &   $101.2 \pm 1.7 $ &     $57.8 \pm 1.0$ &        $51 \pm 4 $ &   $25.6  \pm 0.4 $ &       $222 \pm 7 $ &           \\
\vmic\,(\kms) &     $1.7 \pm 1.0 $ &         $\leq 4  $ &      $3.2 \pm 1.1$ &    $2.97 \pm 0.24$ &     $1.3 \pm 1.1 $ &    $3.7  \pm 0.4 $ &         $\leq 2  $ &           \\
\hline                                                                                                                  
           He &   $-0.91 \pm 0.14$ &  $-1.24 \pm 0.20$* &     $            $ &           $      $ &           $      $ &    $             $ &  $-1.21 \pm 0.37$* &  $-1.11 $ \\
            C &     $\leq -3.4  $* &   $-3.33 \pm 0.22$ &   $-3.67 \pm 0.07$ &   $-3.61 \pm 0.16$ &   $-3.27 \pm 0.18$ &   $-3.75 \pm 0.22$ &   $-3.63 \pm 0.29$ &  $-3.65 $ \\
            N &  $-4.03 \pm 0.15$* &   $-3.9 \pm 0.3 $* &     $            $ &           $      $ &   $-3.7 \pm 0.3 $* &   $-4.17 \pm 0.13$ &   $-4.0 \pm 0.3 $* &  $-4.26 $ \\
            O &   $-3.16 \pm 0.13$ &   $-3.27 \pm 0.20$ &  $-3.28 \pm 0.05$* &  $-3.18 \pm 0.10$* &   $-3.17 \pm 0.10$ &  $-3.33 \pm 0.10$* &   $-3.05 \pm 0.10$ &  $-3.38 $ \\
           Ne &   $-3.76 \pm 0.28$ &           $      $ &           $      $ &           $      $ &           $      $ &    $             $ &    $             $ &  $-4.20 $ \\
           Na &  $-5.30 \pm 0.10$* &     $            $ &  $-5.72 \pm 0.25$* &  $-5.58 \pm 0.15$* &   $-6.2 \pm 0.4 $* &   $-6.14 \pm 0.12$ &     $\leq -5.2  $* &  $-5.87 $ \\
           Mg &   $-4.43 \pm 0.15$ &   $-4.89 \pm 0.33$ &   $-4.12 \pm 0.13$ &   $-4.23 \pm 0.14$ &   $-5.11 \pm 0.14$ &   $-4.74 \pm 0.07$ &   $-4.90 \pm 0.11$ &  $-4.51 $ \\
           Al &   $-5.81 \pm 0.17$ &           $      $ &           $      $ &           $      $ &           $      $ &    $             $ &   $-5.6 \pm 0.5 $* &  $-5.67 $ \\
           Si &   $-4.55 \pm 0.14$ &   $-4.76 \pm 0.22$ &   $-4.29 \pm 0.19$ &   $-4.58 \pm 0.11$ &   $-5.36 \pm 0.15$ &   $-5.10 \pm 0.13$ &   $-4.99 \pm 0.31$ &  $-4.53 $ \\
            P &   $-6.43 \pm 0.14$ &            $     $ &           $      $ &           $      $ &           $      $ &         $        $ &           $      $ &  $-6.68 $ \\
            S &   $-4.99 \pm 0.08$ &            $     $ &   $-4.6 \pm 0.3 $* &  $-4.79 \pm 0.16$* &   $-4.6 \pm 0.3 $* &   $-5.18 \pm 0.15$ &           $      $ &  $-4.90 $ \\
           Ar &   $-5.6 \pm 0.3 $* &            $     $ &           $      $ &           $      $ &           $      $ &         $        $ &           $      $ &  $-5.86 $ \\
            K &          $       $ &           $      $ &           $      $ &           $      $ &           $      $ &  $-7.29 \pm 0.25$* &     $            $ &  $-6.96 $ \\
           Ca &   $-5.7 \pm 0.2 $* &   $-6.10 \pm 0.17$ &   $-5.38 \pm 0.23$ &   $-5.59 \pm 0.16$ &   $-6.47 \pm 0.23$ &   $-6.18 \pm 0.16$ &   $-6.34 \pm 0.38$ &  $-5.73 $ \\
           Sc &          $       $ &   $-9.04 \pm 0.34$ &   $-8.72 \pm 0.15$ &   $-8.90 \pm 0.26$ &   $-9.49 \pm 0.26$ &   $-9.39 \pm 0.15$ &   $-9.18 \pm 0.37$ &  $-8.87 $ \\
           Ti &   $-7.37 \pm 0.15$ &   $-7.47 \pm 0.27$ &   $-6.82 \pm 0.09$ &   $-6.98 \pm 0.20$ &   $-7.71 \pm 0.25$ &   $-7.51 \pm 0.14$ &   $-7.74 \pm 0.32$ &  $-7.14 $ \\
            V &          $       $ &           $      $ &           $      $ &   $-7.8 \pm 0.4 $* &     $            $ &   $-8.4 \pm 0.5 $* &     $            $ &  $-8.04 $ \\
           Cr &   $-6.57 \pm 0.16$ &   $-6.67 \pm 0.27$ &   $-6.09 \pm 0.17$ &   $-6.30 \pm 0.17$ &   $-6.98 \pm 0.24$ &   $-6.80 \pm 0.11$ &   $-7.31 \pm 0.31$ &  $-6.40 $ \\
           Mn &   $-6.6 \pm 0.3 $* &     $            $ &   $-6.59 \pm 0.08$ &   $-6.70 \pm 0.16$ &   $-6.4 \pm 0.5 $* &   $-7.43 \pm 0.19$ &           $      $ &  $-6.65 $ \\
           Fe &   $-4.70 \pm 0.08$ &   $-4.91 \pm 0.25$ &   $-4.47 \pm 0.13$ &   $-4.49 \pm 0.14$ &   $-5.16 \pm 0.22$ &   $-5.07 \pm 0.13$ &   $-5.25 \pm 0.32$ &  $-4.59 $ \\
           Co &          $       $ &           $      $ &          $       $ &           $      $ &           $      $ &         $        $ &           $      $ &  $-7.12 $ \\
           Ni &   $-5.85 \pm 0.06$ &   $-6.16 \pm 0.27$ &   $-5.66 \pm 0.05$ &   $-5.76 \pm 0.20$ &   $-6.25 \pm 0.32$ &   $-6.33 \pm 0.14$ &  $\leq -6.0     $* &  $-5.81 $ \\
           Cu &          $       $ &     $            $ &           $      $ &      $\leq -7.5 $* &     $            $ &   $-8.6 \pm 0.4 $* &     $            $ &  $-7.83 $ \\
           Zn &          $       $ &     $            $ &           $      $ &   $-7.8 \pm 0.4 $* &     $            $ &   $-8.3 \pm 0.3 $* &     $            $ &  $-7.44 $ \\
           Sr &          $       $ &   $-9.6 \pm 0.3 $* &     $            $ &   $-9.4 \pm 0.3 $* &     $            $ &         $        $ &  $-10.5 \pm 0.5 $* &  $-9.12 $ \\
            Y &          $       $ &     $            $ &   $-9.57 \pm 0.30$ &   $-9.56 \pm 0.17$ &           $      $ &   $-10.23\pm 0.15$ &           $      $ &  $-9.83 $ \\
           Zr &          $       $ &           $      $ &           $      $ &   $-9.3 \pm 0.3 $* &     $            $ &     $            $ &           $      $ &  $-9.48 $ \\
           Ba &          $       $ &   $-9.7 \pm 0.4 $* &  $-9.56 \pm 0.24$* &   $-9.46 \pm 0.29$ &  $-10.3 \pm 0.4 $* &   $-10.29\pm 0.21$ &   $-9.2 \pm 0.8 $* &  $-9.87 $ \\
           La &                    &                    &                    &                    &                    &       $          $ &     $            $ &  $-10.91$ \\
           Ce &                    &                    &                    &                    &                    &   $\leq -10.5   $* &     $            $ &  $-10.34$ \\
           Nd &                    &                    &                    &                    &                    &       $          $ &     $            $ &  $-10.59$ \\
           Eu &                    &                    &                    &                    &                    &         $        $ &     $            $ &  $-11.52$ \\
\hline\hline
              &          HD 142666 &          HD 144432 &          HD 163296 &          HD 169142 &          HD 176386 &          HD 179218 &          HD 244604 &  Solar    \\
\hline                                                                                                                                                                       
   \teff\,(K) &    $7500 \pm 200 $ &    $7400 \pm 200 $ &    $9200 \pm 300 $ &    $7500 \pm 200 $ &   $11000 \pm 400 $ &    $9640 \pm 250 $ &    $8700 \pm 220 $ &           \\
  \lgg\,(cgs) &     $3.9 \pm 0.3 $ &     $3.9 \pm 0.3 $ &     $4.2 \pm 0.3 $ &     $4.3  \pm 0.2$ &     $4.1 \pm 0.3 $ &     $3.9 \pm 0.2 $ &     $4.0 \pm 0.2 $ &           \\
  \vs\,(\kms) &    $68.0 \pm 0.9 $ &    $80.3 \pm 1.0 $ &       $122 \pm 3 $ &    $51.6 \pm 0.5 $ &   $169.0 \pm 1.5 $ &        $70 \pm 4 $ &       $101 \pm 5 $ &           \\
\vmic\,(\kms) &    $3.55 \pm 0.31$ &    $3.62 \pm 0.23$ &     $1.5 \pm 1.2 $ &    $2.09 \pm 0.47$ &     $1.7 \pm 0.7 $ &     $2.0 \pm 0.5 $ &     $1.9 \pm 0.4 $ &           \\
\hline                                                                                                                                                                       
           He &    $             $ &    $             $ &    $             $ &    $             $ &  $-1.17 \pm 0.20$* &  $-1.15 \pm 0.15$* &    $             $ &  $-1.11 $ \\
            C &   $-3.62 \pm 0.15$ &   $-3.79 \pm 0.18$ &   $-3.82 \pm 0.25$ &   $-3.55 \pm 0.12$ &  $-3.38 \pm 0.29$* &   $-3.45 \pm 0.16$ &   $-3.69 \pm 0.17$ &  $-3.65 $ \\
            N &    $             $ &    $             $ &    $             $ &    $             $ &   $-3.9 \pm 0.3 $* &   $-3.5 \pm 0.4 $* &   $-4.2 \pm 0.3 $* &  $-4.26 $ \\
            O &  $-3.18 \pm 0.15$* &  $-3.17 \pm 0.10$* &   $-3.31 \pm 0.15$ &   $-3.38 \pm 0.13$ &  $-3.20 \pm 0.10$* &   $-3.10 \pm 0.13$ &   $-3.23 \pm 0.08$ &  $-3.38 $ \\
           Ne &    $             $ &    $             $ &    $             $ &    $             $ &    $             $ &    $             $ &    $             $ &  $-4.20 $ \\
           Na &  $-5.77 \pm 0.15$* &   $-5.86 \pm 0.09$ &   $-5.6 \pm 0.5 $* &  $-6.18 \pm 0.08$* &     $            $ &   $-5.5 \pm 0.3 $* &   $-5.4 \pm 0.3 $* &  $-5.87 $ \\
           Mg &   $-4.61 \pm 0.06$ &   $-4.46 \pm 0.04$ &   $-4.12 \pm 0.15$ &   $-4.89 \pm 0.06$ &   $-4.28 \pm 0.15$ &   $-4.73 \pm 0.17$ &   $-4.03 \pm 0.22$ &  $-4.51 $ \\
           Al &     $            $ &           $      $ &   $-5.5 \pm 0.4 $* &     $            $ &   $-5.5 \pm 0.3 $* &   $-6.0 \pm 0.4 $* &           $      $ &  $-5.67 $ \\
           Si &   $-4.91 \pm 0.20$ &   $-4.81 \pm 0.24$ &   $-4.24 \pm 0.14$ &   $-5.13 \pm 0.10$ &   $-4.44 \pm 0.14$ &   $-4.82 \pm 0.22$ &   $-4.34 \pm 0.06$ &  $-4.53 $ \\
            P &           $      $ &           $      $ &           $      $ &           $      $ &           $      $ &           $      $ &           $      $ &  $-6.68 $ \\
            S &   $-4.70 \pm 0.15$ &   $-4.82 \pm 0.05$ &           $      $ &   $-5.10 \pm 0.12$ &           $      $ &   $-4.2 \pm 0.4 $* &  $-4.48 \pm 0.15$* &  $-4.90 $ \\
           Ar &           $      $ &           $      $ &           $      $ &           $      $ &           $      $ &           $      $ &           $      $ &  $-5.86 $ \\
            K &           $      $ &           $      $ &           $      $ &          $       $ &           $      $ &           $      $ &           $      $ &  $-6.96 $ \\
           Ca &   $-5.98 \pm 0.18$ &   $-5.78 \pm 0.15$ &   $-5.54 \pm 0.30$ &   $-6.15 \pm 0.12$ &  $-6.27 \pm 0.19$* &   $-6.25 \pm 0.38$ &   $-5.42 \pm 0.24$ &  $-5.73 $ \\
           Sc &   $-9.23 \pm 0.23$ &   $-9.09 \pm 0.15$ &   $-8.76 \pm 0.08$ &   $-9.58 \pm 0.24$ &   $-9.0 \pm 0.4 $* &   $-9.40 \pm 0.16$ &   $-8.75 \pm 0.24$ &  $-8.87 $ \\
           Ti &   $-7.39 \pm 0.20$ &   $-7.26 \pm 0.16$ &   $-6.92 \pm 0.10$ &   $-7.65 \pm 0.10$ &   $-7.12 \pm 0.29$ &   $-7.57 \pm 0.12$ &   $-6.81 \pm 0.31$ &  $-7.14 $ \\
            V &   $-8.4 \pm 0.4 $* &   $-8.3 \pm 0.4 $* &     $            $ &  $-8.49 \pm 0.16$* &     $            $ &           $      $ &           $      $ &  $-8.04 $ \\
           Cr &   $-6.60 \pm 0.21$ &   $-6.51 \pm 0.23$ &   $-5.95 \pm 0.20$ &   $-6.98 \pm 0.13$ &   $-6.46 \pm 0.21$ &   $-6.97 \pm 0.13$ &   $-6.07 \pm 0.16$ &  $-6.40 $ \\
           Mn &   $-7.08 \pm 0.15$ &   $-6.96 \pm 0.15$ &   $-7.0 \pm 0.5 $* &   $-7.57 \pm 0.20$ &           $      $ &        $         $ &   $-6.91 \pm 0.23$ &  $-6.65 $ \\
           Fe &   $-4.84 \pm 0.11$ &   $-4.70 \pm 0.09$ &   $-4.39 \pm 0.15$ &   $-5.13 \pm 0.11$ &   $-4.48 \pm 0.27$ &   $-5.03 \pm 0.13$ &   $-4.35 \pm 0.24$ &  $-4.59 $ \\
           Co &   $-6.8 \pm 0.4 $* &   $-7.1 \pm 0.5 $* &     $            $ &     $\leq -7.6  $* &     $            $ &        $         $ &           $      $ &  $-7.12 $ \\
           Ni &   $-6.16 \pm 0.13$ &   $-6.00 \pm 0.15$ &   $-5.72 \pm 0.12$ &   $-6.40 \pm 0.16$ &           $      $ &     $\leq -5.9  $* &   $-6.00 \pm 0.18$ &  $-5.81 $ \\
           Cu &           $      $ &     $\leq -7.7  $* &     $            $ &   $-8.85 \pm 0.28$ &           $      $ &           $      $ &            $     $ &  $-7.83 $ \\
           Zn &   $-7.9 \pm 0.3 $* &   $-7.57 \pm 0.20$ &           $      $ &  $-8.71 \pm 0.08$* &     $            $ &           $      $ &            $     $ &  $-7.44 $ \\
           Sr &  $-9.56 \pm 0.19$* &   $-9.4 \pm 0.3 $* &     $            $ &  $-9.58 \pm 0.11$* &     $            $ &  $-10.2 \pm 0.2 $* &            $     $ &  $-9.12 $ \\
            Y &   $-9.94 \pm 0.22$ &   $-9.80 \pm 0.13$ &           $      $ &  $-10.18 \pm 0.14$ &           $      $ &  $-10.0 \pm 0.4 $* &   $-9.8 \pm 0.4 $* &  $-9.83 $ \\
           Zr &  $-9.36 \pm 0.26$* &   $-9.3 \pm 0.4 $* &     $            $ &  $-9.65 \pm 0.13$* &     $            $ &           $      $ &           $      $ &  $-9.48 $ \\
           Ba &   $-9.81 \pm 0.18$ &   $-9.69 \pm 0.16$ &  $-10.0 \pm 0.4 $* &   $-9.96 \pm 0.21$ &     $\leq -9.0  $* &  $-10.4 \pm 0.4 $* &   $-9.6 \pm 0.2 $* &  $-9.87 $ \\
           La &    $\leq -10.8  $* &  $-11.2 \pm 0.4 $* &     $            $ &           $      $ &           $      $ &                    &                    &  $-10.91$ \\
           Ce &           $      $ &           $      $ &           $      $ &           $      $ &           $      $ &                    &                    &  $-10.34$ \\
           Nd &           $      $ &           $      $ &           $      $ &  $-10.64\pm 0.14$* &     $            $ &                    &                    &  $-10.59$ \\
           Eu &           $      $ &           $      $ &           $      $ &  $-12.19\pm 0.10$* &           $      $ &                    &                    &  $-11.52$ \\
\hline
\end{tabular}
}

\label{abun-tab}
\end{table*}
\end{centering}

\begin{centering}
\begin{table*}
\contcaption{Best fit parameters for the stars in our sample. Chemical abundances are in units of $\log(N_{X}/N_{tot})$. 
Elements marked with an asterisk are based on less then $\sim$4 useful lines 
and have uncertainties estimated by eye.   Microturbulence is given by \vmic. }
\scriptsize{
\begin{tabular}{lcccccccccccc}
\hline\hline

              &          HD 245185 &          HD 278937 &              T Ori &          HD 101412 &          HD 190073 &         V380 Ori A &         V380 Ori B &  Solar    \\
              &                    &          (IP Per)  &      (BD -05 1329) &                    &                    &                    &                    &           \\
\hline                                                                                                                                                                       
   \teff\,(K) &    $9500 \pm 750 $ &    $8000 \pm 250 $ &    $8500 \pm 300 $ &    $8600 \pm 300 $ &   $9230 \pm 260  $ &  $12600 \pm 1000 $ &   $5800 \pm 350  $ &           \\
  \lgg\,(cgs) &     $4.0 \pm 0.4 $ &     $4.1 \pm 0.2 $ &     $4.2 \pm 0.3 $ &     $4.0 \pm 0.5 $ &     $3.7 \pm 0.3 $ &    $4.0 \pm 0.5  $ &     $4.1 \pm 0.3 $ &           \\
  \vs\,(\kms) &      $136 \pm 10 $ &    $83.8 \pm 4.6 $ &      $163 \pm 11 $ &     $6.8 \pm 0.4 $ &   $8.50 \pm 0.23 $ &     $9.9 \pm 1.0 $ &    $24.7 \pm 1.9 $ &           \\
\vmic\,(\kms) &      $\leq 4     $ &     $2.0 \pm 0.9 $ &     $2.6 \pm 1.1 $ &   $\leq 2        $ &   $\leq 2        $ &  $\leq 3         $ &    $\leq 2       $ &           \\
\hline                                                                                                                                                                       
           He &   $-0.9 \pm 0.3 $* &     $            $ &   $-1.4 \pm 0.5 $* &   $-0.9 \pm 0.3 $* &  $-0.90 \pm 0.20$* &  $-0.67 \pm 0.20$* &          $       $ &  $-1.11 $ \\
            C &   $-3.72 \pm 0.15$ &   $-3.57 \pm 0.17$ &   $-3.59 \pm 0.19$ &   $-3.58 \pm 0.12$ &  $-3.72 \pm 0.14 $ &   $-3.4 \pm 0.4 $* &          $       $ &  $-3.65 $ \\
            N &   $-3.8 \pm 0.4 $* &   $-4.0 \pm 0.3 $* &   $-4.2 \pm 0.4 $* &    $             $ &  $-3.60 \pm 0.20$* &          $       $ &          $       $ &  $-4.26 $ \\
            O &   $-3.17 \pm 0.17$ &   $-3.41 \pm 0.05$ &   $-3.16 \pm 0.12$ &   $-3.12 \pm 0.09$ &  $-3.26 \pm 0.11 $ &  $-3.00 \pm 0.16 $ &          $       $ &  $-3.38 $ \\
           Ne &           $      $ &           $      $ &           $      $ &    $             $ &   $-3.8 \pm 0.3 $* &  $-3.93 \pm 0.36 $ &          $       $ &  $-4.20 $ \\
           Na &     $            $ &   $-6.3 \pm 0.3 $* &   $-6.1 \pm 0.3 $* &  $-5.56 \pm 0.15$* &  $-5.13 \pm 0.15$* &          $       $ &  $-5.95 \pm 0.30$* &  $-5.87 $ \\
           Mg &   $-5.39 \pm 0.23$ &   $-4.91 \pm 0.11$ &   $-5.05 \pm 0.23$ &   $-4.95 \pm 0.15$ &  $-4.38 \pm 0.17 $ &  $-4.19 \pm 0.14 $ &   $-4.7 \pm 0.7 $* &  $-4.51 $ \\
           Al &   $-5.4 \pm 0.5 $* &     $            $ &   $-5.2 \pm 0.4 $* &  $-5.99 \pm 0.20$* &  $-5.58 \pm 0.17$* &  $-5.07 \pm 0.21$* &          $       $ &  $-5.67 $ \\
           Si &   $-4.95 \pm 0.16$ &   $-4.97 \pm 0.19$ &   $-4.89 \pm 0.20$ &   $-5.30 \pm 0.19$ &  $-4.48 \pm 0.06 $ &  $-4.11 \pm 0.09 $ &  $-4.44 \pm 0.26 $ &  $-4.53 $ \\
            P &            $     $ &           $      $ &           $      $ &           $      $ &          $       $ &  $-5.86 \pm 0.30$* &          $       $ &  $-6.68 $ \\
            S &      $           $ &   $-4.83 \pm 0.14$ &   $-4.1 \pm 0.3 $* &   $-4.96 \pm 0.12$ &  $-4.57 \pm 0.20$* &  $-4.93 \pm 0.26 $ &          $       $ &  $-4.90 $ \\
           Ar &            $     $ &           $      $ &     $            $ &           $      $ &          $       $ &          $       $ &          $       $ &  $-5.86 $ \\
            K &           $      $ &           $      $ &        $         $ &           $      $ &          $       $ &          $       $ &          $       $ &  $-6.96 $ \\
           Ca &   $-6.12 \pm 0.25$ &   $-6.28 \pm 0.18$ &   $-6.00 \pm 0.31$ &   $-6.37 \pm 0.20$ &  $-5.61 \pm 0.12 $ &  $\leq -5.6     $* &  $-5.78 \pm 0.40 $ &  $-5.73 $ \\
           Sc &   $-9.6 \pm 0.7 $* &   $-9.52 \pm 0.21$ &   $-9.4 \pm 0.3 $* &   $-9.33 \pm 0.17$ &  $-8.85 \pm 0.20 $ &          $       $ &          $       $ &  $-8.87 $ \\
           Ti &   $-7.93 \pm 0.38$ &   $-7.81 \pm 0.10$ &   $-7.55 \pm 0.17$ &   $-7.75 \pm 0.23$ &  $-7.16 \pm 0.12 $ &          $       $ &  $-6.78 \pm 0.28 $ &  $-7.14 $ \\
            V &     $\leq -7.5  $* &     $            $ &           $      $ &   $-8.1 \pm 0.3 $* &  $-7.99 \pm 0.25$* &          $       $ &  $-7.50 \pm 0.79 $ &  $-8.04 $ \\
           Cr &   $-7.27 \pm 0.34$ &   $-7.04 \pm 0.18$ &   $-6.82 \pm 0.07$ &   $-6.69 \pm 0.29$ &  $-6.13 \pm 0.13 $ &  $\leq -4.8     $* &  $-5.92 \pm 0.26 $ &  $-6.40 $ \\
           Mn &   $-7.0 \pm 0.6 $* &   $-7.3 \pm 0.4 $* &      $\leq -6.0 $* &   $-6.98 \pm 0.09$ &  $-6.54 \pm 0.08 $ &  $-5.92 \pm 0.16 $ &  $-5.97 \pm 0.43 $ &  $-6.65 $ \\
           Fe &   $-5.27 \pm 0.33$ &   $-5.12 \pm 0.16$ &   $-4.98 \pm 0.10$ &   $-5.08 \pm 0.19$ &  $-4.42 \pm 0.06 $ &  $-3.92 \pm 0.12 $ &  $-4.25 \pm 0.16 $ &  $-4.59 $ \\
           Co &           $      $ &           $      $ &           $      $ &           $      $ &          $       $ &          $       $ &   $-6.6 \pm 0.6 $* &  $-7.12 $ \\
           Ni &   $-6.5 \pm 0.7 $* &   $-6.37 \pm 0.22$ &           $      $ &   $-6.13 \pm 0.27$ &  $-5.69 \pm 0.17 $ &  $-5.19 \pm 0.18 $ &  $-5.58 \pm 0.32 $ &  $-5.81 $ \\
           Cu &            $     $ &      $\leq -7.5 $* &           $      $ &   $-8.6 \pm 0.3 $* &          $       $ &              $   $ &   $-7.4 \pm 0.7 $* &  $-7.83 $ \\
           Zn &            $     $ &     $\leq -8.0  $* &           $      $ &           $      $ &  $-7.52 \pm 0.25$* &              $   $ &   $-7.4 \pm 0.6 $* &  $-7.44 $ \\
           Sr &            $     $ &           $      $ &  $-9.3  \pm 0.2 $* &           $      $ &          $       $ &              $   $ &          $       $ &  $-9.12 $ \\
            Y &      $           $ &  $-10.4 \pm 0.4 $* &           $      $ &  $-10.12 \pm 0.25$ &  $-9.70 \pm 0.20 $ &              $   $ &  $\leq -8.2     $* &  $-9.83 $ \\
           Zr &            $     $ &           $      $ &           $      $ &   $-9.8 \pm 0.3 $* &   $-9.2 \pm 0.3 $* &              $   $ &          $       $ &  $-9.48 $ \\
           Ba &  $-10.0 \pm 0.7 $* &  $-10.2 \pm 0.3 $* &  $-10.3 \pm 0.3 $* &  $-10.71 \pm 0.24$ &  $-9.71 \pm 0.13$* &              $   $ &  $-8.89 \pm 0.40 $ &  $-9.87 $ \\
           La &                    &                    &                    &           $      $ &          $       $ &              $   $ &          $       $ &  $-10.91$ \\
           Ce &                    &                    &                    &           $      $ &  $\leq -9.6     $* &              $   $ &    $\leq -8.5   $* &  $-10.34$ \\
           Nd &                    &                    &                    &           $      $ &  $-10.2 \pm 0.3 $* &              $   $ &              $   $ &  $-10.59$ \\
           Eu &                    &                    &                    &           $      $ &          $       $ &              $   $ &              $   $ &  $-11.52$ \\
\hline
\end{tabular}
}
\label{abun-tab-pt2}
\end{table*}
\end{centering}

\section{Trends}
\label{trends}
In order to investigate trends in chemical abundance relative to other stellar parameters, 
we constructed a peculiarity index to describe whether a star is chemically normal, 
more like a $\lambda$ Boo star, or more like an Ap/Bp star.  
This index is designed to be positive for stars 
with overabundances of iron peak elements, zero for chemically normal stars, 
and negative for stars underabundant in iron peak elements.  
The peculiarity index is the difference between the average of the Cr, Fe, and Ni 
abundances relative to solar and the average of the C, N, and O abundances 
relative to solar.  
Specifically, the index is calculated as: 
\begin{equation}
[P] = \frac{1}{3} \big( ([\mathrm{Cr}]+[\mathrm{Fe}]+[\mathrm{Ni}]) - ([\mathrm{C}]+[\mathrm{N}]+[\mathrm{O}]) \big) , 
\end{equation}
where the abundances are relative to solar: 
$[\mathrm{X}] = \log(N_{X,\star}/N_{tot,\star}) - \log(N_{X,\odot}/N_{tot,\odot})$.  
These particular elements were chosen because they tend to have peculiar abundances 
in chemically peculiar stars (for Cr, Fe, and Ni) or normal abundances in chemically normal stars (for C, N and O). 
These elements were also chosen because reliable abundances for them are available for most stars in our sample.  
In the few cases in which we could not determine an abundance for an element 
used in the peculiarity index, that element was left out of the average.  

\begin{figure*}
\centering
\includegraphics[width=3.0in]{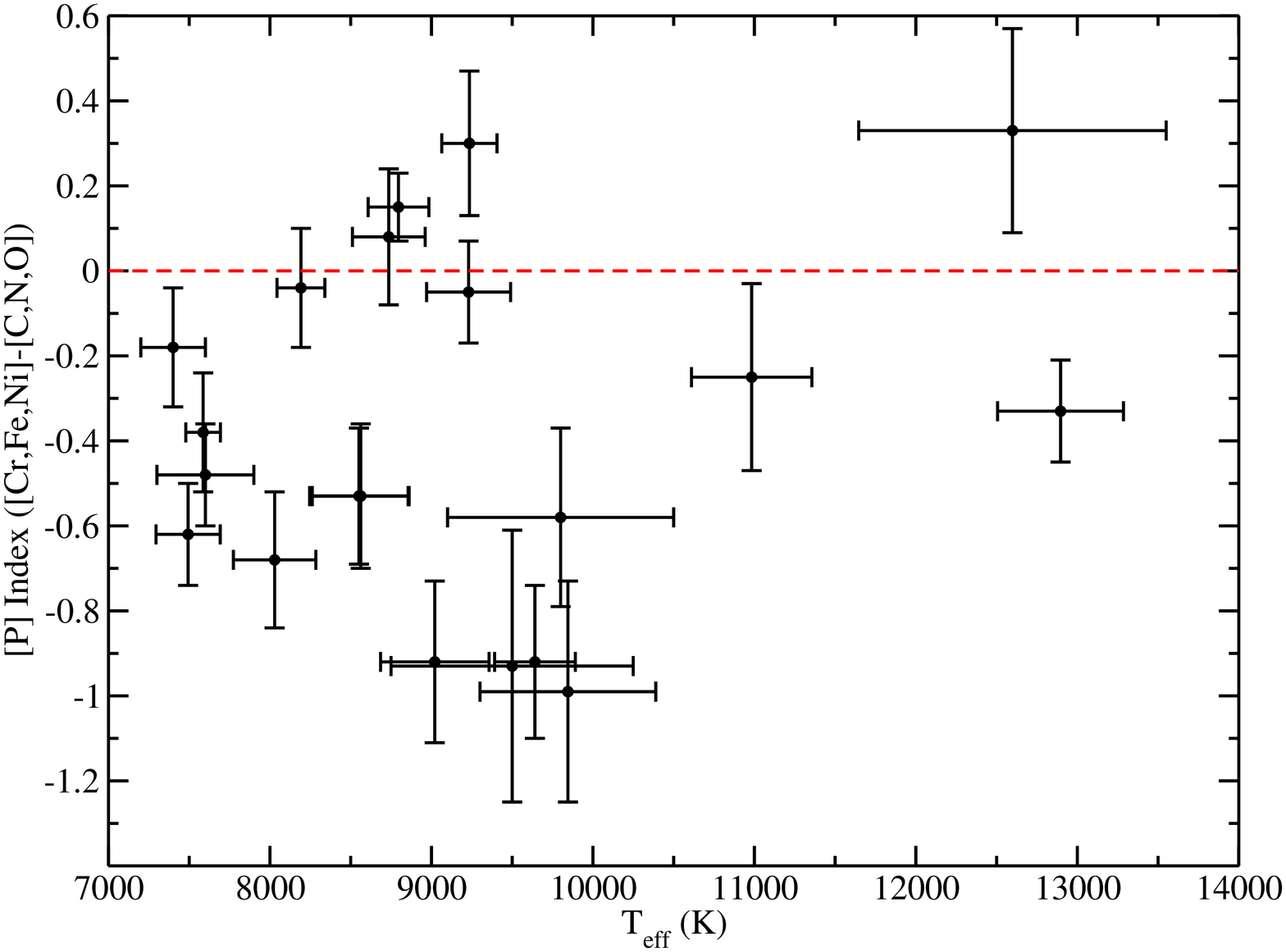}
\includegraphics[width=3.0in]{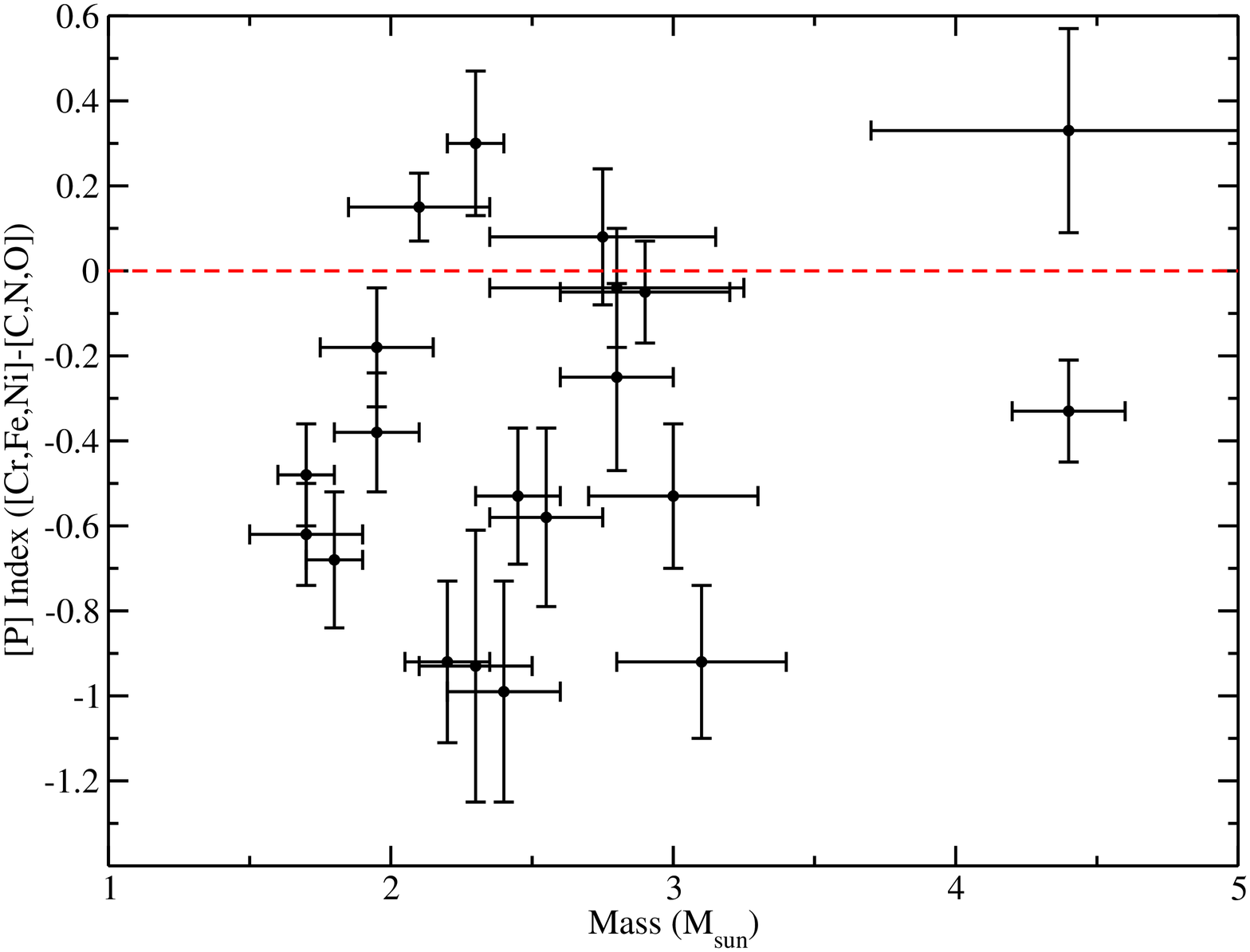}
\includegraphics[width=3.0in]{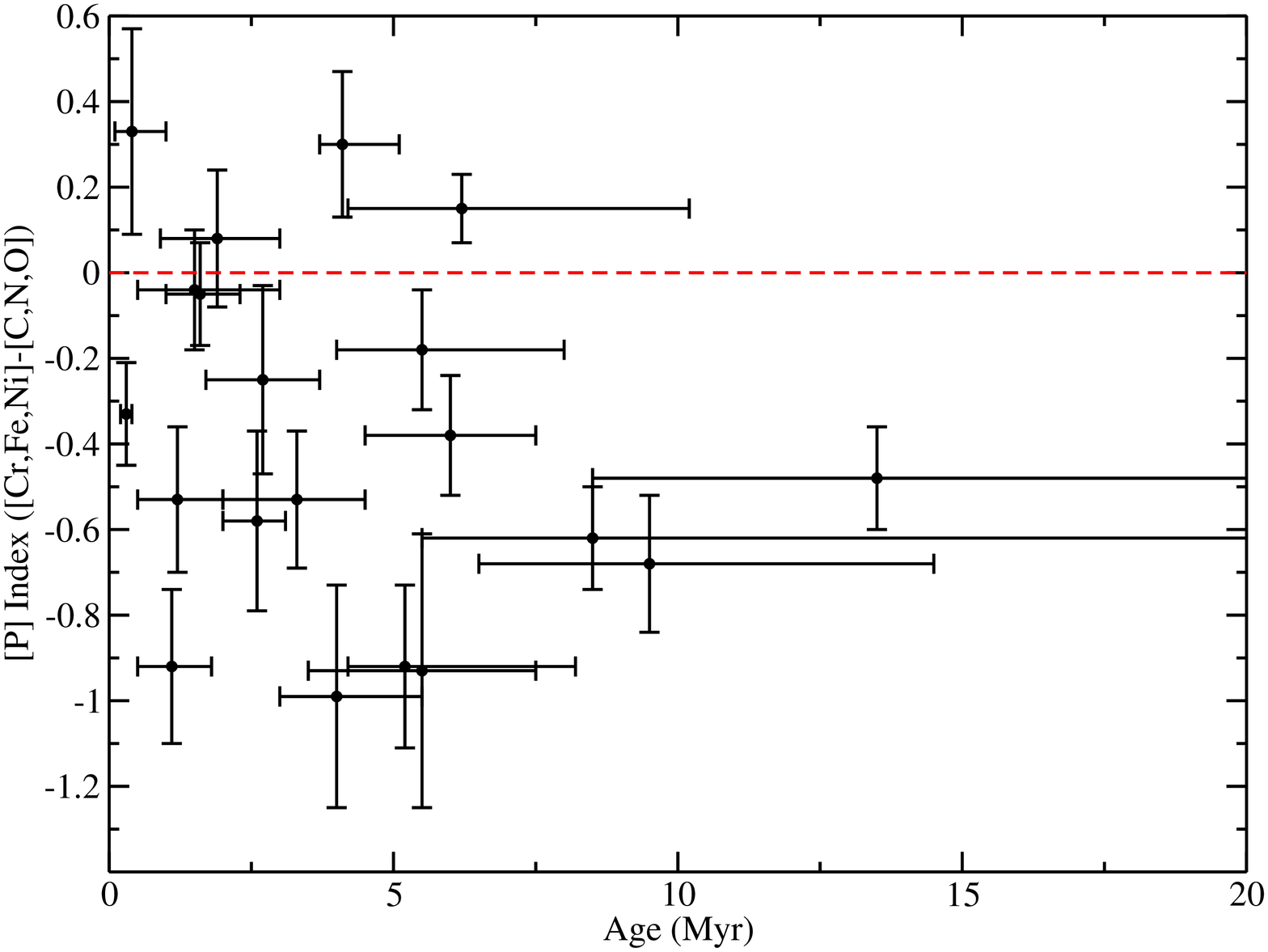}
\includegraphics[width=3.0in]{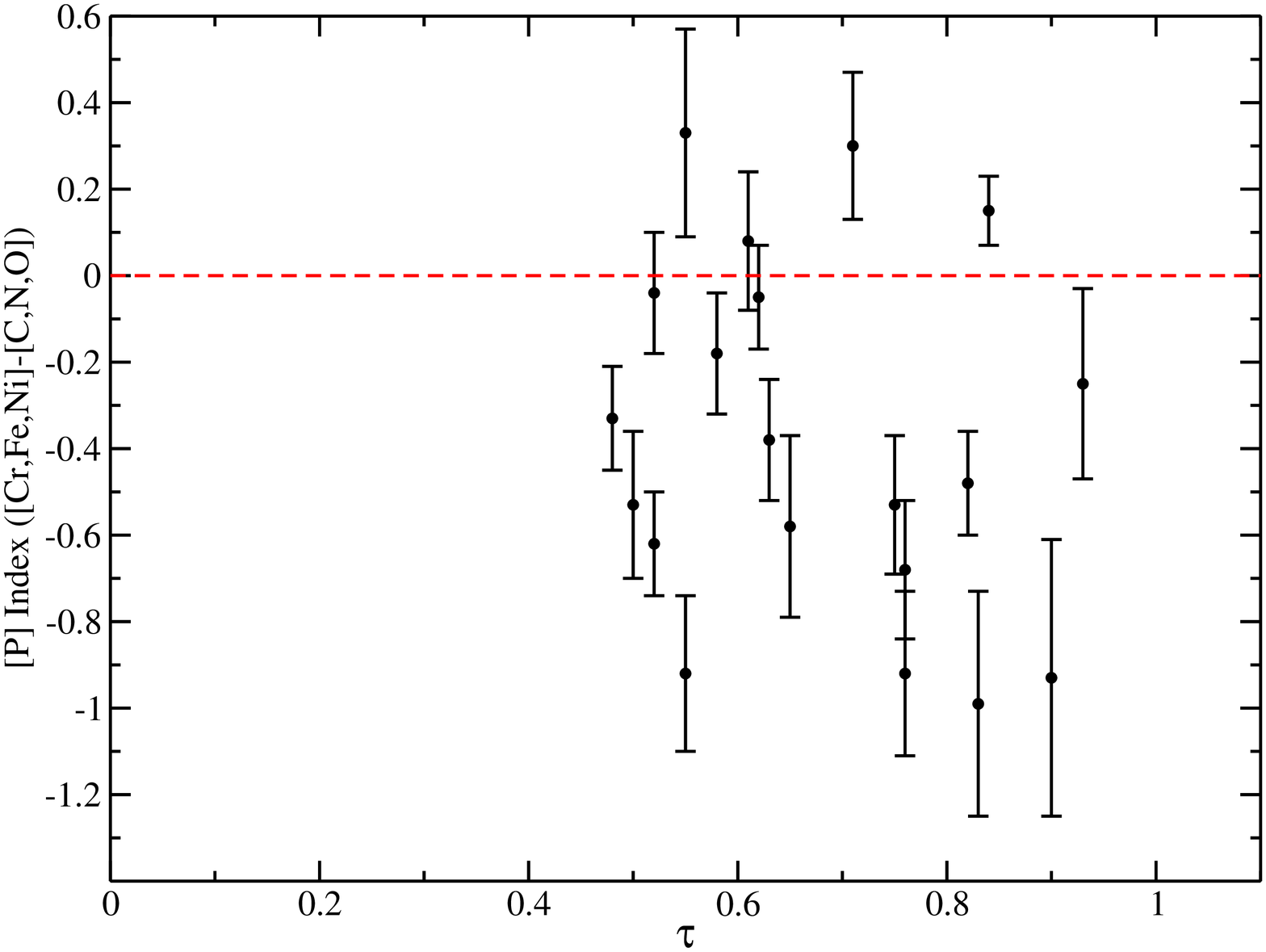}
\caption{Trends in stellar parameters with the peculiarity index.  
The index is the average of Cr, Fe and Ni abundances each relative to solar, 
minus the average of the C, N and O abundances relative to solar.  
Plots are included for effective temperature, mass, age, 
and fractional pre-main sequence age ($\tau$).  }
\label{index-trend-plots}
\end{figure*}

We plotted the index $[P]$ against effective temperature, mass, age, and fractional pre-main 
sequence age, as shown in Fig. \ref{index-trend-plots}.  No clear trends were found, 
with a large scatter in chemical abundance at most temperatures, masses, and ages.  
This peculiarity index was also compared to stellar radius, \vs, and microturbulence, 
but again no clear trends were found.  
Thus we conclude that the $\lambda$ Boo peculiarities that we see are not 
restricted to a narrow mass range, and that they are present throughout 
the pre-main sequence.

\section{Discussion}

\subsection{$\lambda$ Boo peculiarities}

The most striking result of this study is the discovery of a large number of HAeBe stars with 
moderate $\lambda$ Boo type chemical peculiarities.  This trend seems to be present in 
both stars with and without detected magnetic fields, as shown by HD 101412.  
We find 7 stars showing fairly strong $\lambda$ Boo peculiarities, 
and another 4 stars with weaker $\lambda$ Boo peculiarities, while 8 stars 
are chemically normal and 1 is rather uncertain but appears to be normal. 
With the derived abundances, we see a continuous distribution 
of $\lambda$ Boo peculiarities, from fairly strong to nearly undetectable.  
This incidence of roughly 50\% is much higher than the incidence of 
$\lambda$ Boo stars on the main sequence of roughly 2\% 
\citep{Paunzen2001-lambdaBoo-survey-incidence}.  
While a detection rate of 50\% may be an overestimate 
(since our sample is biased towards stars with limited emission in their optical spectra), 
the true incidence of $\lambda$ Boo peculiarities in HAeBe stars certainly is 
much higher than observed on the main sequence, suggesting an important connection 
with the pre-main sequence evolutionary phase.  

The sample of stars in our study is biased towards stars with 
weaker circumstellar emission or absorption in their spectra, since these stars 
allow for a more complete analysis of their photospheric spectra.  However, the stars were drawn 
from the larger sample of \citet{Alecian2012-big-HAeBe-magnetism} 
which did not have this bias.  We found 65\% of the stars in their sample had 
modest amounts of circumstellar contamination, and would have been 
suitable for our photospheric analysis. 
Thus, as a lower limit, 33\% of HAeBe stars should show $\lambda$ Boo peculiarities, 
assuming that all the unsuitable stars in the sample of 
\citet{Alecian2012-big-HAeBe-magnetism} are chemically normal, 
and using our derived incidence of peculiarities for the suitable stars. 
The true incidence is likely higher than this lower limit.  

There are almost certainly more undetected $\lambda$ Boo stars in the sample 
of \citet{Alecian2012-big-HAeBe-magnetism} that we did not analyse, 
simply because analysing their complete sample was beyond the scope of this study.  
Indeed \citet{Alecian2012-big-HAeBe-magnetism} mention that HD 34282 appears 
to have weak lines of iron peak elements, but a normal O line.  
To roughly estimate the number of potential $\lambda$ Boo stars in the remaining sample, 
we compared synthetic spectra computed using literature values for their atmospheric parameters 
and assuming solar abundances with the ESPaDOnS observations.  
In doing so we identified 5 additional stars that have suspiciously weak metal lines 
for their literature temperatures: HD 37806, HD 98922, HD 144668, HD 76534, and VX Cas.  
If all these stars do indeed show $\lambda$ Boo peculiarities, 
35\% of the stars observed with ESPaDOnS by \citet{Alecian2012-big-HAeBe-magnetism} 
would be $\lambda$ Boo stars, which is consistent with our results.  

This large incidence of $\lambda$ Boo type peculiarities in HAeBe stars is 
qualitatively consistent with 
a selective accretion model of $\lambda$ Boo stars.  
HAeBe stars have recently undergone accretion, and may still be accreting, 
so it is reasonable that a process which depends on accretion is seen 
more frequently in these stars.  Thus we consider our results supportive of 
the selective accretion model of $\lambda$ Boo stars. 

The large number of pre-main sequence $\lambda$ Boo stars we find suggests that the 
majority of $\lambda$ Boo stars may develop their peculiarities on the pre-main sequence 
by accreting their circumstellar material.  
However, we cannot rule out the existence more of evolved $\lambda$ Boo stars that only 
develop peculiarities on the main sequence, by accreting from diffuse interstellar clouds.  
Due to the relatively short lifetime of $\lambda$ Boo peculiarities after accretion has halted, 
the presence of such peculiarities later in the main sequence 
\citep[e.g.][]{Paunzen2002-lambdaBoo-hipparcos} may require a different 
source of accreted material from that of pre-main sequence $\lambda$ Boo stars.

\citet{Turcotte1993-accreation-diffusion} modelled the diffusion 
of chemical elements in the atmosphere of a star (with \teff~$=8000$ K and \lgg~$=4.3$)
that was accreting gas depleted in iron peak elements.  
They found that $\lambda$ Boo peculiarities can be generated quickly ($\sim$0.1 Myr), 
but also dissipate quickly after accretion has stopped ($\sim$1 Myr). 
This implies that, in the context of selective accretion, 
there is a good chance that HAeBe stars with $\lambda$ Boo peculiarities are still 
accreting, or were very recently accreting.  

The range in strengths of the $\lambda$ Boo peculiarities seen in the stars 
in our sample might be interpreted as the consequence of different accretion rates in the stars.  
However, \citet{Turcotte1993-accreation-diffusion} do not find a strong impact 
on surface abundances from different accretion rates, as long as they are 
sufficient to overwhelm diffusion and rotational mixing.  

Alternately, the differences in the strengths of peculiarities could 
reflect the abundances of the accreted gas.  These differences being due, 
for example, to the efficiency with which iron peak elements are bound into 
dust grains, or to the degree to which dust is blown away from the star 
while gas is accreted.  

In this context, the chemically normal stars are most likely not accreting 
significant amounts of material, and thus any $\lambda$ Boo peculiarities 
have dissipated.  However, it is possible that accretion is ongoing 
but the selection process in the circumstellar material has broken down.  
For example, dust grain formation may not be proceeding efficiently, 
and thus the accreted gas would not be depleted in iron peak elements

\begin{figure*}
\centering
\includegraphics[width=3.0in]{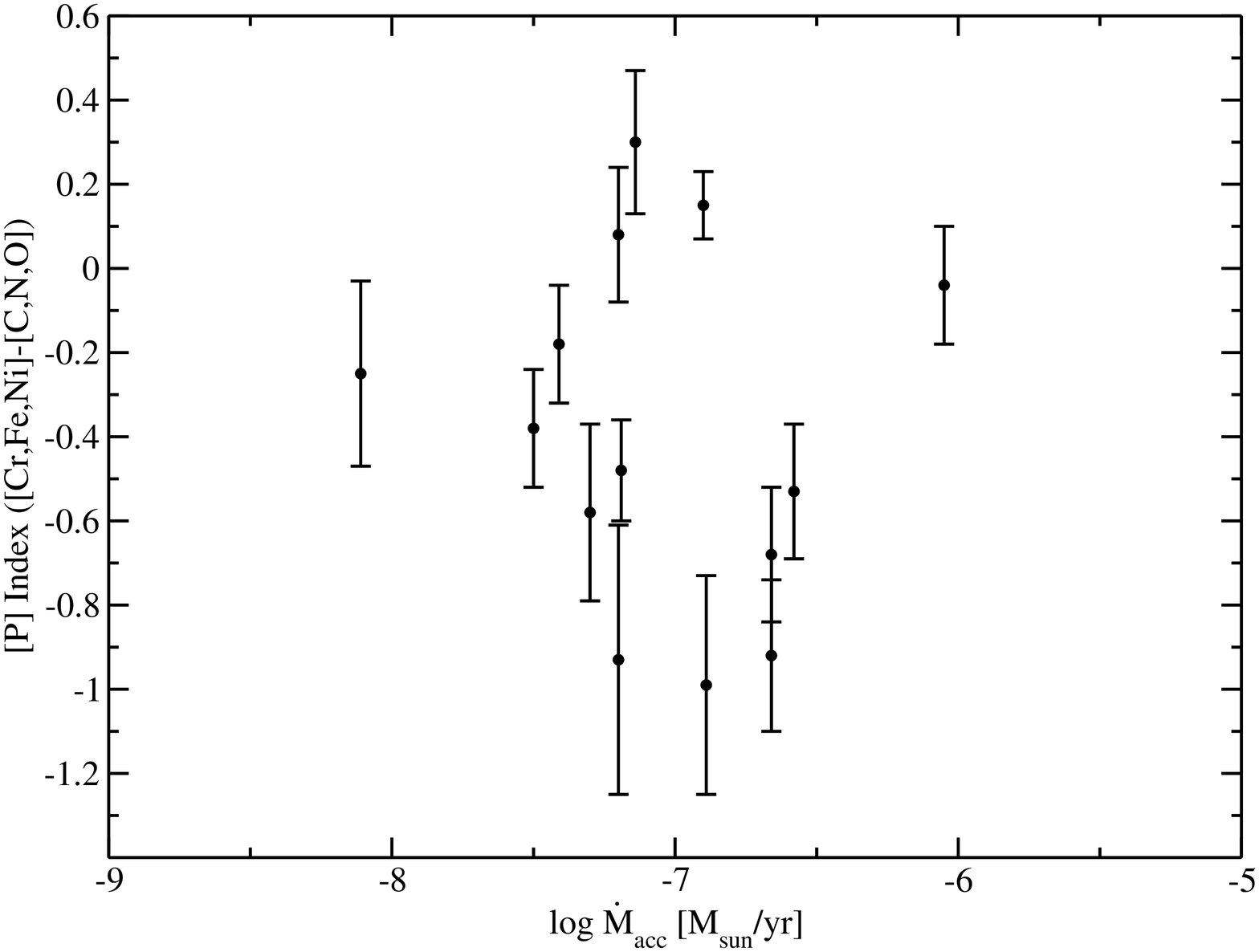}
\includegraphics[width=3.0in]{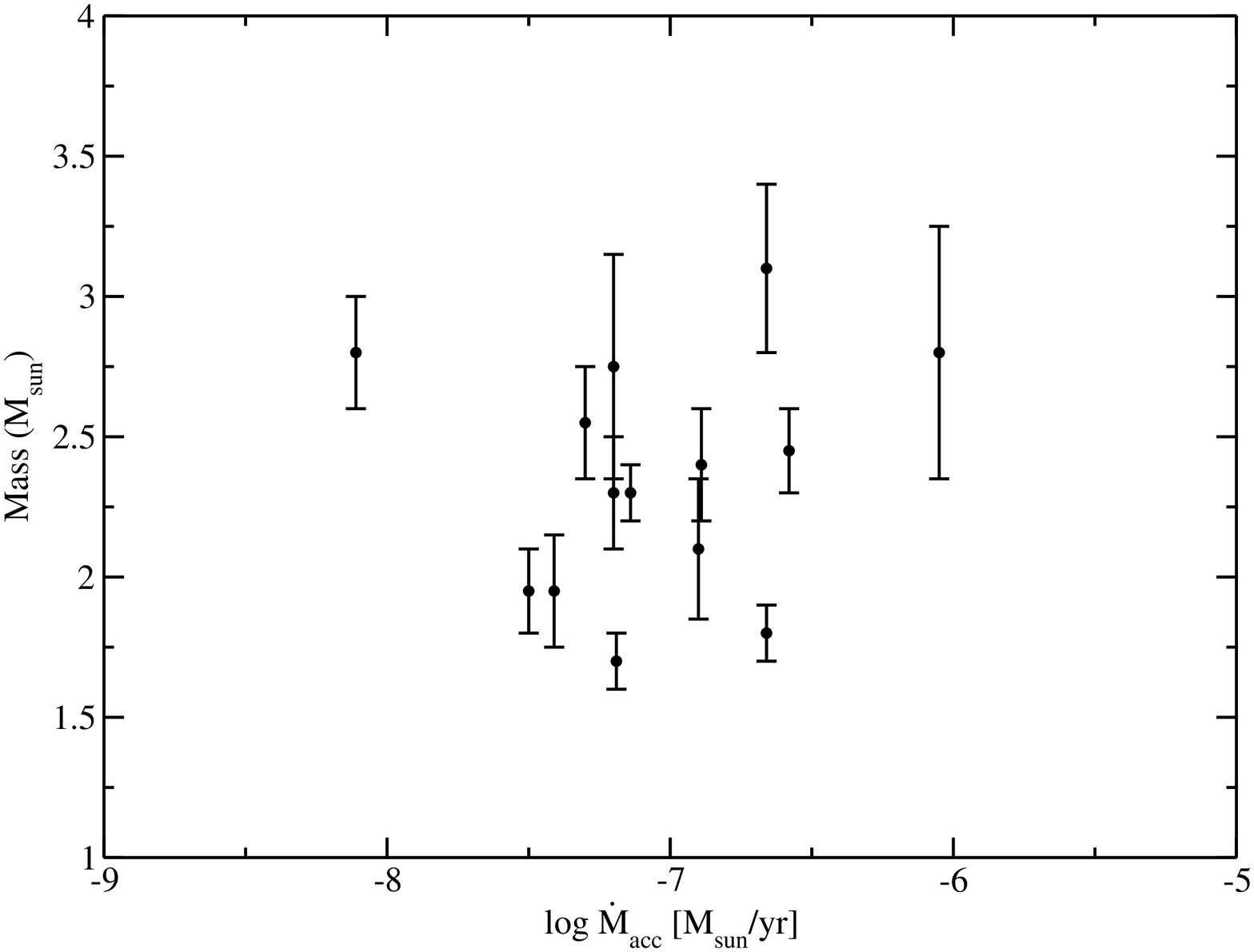}
\includegraphics[width=3.0in]{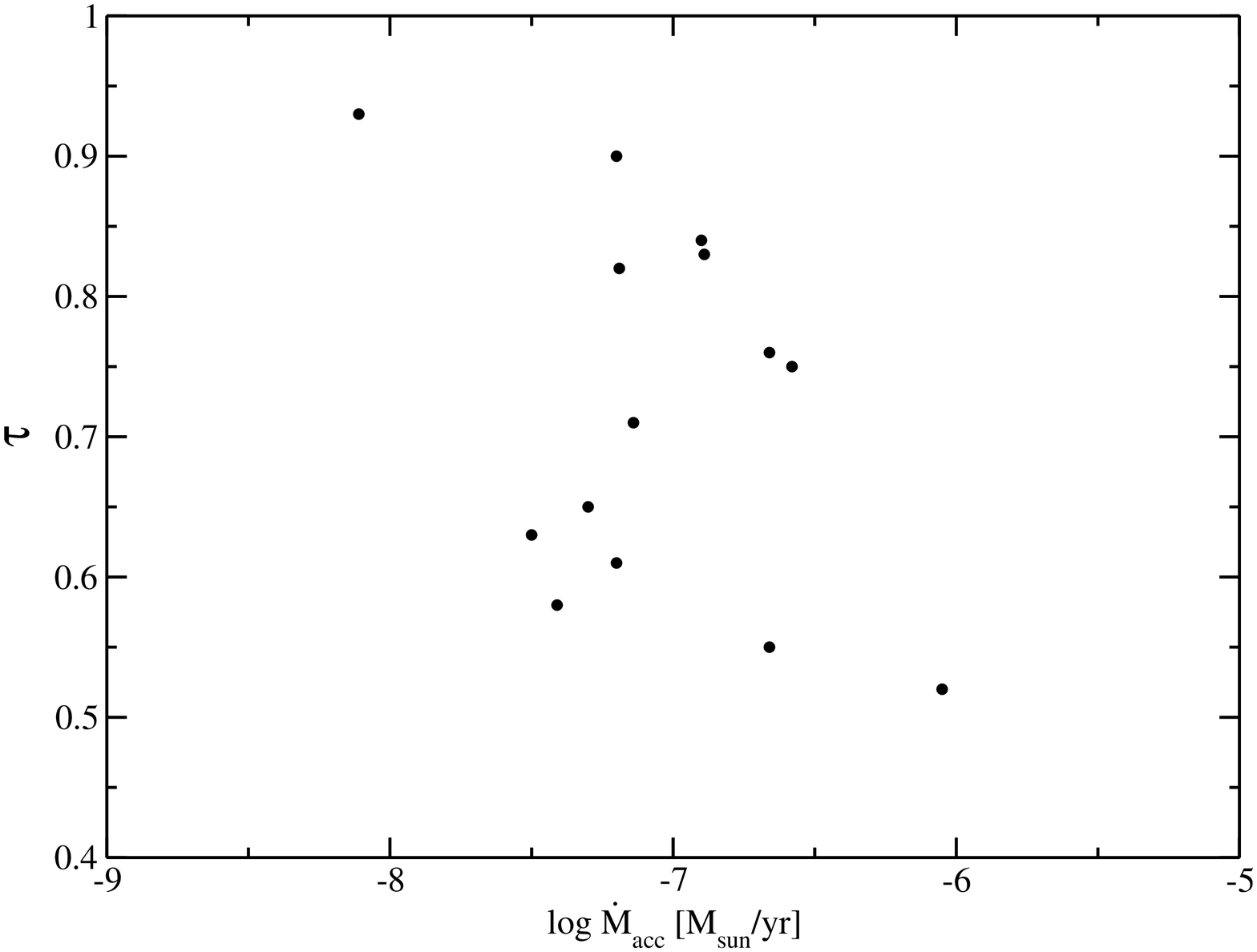}
\caption{Trends in stellar parameters with literature accretion rates.  
Plots are included for our chemical peculiarity index, stellar mass, 
and fractional pre-main sequence age ($\tau$).  }
\label{accretion-trend-plots}
\end{figure*}

Unfortunately, deriving accretion rates for HAeBe stars is not a simple process.  
For example, \citet{Donehew2011-HAeBe-accretion-rates} measured accretion 
rates based on Balmer jump excess fluxes.  They obtained values between 
a few $10^{-7}$ and $10^{-8}$ \acc.   These values rely on models from 
\citet{Muzerolle2004-magnetospheric-accretion-model}, which are constructed 
assuming magnetospheric accretion.  Given the absence of detectable magnetic fields 
in most of the stars in our study, magnetospheric accretion may not be an 
accurate model \citep[see][]{Wade2007-HAeBe_survey}, 
and thus these accretion rates may not be accurate either.  
Generally, accretion rates derived for HAeBe stars are based on rather simple models, 
and thus it is not clear how reliable those accretion rates are.  
Eleven of the stars in our sample have measured accretion rates by 
\citet{Donehew2011-HAeBe-accretion-rates}.  
Supplementing these results with accretion rates from 
\citet{Mendigutia2011-accretion-HAeBe}, who used a similar methodology,
and from \citet{GarciaLopez2006-accretion-HAeBe}, who based their accretion rates on 
Br$\gamma$ emission, provides literature values for 15 stars in our sample.  
Worryingly, the literature accretion rates differ for five of these stars 
by between 0.5 and 1 dex.  Whether this apparent disagreement is due to 
intrinsic stellar variability, differences in methodology, or simply 
large uncertainties is not clear \citep{Mendigutia2011-accretion-HAeBe}. 
We compare these accretion rates with our peculiarity index $[P]$ (defined in Sect. \ref{trends}), 
stellar mass, and fractional pre-main sequence age in Fig. \ref{accretion-trend-plots}.  
In cases where multiple accretion rates are available the average value was taken.  
V380 Ori was left out of the comparison due to its binarity. 
We see no correlation between accretion rate and chemical peculiarity or mass.   
There is possibly a weak correlation with fractional pre-main sequence age, 
but due to the large uncertainties on both the accretion rates and 
the fractional ages it is not entirely clear. 

Our results provide further evidence against the hypothesis that $\lambda$ Boo stars 
are formed by radiative diffusion with mass loss, as proposed by 
\citet{Michaud1986-lambdaBoo-diffusion-massloss}.  
In their calculations, \citet{Michaud1986-lambdaBoo-diffusion-massloss} find that 
it takes $10^8$ or $10^9$ years to build up $\lambda$ Boo peculiarities by 
diffusion in the presence of mass loss.  This is much greater than the 
ages of the $\lambda$ Boo-like stars in our sample, which are only 
a few $\times 10^6$ years old.  Thus the time scale for this mechanism is 
dramatically incompatible with our observations.  This is consistent with 
the generally accepted view that diffusion with mass loss is insufficient 
for producing $\lambda$ Boo peculiarities \citep[e.g.][]{Charbonneau1993-lambdaBoo-diffusion-massloss} .

\subsection{Magnetic stars}

The one possible Bp star in our sample is V380 Ori A.  
This star shows modest overabundances of Fe, Mn, and Ni of $\sim$0.6 dex, 
and Si is also enhanced relative to solar.  
C, Ne, and S are consistent with solar, as are O and Mg at 2$\sigma$.  
This pattern is characteristic of Ap/Bp stars.  
Arguably, results are marginally consistent with the star simply having a 
high metallicity ($\sim$0.4 dex above solar).  However, since Fe, Mn and Ni are 
more enhanced than 0.4 dex, and C, O, Ne, Mg and S are less enhanced than this, 
we consider it more likely that the star is weakly chemically peculiar.  
If the star is a Bp star, the star has begun developing Ap/Bp chemical peculiarities, 
but they have not yet reached the degree seen in most main sequence Ap and Bp stars. 

Atomic diffusion producing chemical peculiarities was modelled in V380 Ori A  
by \citet{Vick2011-model-pms-peculiarities}, using a model that includes 
stellar evolution, as well as mass loss and atomic diffusion.  
They used the stellar parameters determined by \citet{Alecian2009-v380ori}, 
who find a somewhat lower mass than we do (mostly due to assuming $R_V=3$ rather than 5, 
thus using less extinction).  Nevertheless, the abundances they derive 
should be roughly representative.  \citet{Vick2011-model-pms-peculiarities} find 
Mn, Fe and Ni to be overabundant by $\sim$0.5 dex, which agrees nicely with our results.  
They find roughly solar abundances for C, O and S, which agrees with our results.  
The overabundance of P that they find is also roughly consistent with our results.  
However, the underabundant Mg, and solar Si abundance that they find are 
inconsistent with our results.  
Thus, while this model is imperfect, we find it does provide a reasonable approximation 
of the majority of the observed peculiarities in V380 Ori A.  

Interestingly, HD 190073 and HD 101412 are both magnetic but do not show 
Ap/Bp chemical peculiarities.  HD 190073 is chemically normal, while 
HD 101412 shows $\lambda$ Boo type chemical peculiarities.  
Thus, unlike on the main sequence, not all pre-main sequence magnetic A and B stars 
are chemically peculiar.  Atomic diffusion theory suggests that enough 
time must elapse after accretion and deep surface convection 
have halted for significant peculiarities to develop. 
The dissipation of a large convective envelope can be 
modelled easily enough, and in most models dissipates early in pre-main sequence 
life of A and B stars.  However, determining when accretion has dropped 
sufficiently for atomic diffusion to dominate is much harder.  
Indeed it may depend on the circumstellar environment of the particular star, 
and thus not be reliably predictable.  

The $\lambda$ Boo peculiarities of HD 101412 presumably simply reflect the 
high incidence of $\lambda$ Boo peculiarities in HAeBe stars in general.  
Thus, we suspect that these peculiarities are not a consequence of 
the magnetic field of HD 101412.

Modelling by \citet{Vick2011-model-pms-peculiarities} suggests that HD 190073 
has not yet had time to develop Ap/Bp chemical peculiarities.  In their models, 
a star of this mass requires $\sim$2 Myr to develop peculiarities by atomic diffusion, 
while we find HD 190073 is only $1.6^{+0.7}_{-0.6}$ Myr old.  
HD 101412 has almost the same mass, and we find an age of only $1.2^{+0.8}_{-0.7}$ Myr, 
thus by the models of \citet{Vick2011-model-pms-peculiarities} it is also too young 
to have developed Ap/Bp peculiarities.  The younger age of HD 101412 is consistent 
with the picture of its $\lambda$ Boo type chemical peculiarities being due to 
selective accretion, since accretion is more likely to be ongoing earlier in a stars life.  

It is interesting that the two stars in this sample with the largest amount 
of emission in their optical spectra are V380 Ori A and HD 190073, 
both of which are magnetic.  In particular, the emission mostly appears in 
lines that would be strongly in absorption for a purely photospheric spectrum 
of the star.  HD 101412, the third magnetic star, displays a fairly modest 
amount of emission.  
Additionally, the measured accretion rates for V380 Ori and HD 190073 are 
both substantially higher than for other stars in our sample 
\citep[$\log \dot M_{acc} = -5.6$ $\lbrack$\acc$\rbrack$ for V380 Ori, $\log \dot M_{acc} = -5.0$ $\lbrack$\acc$\rbrack$ for HD 190073;][]{Donehew2011-HAeBe-accretion-rates,Mendigutia2011-accretion-HAeBe}.  
The impact of the secondary in V380 Ori does not appear to be considered 
in the measurement of the star's accretion rate, and thus it may be somewhat inaccurate.  
In classical T Tauri stars, the strong emission in their spectra 
is thought to result from magnetically channelled accretion flows falling 
onto the star.  We speculate that the strong emission in V380 Ori A 
and HD 190073 might be produced by a similar process.  
However, it is not clear whether this is the case, both because of the 
small sample size, and because the non-magnetic stars in the sample are 
biased towards lower emission stars (since large amounts of emission 
limit the abundance analysis).  
Magnetic fields are not detected in individual emission lines for any of our stars, 
thus we cannot use those lines to assess the presence of magnetospheric accretion directly.  

The three stars in our sample with confirmed magnetic fields also have the lowest \vs\, 
values \citep[for more analysis see][]{Alecian2012-vsini-HAeBe-magnetism}. 
In this respect these three magnetic HAeBe star are similar to Ap/Bp stars, which 
rotate slower than normal non-magnetic A and B stars.  
The low \vs\, of the magnetic HAeBe stars HD 200775A \citep[\vs~$=26\pm2$ \kms,][]{Alecian2008-HD200775} and  
of NGC 2244 OI 201 \citep[\vs~$=23.5\pm0.5$ \kms,][]{Alecian2008-ClusterHAeBeLetter}, 
and the moderate \vs\, of HD 72106 A \citep[\vs~$=41.0\pm0.6$ \kms,][]{Folsom2008-hd72106} 
support this observation.  However, the very high \vs\, of NGC 6611 W610 
\citep[\vs~$=190\pm10$ \kms,][]{Alecian2008-ClusterHAeBeLetter} indicates that  
not all magnetic HAeBe stars rotate slowly.  
Nevertheless, the large number of low \vs\, magnetic HAeBe stars suggests 
that magnetic braking may have occurred in many of these stars 
\citep[see][]{Alecian2012-vsini-HAeBe-magnetism}.

Several other very young magnetic A and B stars are known to have chemical peculiarities.  
\citet{Folsom2008-hd72106} examined the magnetic star HD 72106 A, and found Ap/Bp peculiarities.  
They find that the star has a mass of 2.4 \Msun. 
While they were unable to firmly conclude that HD 72106 A is on the pre-main sequence, 
they do find that the system is between 6 and 13 Myr from the birthline, 
based on the H-R diagram position of the secondary in the binary system.  
The pattern of abundances seen in HD 72106 A is similar to that seen in V380 Ori A, 
but more extreme.  Overabundances of iron peak elements 
and a weak overabundance of Si is seen in both stars.  However, the strong 
He underabundance seen in HD 72106 A is not seen in V380 Ori A. 
The differences in abundances between the stars could be due to V380 Ori A 
being hotter and more massive HD 72106 A, or due to HD 72106 A being more evolved.  

\citet{Bagnulo2004-B_NGC2244-334} investigated the main sequence magnetic star NGC 2244-334, 
which is a member of the open cluster NGC 2244, and found Ap/Bp peculiarities. 
They found that the star has a mass of $\sim$4 \Msun\, and commented that, 
as a cluster member, the star is likely $\sim$2 Myr old as measured from the ZAMS.  
This is a similar mass to V380 Ori A, though NGC 2244-334 is somewhat older.  
The peculiarities they found are very similar to those of HD 72106 A, 
with overabundances in iron peak elements of 1-2 dex, a somewhat weaker overabundance of Si, 
and a strongly underabundant He.  This is again similar to what we see in V380 Ori A, 
though much stronger, and with a He underabundance.  

\citet{Alecian2008-ClusterHAeBeLetter} studied the magnetic Herbig Be star NGC 6611 W601, 
and noted peculiarly strong He and Si lines in its spectrum.  
This suggest the star likely represents a pre-main sequence He-strong star, 
but a detailed abundance analysis has yet to be performed.  
While this pattern of peculiarities would be different from that of V380 Ori A, 
the star has a much higher temperature of $22500\pm2500$ K and a much 
higher mass of $10.2^{+1.2}_{-0.7}$ \Msun \citep{Alecian2008-ClusterHAeBeLetter}.  
For a main sequence magnetic star in this temperature and mass range, 
one would expect to see He-strong 
peculiarities, rather than Ap/Bp peculiarities.  
\citet{Alecian2008-ClusterHAeBeLetter} estimate the age of NGC 6611 W601 to be less than 0.06 Myr 
from the birthline of \citet[][for an accretion rate of $10^{-4}$ \acc]{Palla1993-PMS-Evol}, 
based on its H-R diagram position.  
While this is very young, there are no models from \citet{Vick2011-model-pms-peculiarities} 
for a comparable mass.  We do note, however, that \citet{Vick2011-model-pms-peculiarities} 
find peculiarities developing faster for more massive models.   

Thus, while the peculiarities seen in V380 Ori A are weak, 
the pattern of peculiarities is consistent with those seen in other young intermediate mass 
magnetic stars.  This is consistent with the hypothesis that the weak peculiarities 
we see in this star will develop into the stronger peculiarities seen in most 
main sequence Ap/Bp stars.

\subsection{Chemically normal stars}

We find no hints of any chemical peculiarities in the stars of our sample other 
than $\lambda$ Boo peculiarities and the weak Bp peculiarities in V380 Ori A.  
Thus we find no evidence for Am or HgMn stars on the pre-main sequence.  
Our sample is not large enough to definitively exclude the 
existence of Am and HgMn stars at the same incidence as on the main sequence. 
However, if the roughly 10-20\% incidence \citep[e.g.][]{Smith1996-peculiar-statistics} 
of these stars on the main sequence is extended to the pre-main sequence, 
we would expect to have seen two to four such stars.  Thus it is possible that 
these peculiarities do not form on the pre-main sequence, unlike for Ap/Bp stars.  
The atomic diffusion thought to give rise to these peculiarities may be disrupted 
by weak accretion more readily than the diffusion in the presence of a 
magnetic field that produces Ap/Bp stars.  The presence of a magnetic field 
may stabilise a stellar atmosphere and increase diffusion velocities, 
potentially making diffusion in the presence of 
a magnetic field more robust \citep{Michaud1970-diffusion,Michaud1981-diffusion_magneticApBp}.  
Alternately, diffusion in non-magnetic stars may simply require more time than 
the pre-main sequence lifetime of these stars to produce significant peculiarities.

The large majority of young A and B stars analysed, both on the main sequence and 
on the pre-main sequence \citep[e.g.][]{Acke2004-HAeBe_Abun}, have been found to be chemically normal. 
A very precise study of A and B star abundances by \citet{Fossati2009-normal-star} 
investigated what `normal' abundances are for A and B stars. They found that solar 
abundances are a good approximation, though there were small deviations for specific elements,
which appeared to depend on the temperature of the star.  
Studies of nearby open clusters support this conclusion 
\citep[e.g.][]{Fossati2008-praesepe-abun, Villanova2009-NGC6475-abun, Fossati2011-NGC5460-abun}.

\citet{Acke2004-HAeBe_Abun} performed an abundance analysis of 24 pre-main sequence 
and very young main sequence stars, based on equivalent widths.  
A comparison of individual results for the 7 stars we have in common with their 
study is discussed in Sect. \ref{individual-stars}.  
Our results are generally consistent, although we are able to determine abundances 
for a wider range of elements, and in many cases provide more precise abundances.  
The one notable difference in our results is for HD 139614.  
For this star our more accurate \teff\, allows us to classify the star 
as $\lambda$ Boo, while \citet{Acke2004-HAeBe_Abun} could only 
comment that it appeared to be metal weak.  
In total \citet{Acke2004-HAeBe_Abun} found only one Herbig star that was also 
a clear $\lambda$ Boo star (HD 100546), and they noted that AB Aur 
shows hints of $\lambda$ Boo peculiarity.  
In part this difference in detection rates may due to our more accurate \teff\, and \lgg\, values, 
as the case of HD 139614 demonstrates.  Our higher precision, 
due to an improved methodology and higher S/N observations, 
is also a significant part of this difference.  
Finally, both our studies deal with a fairly small number of stars, 
and thus the true incidence of $\lambda$ Boo peculiarities in HAeBe stars 
remains somewhat uncertain.

\subsection{Possible Veiling}
\label{Possible Veiling}
One concern when analysing spectra of T Tauri stars is veiling.  
Veiling is additional continuum emission, usually from an accretion shock 
(or other circumstellar material), which makes spectral lines appear weaker 
than they would in a purely photospheric spectrum.  
In HAeBe stars, veiling can be seen in the UV \citep[e.g.][]{Donehew2011-HAeBe-accretion-rates}, 
but usually is considered to be unimportant in the optical 
\citep[e.g.][]{Boehm1993-AB-Aur,Hernandez2004-Rv-extinction,Cowley2010-hd101412-lambdaBoo,Donehew2011-HAeBe-accretion-rates}.
HAeBe stars are intrinsically much more luminous than T Tauri stars, 
thus much more emission from an accretion shock would be needed to have the same veiling effect.  
Measured accretion rates for HAeBe stars are usually not much larger than for T Tauri stars, 
thus it is unlikely that HAeBe stars would have much larger amounts 
of emission from accretion shocks. 
Consequently, we would not expect veiling to have much impact on HAeBe star spectra.  

If veiling is present in our observations it could make a star appear 
to be metal poor when it is not, thus it is important to check for veiling.  
If it is present in our spectra, veiling should have some wavelength dependence.  
We have compared abundances derived from different wavelength regions 
for individual stars.  No clear correlation of abundance with wavelength was found, 
suggesting that veiling is unimportant in these stars.  
Veiling should affect all lines in a wavelength region similarly.  
Thus, if it is present veiling should affect most elements similarly, 
since the lines for most elements are widely distributed in our spectra.  
In particular, it could not produce underabundances in iron peak elements 
while keeping solar abundances of C and O that we see in many of the stars in our sample.  
Therefore veiling should not affect the {\em relative} abundances of C, N, and O to Fe, 
even if it were present at a significant level in our spectra.  
We conclude it is unlikely that veiling has a significant impact on our derived abundances, 
and even if veiling is present it is not obvious how it could produce peculiar abundance patterns.  

Arguably, the pattern of $\lambda$ Boo underabundances we observe could be due to 
weak emission infilling in all the lines of some elements, 
however this is very unlikely.  We have searched carefully for emission in lines, 
both by looking for unexpected variability and by looking for lines with shapes inconsistent with  
rotational broadening.  Thus we have excluded most if not 
all lines with emission.  More importantly, lines of an element 
with lower excitation potentials are more likely to have stronger emission, 
thus emission would not be uniform across all lines.  
If emission infilling was occurring and otherwise undetected, 
we would observe it as an inability to simultaneously fit a 
wide range of lines of one element with one abundance.  
We do not find this, and consequently can safely conclude that 
the $\lambda$ Boo peculiarities we detect are real, 
and not simply due to emission infilling.

\section{Conclusions}
We have performed a detailed abundance analysis of 21 stars, 20 of which 
are Herbig Ae and Be stars, 1 of which ($\pi$ Cet) is not a Herbig star, 
but may still be on the pre-main sequence.  
Three of these stars have confirmed magnetic fields, while magnetic fields 
do not appear to be present at detectable levels in the remaining 18.   

We find 9 stars that are chemically normal (one of which is rather uncertain), 
11 that show $\lambda$ Bo\"otis chemical peculiarities, and one that is weakly Ap/Bp.  
This is a remarkably large fraction of stars that display $\lambda$ Boo type peculiarities.  
On the main sequence, only $\sim$2\% of stars display $\lambda$ Boo 
peculiarities \citep{Paunzen2001-lambdaBoo-survey-incidence}.  
We interpret this as evidence in favour of a selective accretion hypothesis 
for the development of $\lambda$ Boo peculiarities.  
Since Herbig stars are either accreting or have recently finished accreting, 
a process which depends on accretion should be seen more frequently in these stars. 

Of the three magnetic stars, HD 101412 displays $\lambda$ Boo peculiarities, 
HD 190073 is chemically normal, and V380 Ori is weakly Bp.  
Thus it appears that magnetic Herbig stars can display Ap/Bp type peculiarities, 
but not all magnetic Herbig stars do.  This is in contrast to main sequence 
magnetic A and late B stars, which always display Ap/Bp type peculiarities \citep{Auriere2007}.  
The presence of $\lambda$ Boo peculiarities in a magnetic star appears to simply 
reflect the high incidence of these peculiarities in all Herbig stars.

\section*{Acknowledgements} 
GAW and JDL are supported by Natural Science and Engineering Research Council (NSERC Canada) Discovery Grants, 
and GAW is supported by a Department of National Defence (Canada) ARP grant.
\bibliography{masivebib.bib}{}
\bibliographystyle{mn2e}


\appendix

\section{Results for individual stars}
\label{individual-stars}

\subsection*{Non-magnetic stars}

\subsection{$\pi$ Cet (HD~17081)}
\label{pi-Cet}
$\pi$ Cet (HD~17081) is a cool B star that was studied in detail by \citet{Fossati2009-normal-star}.  
They performed a very precise, detailed abundance analysis, and considered the star to be a 
good example of a `normal' cool B star.   $\pi$ Cet is not a Herbig Be star by the strict definition, 
since it has no strong emission in its optical spectrum.  
However, the star does display an infrared excess \citep{Malfait1998-ir-excess}, 
as well as some unexplained line profile variations \citep{Fossati2009-normal-star},
and an H-R diagram position to the right of the main sequence.  
Thus the star is certainly very young, and possibly still on the pre-main sequence, 
despite the lack of clear emission.  Based on the H-R diagram position of the star, it 
appears to be less than 0.5 Myr from the birth line, and almost certainly is less than 1 Myr old. 
\citet{Malfait1998-ir-excess} suggested this star was a Vega-type star, 
a classification which is consistent with our observations. 

\begin{figure}
\centering
\includegraphics[width=3.3in]{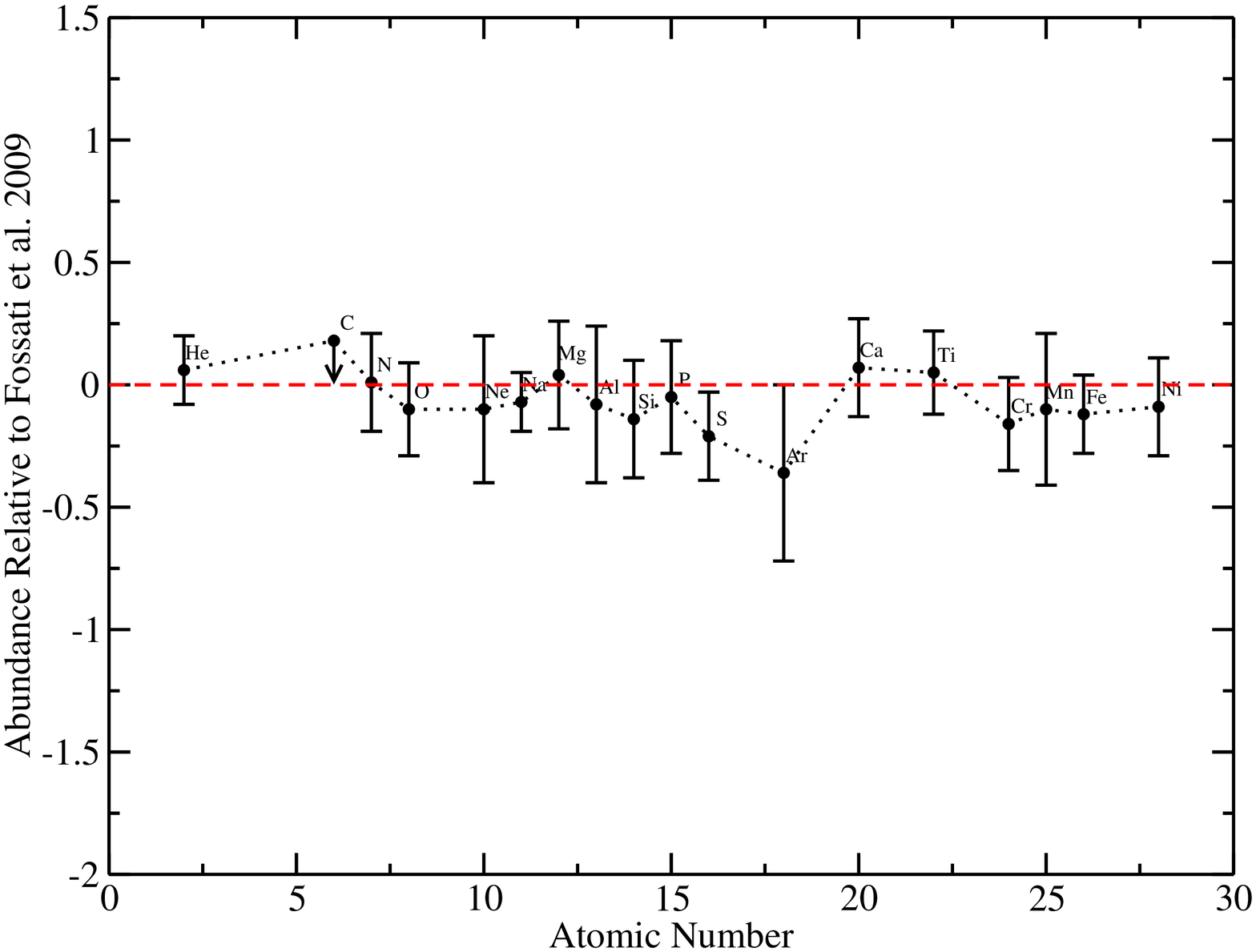}
\caption{The difference between our final abundances and those of 
\citet{Fossati2009-normal-star} for $\pi$ Cet.  A good agreement between the two sets of abundances is found. }
\label{abunplot-wrt-fossati}
\end{figure}

In order to check the accuracy of our methodology, we analysed an observation of $\pi$ Cet 
and compared our results with the results of \citet{Fossati2009-normal-star}.  
Our results are completely consistent with those of \citet{Fossati2009-normal-star}, 
for temperature, gravity, \vs, and abundances.  Our atmospheric parameters and abundances 
are compared with those of \citet{Fossati2009-normal-star} in Table \ref{pi-cet-with-fossati}, 
and the differences between our abundances and those of \citet{Fossati2009-normal-star} 
are plotted in Fig. \ref{abunplot-wrt-fossati}.  The excellent agreement between 
our results and \citet{Fossati2009-normal-star} demonstrates the accuracy of 
our methodology.  

$\pi$ Cet was observed spectropolarimetricly by \citet{Wade2007-HAeBe_survey}, 
who did not detect a magnetic field.  \citet{Alecian2012-big-HAeBe-magnetism} also found no magnetic field 
in $\pi$ Cet, and thus we consider the star to be non-magnetic.  

\begin{table}
\centering
\caption{Our results compared with those of \citet{Fossati2009-normal-star} for $\pi$ Cet. }
\label{pi-cet-with-fossati}
\begin{tabular}{ccc}
\hline \hline \noalign{\smallskip}
      &  Fossati et al. (2009)   & This work  \\
\hline
\teff~(K)   & $12800 \pm 200$  & $12900 \pm 400 $ \\
\lgg        & $3.75 \pm 0.1$   & $3.8 \pm 0.2 $   \\
\vs~(\kms)  & $20.2 \pm 0.9$   & $20.9 \pm 1.2 $  \\
\vmic~(\kms)& $1.0 \pm 0.5$    & $1.7 \pm 1.0 $   \\
\hline
He          & $-0.97 \pm 0.04$ & $-0.91 \pm 0.14$ \\
C           & $-3.58 \pm 0.07$ & $\leq -3.4     $ \\
N           & $-4.03 \pm 0.13$ & $-4.03 \pm 0.15$ \\
O           & $-3.06 \pm 0.14$ & $-3.16 \pm 0.13$ \\
Ne          & $-3.66 \pm 0.09$ & $-3.76 \pm 0.28$ \\
Na          & $-5.23 \pm 0.07$ & $-5.3  \pm 0.1 $ \\
Mg          & $-4.47 \pm 0.16$ & $-4.43 \pm 0.15$ \\
Al          & $-5.73 \pm 0.27$ & $-5.81 \pm 0.17$ \\
Si          & $-4.41 \pm 0.2 $ & $-4.55 \pm 0.14$ \\
P           & $-6.38 \pm 0.19$ & $-6.43 \pm 0.14$ \\
S           & $-4.78 \pm 0.16$ & $-4.99 \pm 0.08$ \\
Ar          & $-5.24 \pm 0.19$ & $-5.6  \pm 0.3 $ \\
Ca          & $-5.77         $ & $-5.7  \pm 0.2 $ \\
Ti          & $-7.42 \pm 0.08$ & $-7.37 \pm 0.15$ \\
Cr          & $-6.41 \pm 0.1 $ & $-6.57 \pm 0.16$ \\
Mn          & $-6.5  \pm 0.09$ & $-6.6  \pm 0.3 $ \\
Fe          & $-4.58 \pm 0.14$ & $-4.70 \pm 0.08$ \\
Ni          & $-5.76 \pm 0.19$ & $-5.85 \pm 0.06$ \\
\noalign{\smallskip} \hline \hline
\end{tabular}
\end{table}

\subsection{HD 31293 (AB Aur)}
We derive \teff~$=9800 \pm700$ K and \lgg~$=3.9\pm0.3$ for HD 31293 (AB Aur).  
This \teff\, is consistent with \citet{Acke2004-HAeBe_Abun}, 
but our \lgg\, is inconsistent with their (remarkably high) value of $5.0$.  
The \teff\, and \lgg\, are both consistent with \citet{Boehm1993-AB-Aur}.  
We find solar He and O abundances, C and N are 1$\sigma$ above solar, 
while the iron peak elements are consistently 1$\sigma$ below solar.  
The systematic underabundance of iron peak elements by $\sim$0.3 dex suggests 
the star may be a weak $\lambda$ Boo star.  
This is strengthened somewhat by the observation that the 
iron peak elements are $\sim$0.5 dex below the average C, N, and O abundance.  
Based on these underabundances, we conclude that the star most likely 
is a weak $\lambda$ Boo star, but it is a marginal case.  

 \citet{Acke2004-HAeBe_Abun} derive abundances for O, N, and Si that 
are consistent with ours, but their Fe abundance of 1 dex below solar 
(based on 2 Fe {\sc i} lines) is inconsistent with our value.  
This difference may be due to the higher \lgg\, \citet{Acke2004-HAeBe_Abun} 
used.  Alternately, it is possible their measurement was impacted 
by the large amount of emission in the spectrum of HD 31293.  
Both Fe abundances are consistent with a $\lambda$ Boo 
classification of the star, but we prefer our larger abundance 
since it is based on larger number and a wider range of lines, 
and it is consistent with the neighbouring iron peak element 
abundances we derive.  

\citet{Alecian2012-big-HAeBe-magnetism} find no magnetic field in HD 31293.  
\citet{Wade2007-HAeBe_survey} also find no magnetic field in this star,
thus we consider it to be non-magnetic.  

\subsection{HD 31648}
For HD 31648 we derive an \teff\, of $8800 \pm190$ K and a \lgg\, of $4.1\pm0.2$, 
which is consistent with \citet{Acke2004-HAeBe_Abun}.  
We find most abundances within 1$\sigma$ of solar, and all abundances 
except Mg and Ti within 2$\sigma$.  The abundances are systematically slightly 
above solar and may reflect a slightly enhanced metallicity, 
or a slightly overestimated \teff.  

Our Ca and Fe abundances are consistent with \citet{Acke2004-HAeBe_Abun}, 
our C abundance is consistent at 2$\sigma$, as is our Y abundance 
if we assume the uncertainty on their value is comparable to ours.  
Our Si and Mg abundances are not consistent with \citet{Acke2004-HAeBe_Abun}, 
although their abundances are based on only 1 line and thus are likely uncertain.  
The large \vs\, of this star causes significant blending 
in its spectrum.  This could explain the discrepancies between our results, 
since the \citet{Acke2004-HAeBe_Abun} abundances are based on 
equivalent width measurements, which can easily be influenced by 
unaccounted-for blending.  

HD 31648 was not found to have a magnetic field by \citet{Alecian2012-big-HAeBe-magnetism}.  
No magnetic field was found by \citet{Wade2007-HAeBe_survey} either.  
\citet{Hubrig2006-HAeBe-fields} claimed a weak 
detection of a magnetic field in their observation of HD 31648 
but, in the re-reduction of the observation made by
\citet{Hubrig2007-HAeBe-magnetic-circumstellar} the field is below their detection threshold. 
Surprisingly, \citet{Hubrig2007-HAeBe-magnetic-circumstellar} report a signal 
the H$\beta$, H$\gamma$, H$\delta$, and Ca H \& K lines of their $V/I$ spectrum, 
which they interpret as a ``circumstellar'' magnetic field.  
This result has yet to be independently confirmed.  
A careful examination of the H$\beta$, H$\gamma$, H$\delta$, and Ca H \& K lines 
in our spectra shows no polarisation signature.  
\citet{Bagnulo2011-FORS-rereduction} re-reduced a wide range 
of spectropolarimetric observations from the FORS1 instrument at the VLT, 
including the observations of \citet{Hubrig2006-HAeBe-fields}.
They find no evidence for a magnetic field in HD 31648.  
Considering the lack of evidence for a photospheric magnetic field, 
and the absence of a confirmation of a ``circumstellar'' magnetic field, 
we conclude that HD 31648 is non-magnetic.  

\subsection{HD 36112}
For HD 36112 we find \teff~$=8190 \pm150$ K and \lgg~$=4.1\pm0.4$, 
both of which are somewhat greater than the \teff~$=7750$ and \lgg~$=3.5$ 
of \citet{Acke2004-HAeBe_Abun}.  
Our uncertainty on \teff\, may be overly optimistic, 
though it does reflect the scatter in \teff\, derived from different spectral windows.  
The uncertainties on the calibration of \teff\, become important at this level, 
and thus there may be some systematic uncertainty not included in our error bar.  
Most of our derived abundances are within 1$\sigma$ of solar, 
and all are within 2$\sigma$.  There are no systematic enhancements or depletions, 
thus we conclude HD 36112 is chemically normal, with solar abundances.  

Our abundances are consistent with those of \citet{Acke2004-HAeBe_Abun} 
for C, Si, S, Ca, Sc, Cr, Mn, Fe, and Ni.  Our abundance 
for Ti and Y differ by $\sim$2$\sigma$.  These differences may be 
a consequence of our hotter temperature, or they may be due 
to the small number of lines \citet{Acke2004-HAeBe_Abun} used for 
determining those abundances.  

\citet{Alecian2012-big-HAeBe-magnetism} do not detect a magnetic field in this star.   
In their FORS1 observations \citet{Wade2007-HAeBe_survey} do not detect a magnetic field, 
nor do \citet{Bagnulo2011-FORS-rereduction} in their re-reduction of the FORS1 observations. 

\subsection{HD 68695}
For HD 68695 we derive \teff~$=9000 \pm300$ K and \lgg~$=4.3\pm0.3$.  
We find C, N, O, and S abundances that are $\sim$1$\sigma$ above solar, 
while iron peak abundances are 2 to 3$\sigma$ below solar ($\sim$0.7 dex below solar), 
except for the rather uncertain Mn abundance.  We conclude that this star 
has clear $\lambda$ Boo peculiarities.  

No magnetic field was detected in HD 68695 by either \citet{Alecian2012-big-HAeBe-magnetism} 
or by \citet{Wade2007-HAeBe_survey}, thus we consider the star to be non-magnetic.  

\subsection{HD 139614}
HD 139614 was analysed by \citet{Acke2004-HAeBe_Abun} using an \teff\, of 8000 K.  
They found an overall deficiency of metals, but no clear evidence of the 
selective depletion characteristic of $\lambda$ Boo stars.  
We find HD 139614 to have an \teff\, of $7600 \pm 300$ K, and a \lgg\, of $3.9 \pm 0.3$, 
which is consistent with the results of \teff~$ = 7400 \pm 200$ K and \lgg~$ = 4.0 \pm 0.4$ 
reported by \citet{Guimaraes2006-HAeBe-param}, 
and marginally consistent with \citet{Acke2004-HAeBe_Abun} (at 1.3$\sigma$ in \teff\, and 2$\sigma$ in \lgg).  
We consider our parameters \citep[and][]{Guimaraes2006-HAeBe-param} to be more accurate, 
since \citet{Acke2004-HAeBe_Abun} values were based on the spectral classification of 
 \citet{Malfait1998-ir-excess},  which could be influenced by 
chemical peculiarities in the star.  
In fact, for our parameters we find a clear $\lambda$ Boo pattern of abundances, 
with solar C, N, and O, and underabundant iron peak elements.  

No magnetic field was detected in HD 139614 by \citet{Alecian2012-big-HAeBe-magnetism}.   
\citet{Hubrig2004-HAeBe} claimed a detection of a magnetic field in this star, 
which was not supported by \citet{Wade2005-HAeBe_Discovery}.  
Further observations by \citet{Hubrig2009-several-HAeBe} did not detect 
a magnetic field in the star, though they comment on a possible signal in the Ca H and K lines.  
In their re-reduction of FORS data, \citet{Bagnulo2011-FORS-rereduction} 
find no evidence for a magnetic field in HD 139614, and conclude that the reported 
detection by \citet{Hubrig2004-HAeBe} is likely spurious.
Consequently, we consider HD 139614 not to have a significant magnetic field.  

\subsection{HD 141569}
We find \teff~$= 9800 \pm 500$ K and \lgg~$= 4.2 \pm 0.4$ for HD 141569, 
which is consistent with the parameters found by \citet{Guimaraes2006-HAeBe-param}.
We find $\lambda$ Boo peculiarities in this star, with solar He, C and N, 
possibly overabundant O, and underabundant iron peak elements.  

\citet{Alecian2012-big-HAeBe-magnetism} do not detect a magnetic field in the star. 
\citet{Wade2007-HAeBe_survey} also find no magnetic field, 
thus we conclude HD 141569 does not have a significant magnetic field. 

\subsection{HD 142666}
For HD 142666 we find \teff~$=7500 \pm 200$ K and \lgg~$= 3.9\pm0.3$, 
which is consistent with \citet{Guimaraes2006-HAeBe-param}.  We find weak 
 $\lambda$ Boo peculiarities in this star.  C, O, Na and S have solar abundance, 
while the iron peak elements are consistently $\sim$0.5 dex below solar, 
except for the rather uncertain Co abundance.

\citet{Wade2007-HAeBe_survey} observed the star and found no magnetic field. 
\citet{Alecian2012-big-HAeBe-magnetism} find no magnetic field in the star either.  

\subsection{HD 144432}
For HD 144432 we find \teff~$=7400 \pm200$ K and \lgg~$=3.9\pm0.3$, 
which is consistent with \citet{Guimaraes2006-HAeBe-param}.  
The abundances of all elements we derive are within 2$\sigma$ of solar, 
and most are within 1$\sigma$.  There are no clear trends in abundances 
of different elements, thus we consider the star chemically normal, 
with nearly solar metallicity.  

\citet{Alecian2012-big-HAeBe-magnetism} observed HD 144432 twice and found no magnetic field. 
\citet{Wade2007-HAeBe_survey} also found no magnetic field in the star. 
\citet{Hubrig2007-HAeBe-magnetic-circumstellar} claim a weak magnetic field detection 
in this star in one observation, but \citet{Hubrig2004-HAeBe} 
and \citet{Hubrig2009-several-HAeBe} did not detect a significant magnetic field in the star.  
\citet{Bagnulo2011-FORS-rereduction} re-reduced the one observation of HD 144432 
with a magnetic field detection from \citet{Hubrig2007-HAeBe-magnetic-circumstellar}.  
\citet{Bagnulo2011-FORS-rereduction} do find a magnetic field in the observation 
at a $4.9\sigma$ level ($\langle B_z \rangle =-108\pm22$ G).  However, they caution that the impact of 
instrumental effects on the detection of weak magnetic fields is not well understood for the FORS1 instrument. 
They conclude that for weak magnetic fields, detections at less than $6\sigma$ should not be 
considered definite, and require supporting observations.  
Thus, while we cannot completely exclude the possibility of a magnetic field in HD 144432,
based on the large majority of the evidence (five out of six observations)
we conclude that HD 144432 likely has no significant magnetic field.  

\subsection{HD 163296}
We determine an \teff\, of $9200 \pm300$ K and a \lgg\, of $4.2\pm0.3$, 
which is consistent with \citet{Guimaraes2006-HAeBe-param}.  
We find the abundances of most elements to be within uncertainty of solar, 
though Mg and Cr are both overabundant at slightly more than 2$\sigma$. 
With largely solar abundance and no clear selective depletion or enhancement, 
we conclude that the star is chemically normal.  

\citet{Alecian2012-big-HAeBe-magnetism} do not find a magnetic field in HD 163296.  
\citet{Hubrig2009-several-HAeBe} also do not detect a magnetic field in the star, 
thus we conclude the star is non-magnetic

\subsection{HD 169142}
For HD 169142 we find \teff~$=7500 \pm200$ K, which is consistent 
with \citet{Guimaraes2006-HAeBe-param},  and \lgg~$=4.3\pm0.2$, 
which is 2$\sigma$ greater than \citet{Guimaraes2006-HAeBe-param} who find \lgg~$=3.7\pm0.1$.  
If we were to adopt the \lgg\, of \citet{Guimaraes2006-HAeBe-param}, 
it would decrease our abundances by between $\sim$0.05 and $\sim$0.2 dex depending on the element, 
which is generally within our uncertainties.  
We find a clear $\lambda$ Boo pattern of abundances in this star. 
C and O are solar, and S is only slightly underabundant, 
while iron peak elements are underabundant by between 0.5 and 1 dex.  
Interestingly Ba and Nd are roughly solar, though Eu may be underabundant.  

\citet{Alecian2012-big-HAeBe-magnetism} do not detect a magnetic field in HD 169142, 
nor do \citet{Hubrig2009-several-HAeBe}.  

\subsection{HD 176386}
We derive an \teff\, of $11000\pm400$ K and a \lgg\, of $4.1\pm0.3$ for HD 176386.  
The abundances we derive for the star are consistent with solar, 
with the possible exception of Ca.  Thus we conclude that the star is chemically normal. 

No magnetic field was found for HD 176386 by \citet{Alecian2012-big-HAeBe-magnetism}.  
\citet{Hubrig2009-several-HAeBe} claim a marginal detection at 3$\sigma$ in one observation.  
\citet{Bagnulo2011-FORS-rereduction} re-reduced this observation and found no magnetic field.
Thus, based on the majority of the evidence, we consider HD 176386 to be a non-magnetic star.  

\subsection{HD 179218}
For HD 179218 we derived \teff~$=9640 \pm250$ K and \lgg~$=3.9\pm0.2$, 
both of which are consistent with \citet{Guimaraes2006-HAeBe-param}. 
We find a solar He abundance, abundances of C, O, and Na that are 
marginally enhanced relative to solar, and N and S abundances that 
are uncertain, but apparently enhanced relative to solar.  
Iron peak elements are depleted relative to solar by $\sim$0.5 dex.  
We conclude that HD 179218 displays $\lambda$ Boo peculiarities.   
The overabundances of C, N, O, and S may reflect an above solar intrinsic metallicity in the star, 
which the $\lambda$ Boo peculiarities are superimposed upon.  
However, this is not certain, as He appears to have a nearly solar abundance.  

\citet{Alecian2012-big-HAeBe-magnetism} find no magnetic field in HD 179218, 
which is supported by the non-detection of \citet{Hubrig2009-several-HAeBe}.  

\subsection{HD 244604}
We found \teff~$= 8700 \pm220$ K and \lgg~$=4.0\pm0.2$ for HD 244604, 
which is consistent with the parameters from \citet{Acke2004-HAeBe_Abun}.  
We find most elements have solar abundances or are enhanced by around 1$\sigma$.  
There is no evidence for selective depletion or enhancement.  
We conclude that the star is chemically normal. 
\citet{Acke2004-HAeBe_Abun} were only able to determine an iron abundance 
for HD 244604, though it is consistent with our abundance if we assume 
their uncertainty is similar to ours.  

No magnetic field in HD 244604 is found by \citet{Alecian2012-big-HAeBe-magnetism}. 
\citet{Wade2007-HAeBe_survey} also find no magnetic field in the star, 
thus we consider the star to be non-magnetic.  

\subsection{HD 245185}
For HD 245185 we derive \teff~$=9500 \pm750$ K and \lgg~$=4.0\pm0.4$, 
which is consistent with \citet{Acke2004-HAeBe_Abun}.  
We find He, C, N, and O abundances that are consistent with solar, 
and iron peak abundances that are $\sim$0.8 dex below solar.  
While the uncertainties are relatively large for this star, 
the strong iron peak underabundances indicate $\lambda$ Boo peculiarities.  
\citet{Acke2004-HAeBe_Abun} were unable to determine any abundances 
for this star due to a low S/N in their observation and 
the high \vs\, of the star.  

\citet{Alecian2012-big-HAeBe-magnetism} do not detect a magnetic field in the star, 
nor do \citet{Wade2007-HAeBe_survey}.  Consequently, 
we consider the star to be non-magnetic.  

\subsection{HD 278937 (IP Per)}
For HD 278937 (IP Per) we find \teff~$=8000 \pm250$ K and \lgg~$=4.1\pm0.2$, 
which is consistent with the \teff\, of \citet{Miroshnichenko2001-ipPer-param}, 
though they derive a \lgg\, of $4.4\pm0.1$ which is 1$\sigma$ above our value.  
The star is a known $\delta$ Scuti pulsator \citep{Ripepi2006-ipPer-delta-scuti}.
We find solar abundances for C, N, O, and S, while iron peak 
elements as well as Na, Mg, and Si are between 0.5 and 0.7 dex underabundant. 
This star shows clear $\lambda$ Boo peculiarities.  
\citet{Miroshnichenko2001-ipPer-param} found that HD 278937 had a metallicity 
below solar ($[M/H]=-0.4$), based on a fit to the star's metallic line spectrum.  
This value is most likely a result of the low iron peak abundances in the star, 
and thus consistent with our abundances.  
The solar abundances of C, N, O and S were likely overlooked by 
the scaled solar abundance model of \citet{Miroshnichenko2001-ipPer-param} 
due to the relative scarcity of lines of those elements.  
Therefore, the star is not a low metallicity star, but rather appears 
to have peculiar photospheric abundances.  

No magnetic field is found in HD 278937 by \citet{Alecian2012-big-HAeBe-magnetism}, 
or by \citet{Wade2007-HAeBe_survey}.  

\subsection{T Ori}
We find \teff~$=8500 \pm300$ K and \lgg~$=4.2\pm0.3$ for T Ori.  
The abundances for He, C, N, and O are all consistent with solar, 
while the S abundance is almost 3$\sigma$ above solar.  
The S abundance is based only on two weak lines at 6749 and 6757 \AA, 
and thus may be somewhat more uncertain than the error bar suggests.  
The iron peak abundances are clearly $\sim$0.5 dex below solar.  
We conclude that T Ori has clear $\lambda$ Boo peculiarities

Neither \citet{Alecian2012-big-HAeBe-magnetism} nor \citet{Wade2007-HAeBe_survey} 
detect a magnetic field in T Ori.

\subsection*{Magnetic stars}

\subsection{HD 101412}
HD 101412 was observed by \citet{Wade2005-HAeBe_Discovery} and reanalysed by 
\citet{Wade2007-HAeBe_survey} who found a longitudinal magnetic field of $500 \pm 100$ G.  
The presence of a magnetic field was recently confirmed by 
\citet{Hubrig2009-several-HAeBe,Hubrig2011-HD101412-mag-geom}, and Alecian et al. (in prep.).  
\citet{Cowley2010-hd101412-lambdaBoo} derived chemical abundances for HD 101412, 
and surprisingly discovered it had $\lambda$ Bo\"otis peculiarities.  
In our abundance analysis we find underabundances of iron peak elements, 
and solar C, O, and He, with values that are consistent with those of \citet{Cowley2010-hd101412-lambdaBoo}.  
Thus we confirm that HD 101412 has $\lambda$ Boo peculiarities.  
While in main sequence A and B stars magnetic fields are associated with 
Ap and Bp chemical peculiarities, that is not the case for HD 101412.  

We confirm the presence of an ``anomalous saturation'' in the lines of HD 101412 noted by 
\citet{Cowley2010-hd101412-lambdaBoo}.  Even with a microturbulence of 0 \kms, 
when we fit weaker lines, stronger lines in the synthetic spectrum are often too deep to fit the observation.  
This problem is not consistent across all lines, and often appears to be fairly minor.  
This anomalous saturation does not appear to be common among HAeBe stars, 
as this is the only object for which we have seen this problem.  
A possible explanation for the discrepancy between the observations and the synthetic spectra is 
the presence of chemical stratification in the stellar atmosphere.  
Since stronger lines are often formed higher in the stellar atmosphere, 
if the abundance of some elements was lower higher in the atmosphere 
that would explain the apparent weakness of many stronger lines.  
The possibility of stratification will be investigated in detail in a future paper.  
The presence of veiling in the spectrum of this star would produce a similar discrepancy to what we see, 
however veiling would affect all lines of the same strength similarly, unlike the non-uniform trend we see.  
Thus we conclude veiling is unlikely to be the source of this discrepancy, 
in agreement with \citet{Cowley2010-hd101412-lambdaBoo}.  
The discrepancy is not likely to be a direct effect of the star's magnetic field on line formation, 
which usually desaturates lines, and we do not find the problem in 
the other magnetic stars of our sample.  

Placing HD 101412 on the H-R diagram, we find it has an age of $1.2 ^{+0.8}_{-0.7}$ Myr and 
a mass of $3.0\pm0.3$ \Msun, both of which are consistent with \citet{Wade2007-HAeBe_survey}.  
Since the star has no Hipparcos parallax, 
we adopted the distance of 500-700 pc proposed by \citet{Vieira2003-HAeBe_ID}, 
however this is somewhat uncertain.  Even with this uncertainty, 
the star falls well before the beginning of the main sequence on the H-R diagram, 
thus the star is almost certainly a pre-main sequence object.  

We find similar $\lambda$ Boo type peculiarities in many of the non-magnetic 
HAeBe stars in our sample, thus this feature is not unique to HD 101412, 
or to magnetic HAeBe stars.  Indeed, it appears that the peculiarities in HD 101412 
occur in a significant fraction of all HAeBe stars.

\subsection{HD 190073}
HD 190073 has a magnetic field, discovered and repeatedly detected by \citet{Catala2007-HD190073}, 
with a longitudinal strength of $74 \pm 10$ G that was constant over the course of their observations.  
In our analysis we find that the star is chemically normal, as shown in Fig. \ref{abunplots4}.  
Apparent over abundances relative to solar in N and Na may be due to non-LTE effects, 
or weak undiagnosed blends, since for both elements only a couple of 
very weak lines were usable for our analysis. 

\citet{Acke2004-HAeBe_Abun} analysed HD 190073 and adopted \teff~$=9250$ K and \lgg~$=3.5$, 
which is consistent with our values of \teff~$=9230 \pm260$ K and \lgg~$=3.7\pm0.3$.  
Based on equivalent widths, they found abundances for 
C, N, O, Mg, Si, Ca, Sc, Ti, V, Cr, Fe, Ni Y, Zr, and Ba, all of which are consistent with ours.  
For Al they found an abundance -0.45 dex below solar, based on only one line, 
which is inconsistent with our value.  Our abundance is based on two lines, 
thus we consider it to be the more accurate value.  
Our result leads to a nearly solar abundance, consistent with other elements in this star.

We find that the star is quite young ($1.6 \pm 0.6$ Myr), which agrees with the age 
determined by \citet{Catala2007-HD190073} of $1.2 \pm 0.6$ Myr.  
Thus it is possible that the star has not yet had time to build up significant 
chemical peculiarities on its surface.  
This hypothesis is supported by the modelling of \citet{Vick2011-model-pms-peculiarities}, 
who found that it takes at least $\sim$2 Myr for a 2.9\Msun\, star to 
develop significant chemical peculiarities by radiative diffusion.

\subsection{V380 Ori A \& B}
V380 Ori is a triple star system, with two components detectable spectroscopically \citep{Alecian2009-v380ori}.  
The more massive spectroscopic component (the primary) shows no detectable change in radial velocity, 
while the fainter, less massive (secondary) component's radial velocity varies with a period 
of $\sim$100 days \citep{Alecian2009-v380ori}.  The third component in the system was 
detected by speckle interferometry in the infra red \citep{Leinert1997-v380ori-speckle}, but is not detectable in 
the visible spectrum of the system \citep{Alecian2009-v380ori}. 
The primary has a magnetic field, detected by \citet{Wade2005-HAeBe_Discovery}, 
while the secondary has no detectable magnetic field.  The magnetic field of the primary 
was repeatedly observed by \citet{Alecian2009-v380ori}, who modelled the geometry as a dipole, 
with a strength of $2120 \pm 150$ G, and an obliquity of $66 \pm 5^\circ$. 

Modelling the spectrum of the V380 Ori system was done somewhat differently than 
for the other stars in this study.  Spectral disentangling was inapplicable 
due to the large amounts of variable emission in the spectrum, 
and the lack of radial velocity variation in the primary.  
Instead we fit a composite synthetic spectrum directly to the observation.  
The composite spectrum was constructed by adding the synthetic spectra of both stars, 
in absolute flux, weighted by the squared ratio of radii of the stars 
(the $T_{\rm eff}^4$ dependence is included in the absolute fluxes of the synthetic spectra).  
The combined spectrum was then normalised by the 
sum of the continuum fluxes, also weighted by the ratio of radii squared.  
Using this composite spectrum, the model parameters for one star were simultaneously 
fit by  $\chi^2$ minimisation, while the other star's parameters were held fixed. 
The ratio of radii was simultaneously fit as well, but the radial velocities 
of the two components were determined before the rest of the fitting process.  
The combination of parameters for both stars from one spectral window was determined iteratively, 
alternating between the two stars, and fitting the parameters 
of one while holding the parameters of the other fixed.  
Initial parameters were taken from \citet{Alecian2009-v380ori}.

We find chemical peculiarities in the primary that are consistent with Ap/Bp peculiarities.  
The peculiarities are weak, but detectable.  C, S, and Ne are within uncertainty of solar, 
while Fe, Ni, and Mn are overabundant by more than $3\sigma$.  
Si is also overabundant by more than $4\sigma$ in the primary, and He may be overabundant at $2.2\sigma$. 
The abundances of the secondary are uncertain, but consistent with solar. 
Most elements in the secondary are within $1\sigma$ of solar, 
and all elements are within $2\sigma$ (except for Ba).  
The large uncertainty in the secondary stems mostly from the uncertainty in the ratio of radii 
(and hence ratio of luminosities) for the system, though uncertainties in effective temperature 
also play a role.  While the average abundance in the secondary appears to be slightly above solar, 
it is not clear whether this reflects an intrinsically higher metallicity in the system, 
or if the luminosity ratio has simply been slightly overestimated.  
The abundances derived for the secondary are highly sensitive to the adopted ratio of radii, 
however the abundances for the primary are not, since the primary dominates the luminosity of the system.

We find a ratio of radii for the V380 Ori system of $R_A/R_B = 0.7 \pm 0.1$.  
The relatively large radius of the secondary is a consequence of the star being less massive, 
and thus being slower to contract towards the main sequence than the more massive primary.  
The H-R diagram positions of the two stars are consistent with a single isochrone.  
 
The appearance of chemical peculiarities in the primary of V380 Ori is consistent with 
the modelling by \citet{Vick2011-model-pms-peculiarities}, who discuss this system in some detail.  
Since the primary is more evolved than HD 190073, it could have developed chemical peculiarities 
while HD 190073 has not yet.

\begin{figure*}
\centering
\includegraphics[width=3.4in]{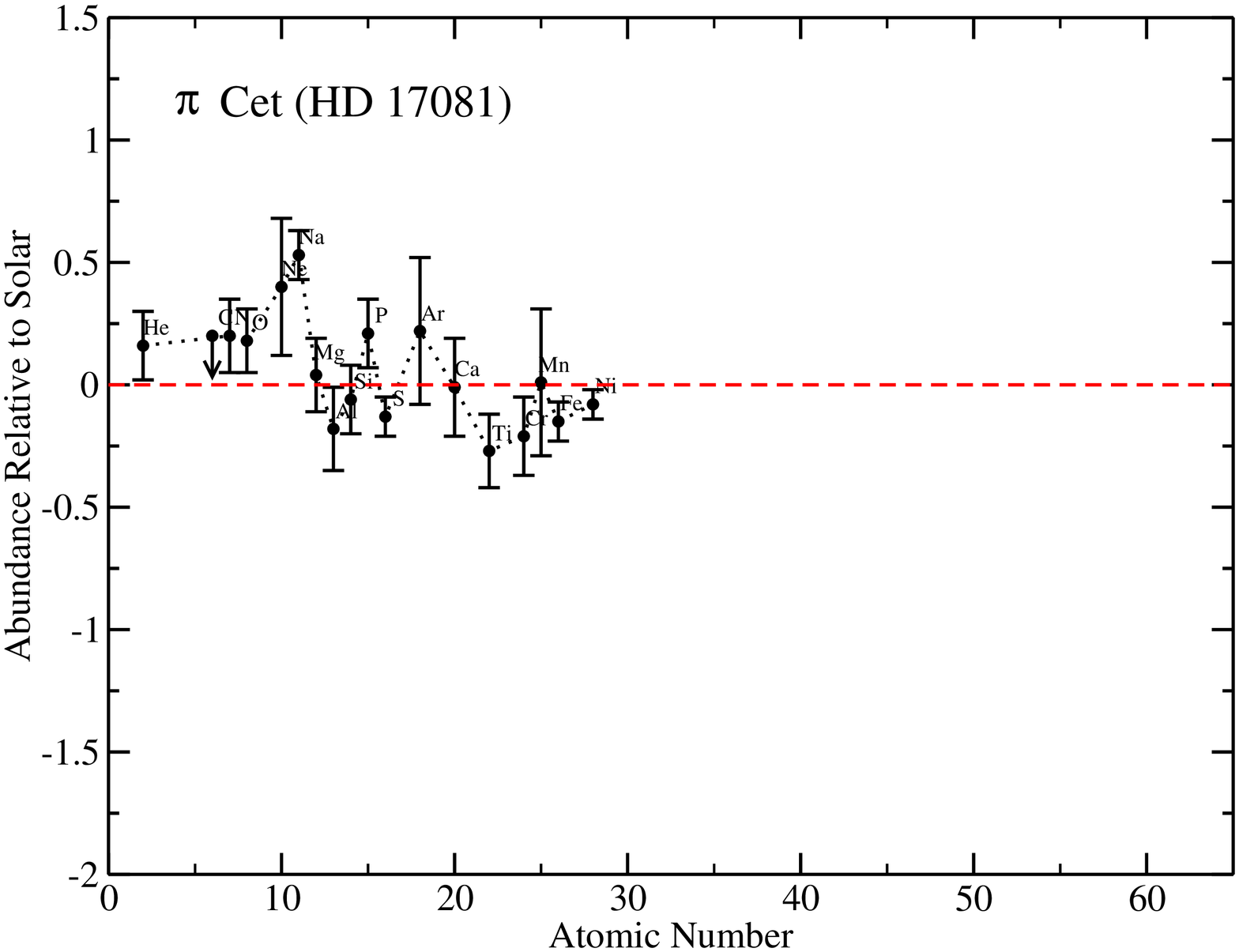}
\includegraphics[width=3.4in]{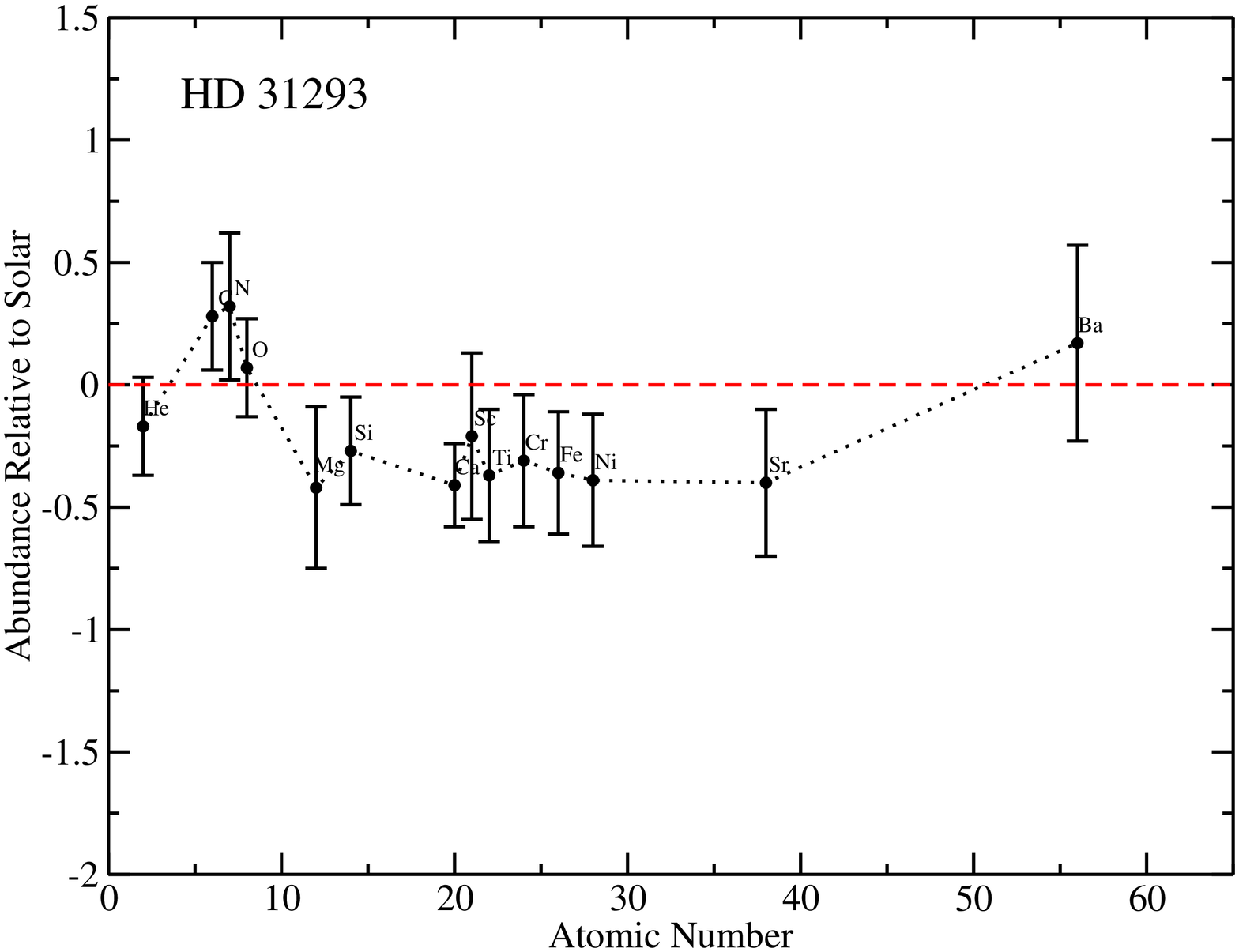}
\includegraphics[width=3.4in]{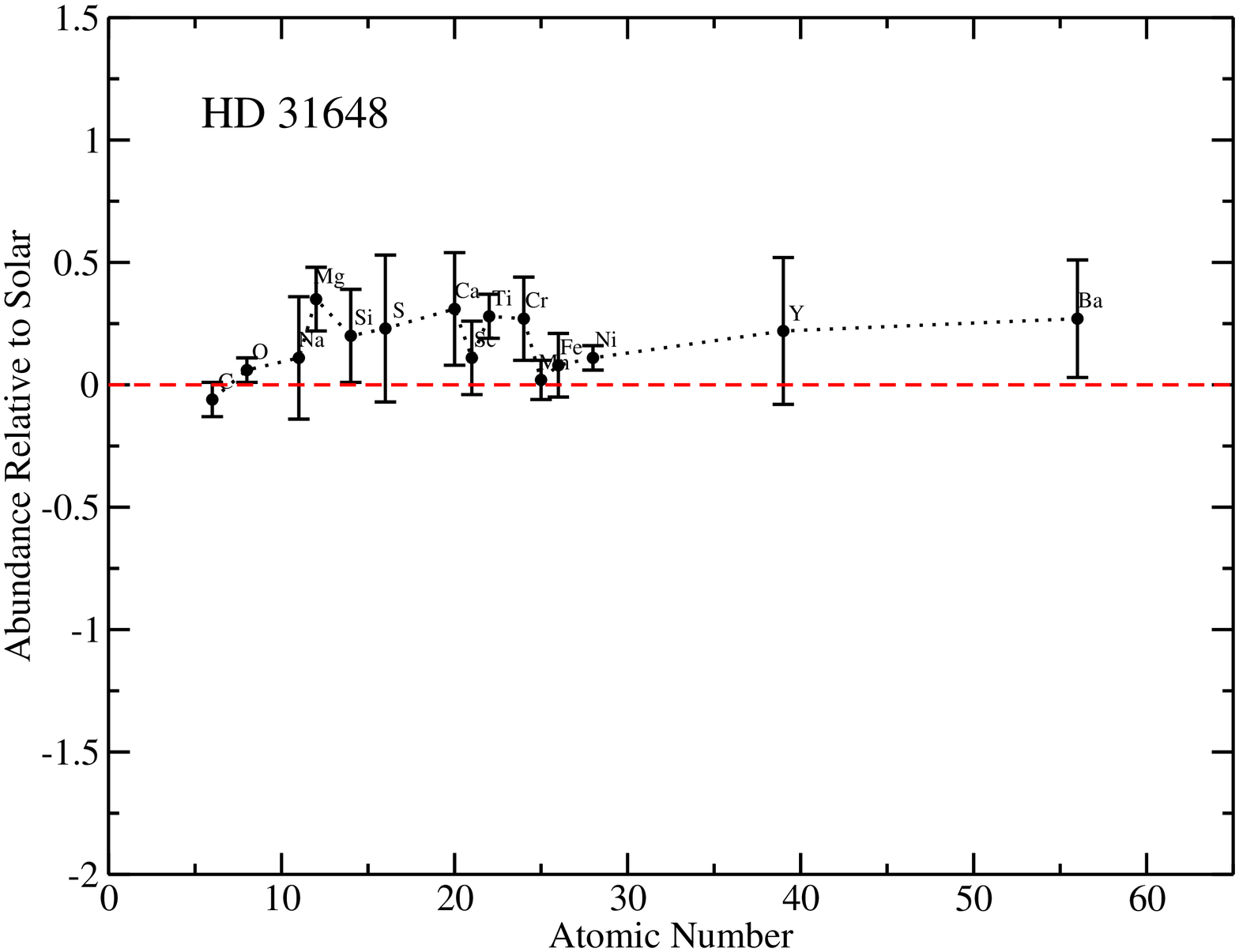}
\includegraphics[width=3.4in]{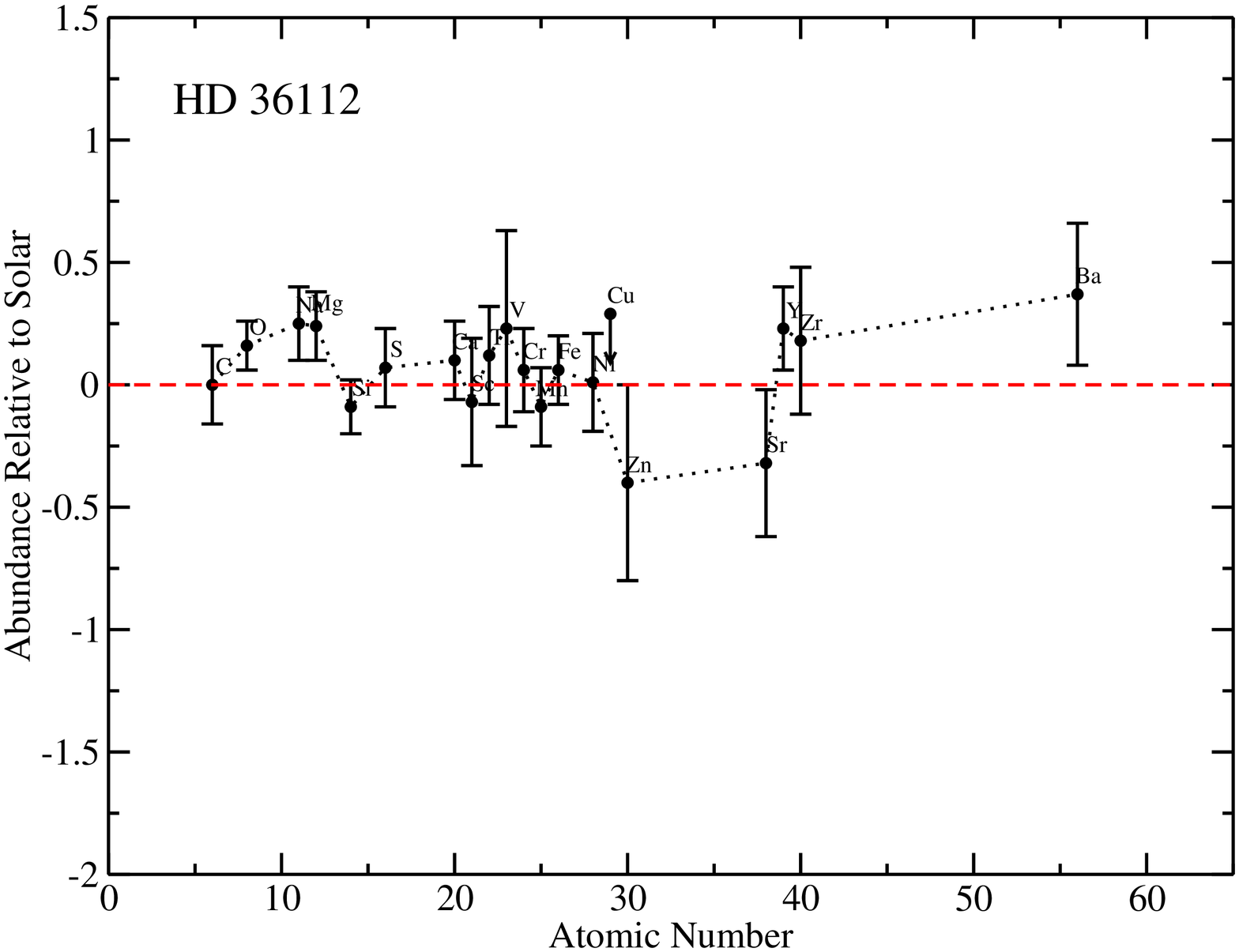}
\includegraphics[width=3.4in]{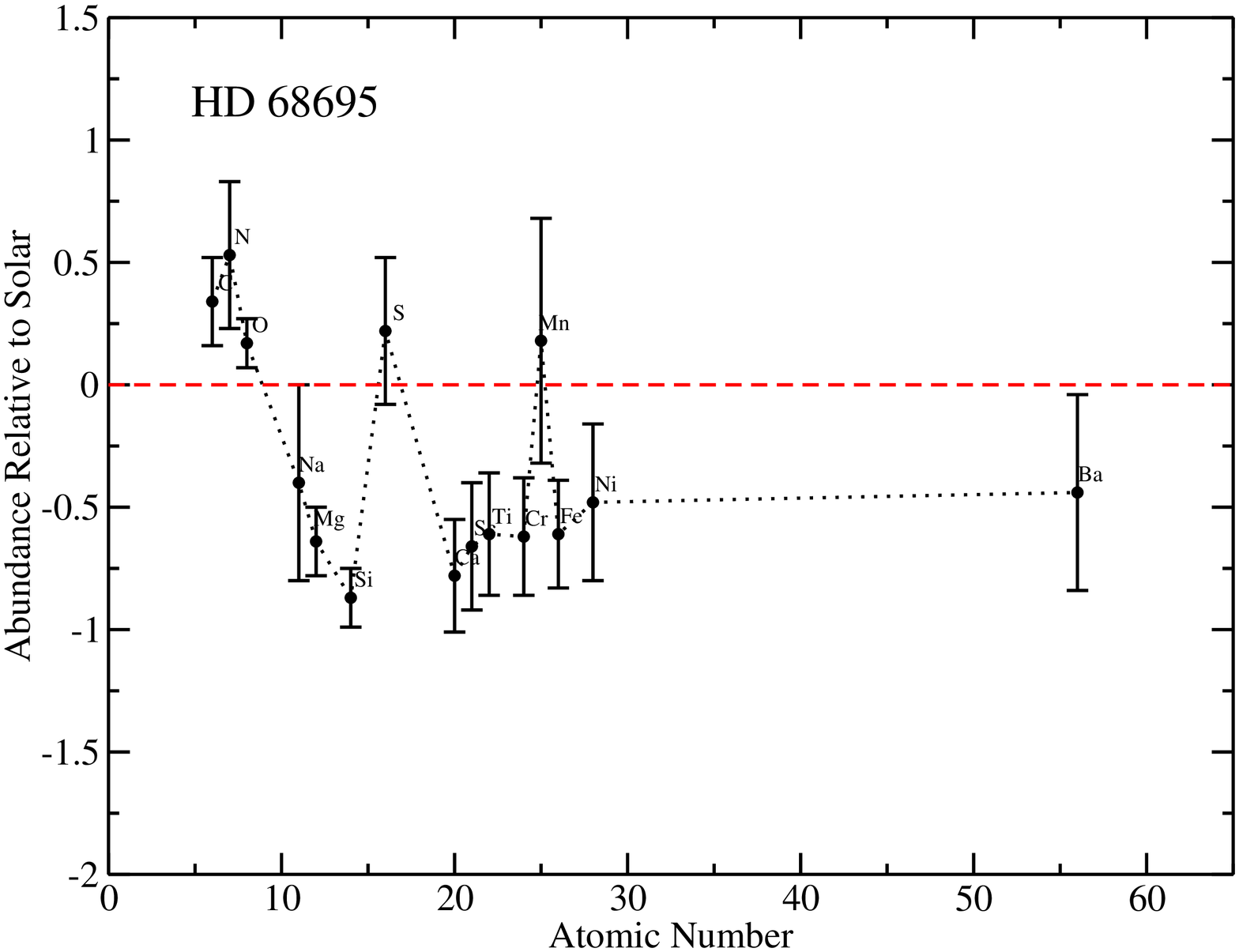}
\includegraphics[width=3.4in]{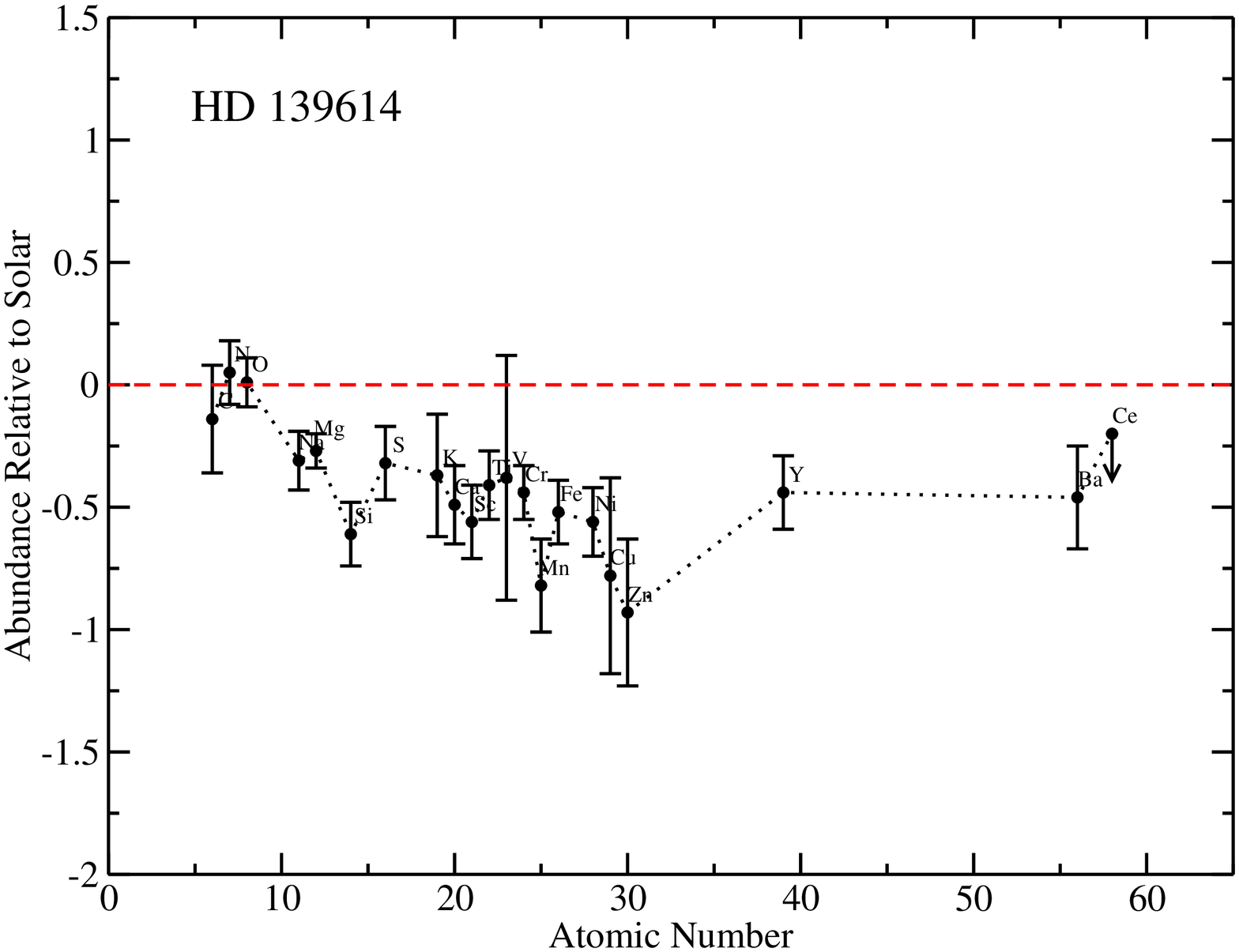}
\caption{Final abundances relative to solar for the stars in this study.  
Solar abundances are taken from \citet{Grevesse2005-solar_abun}.  
Points marked with an arrow represent upper limits only.  }
\label{abunplots1}
\end{figure*}

\begin{figure*}
\centering
\includegraphics[width=3.4in]{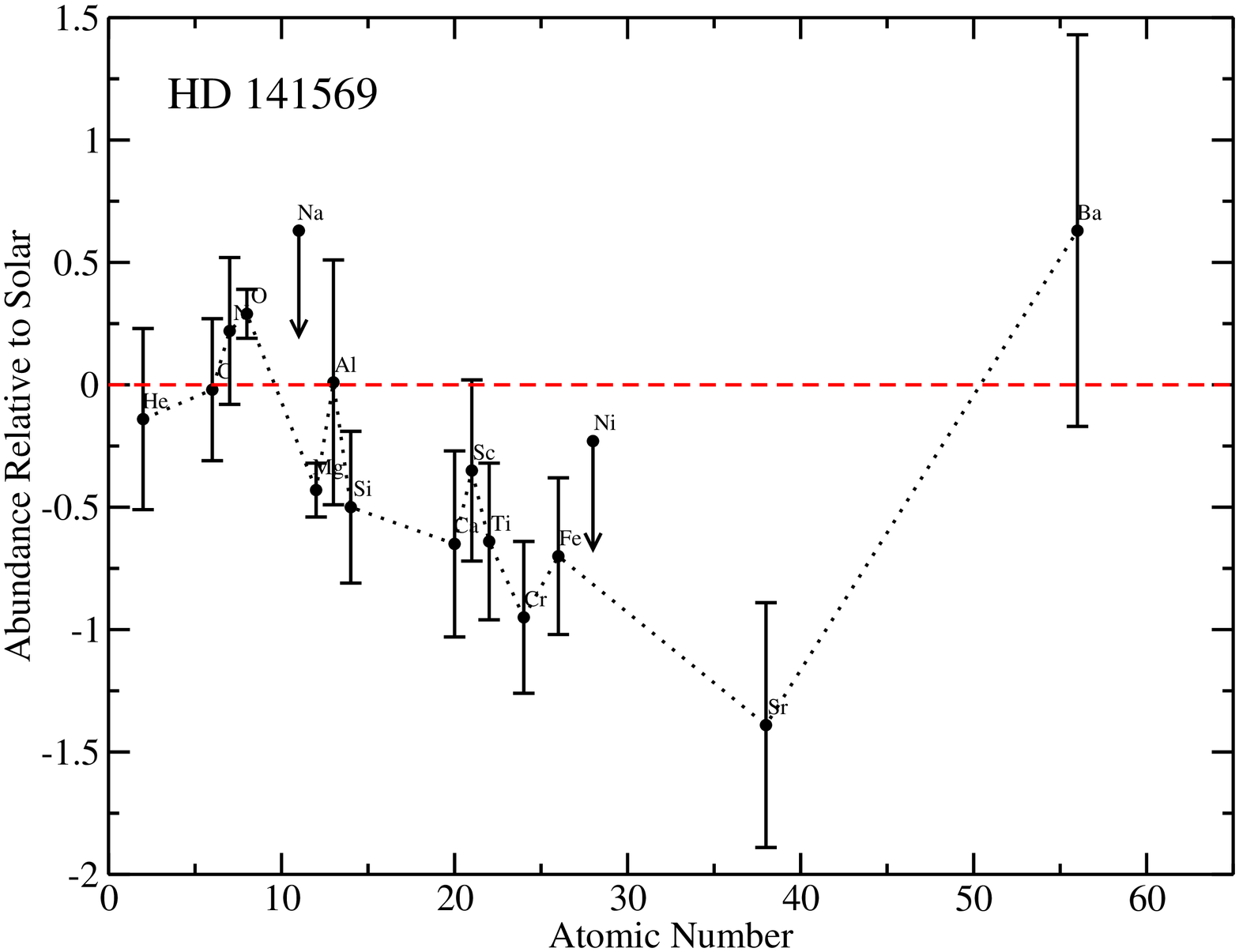}
\includegraphics[width=3.4in]{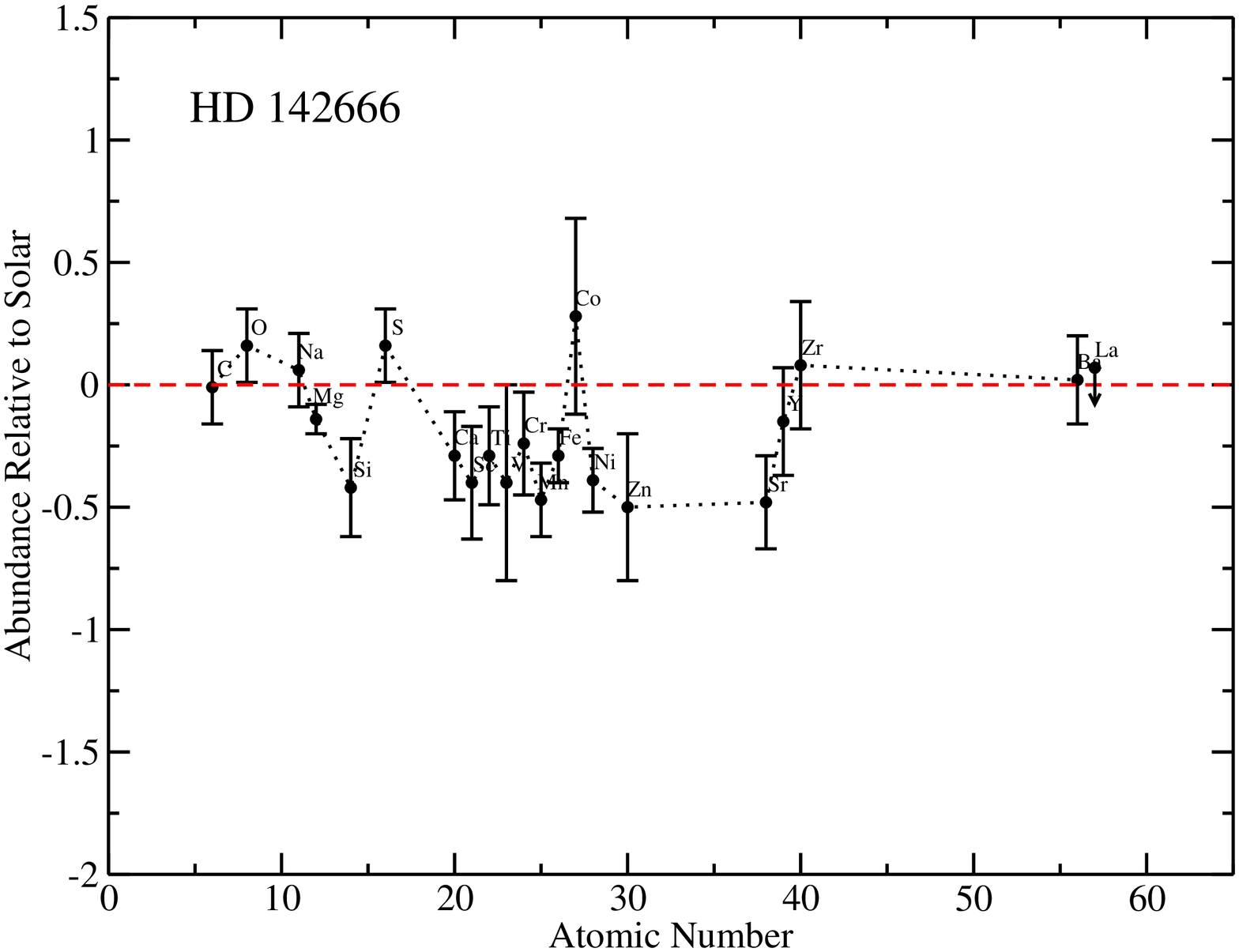}
\includegraphics[width=3.4in]{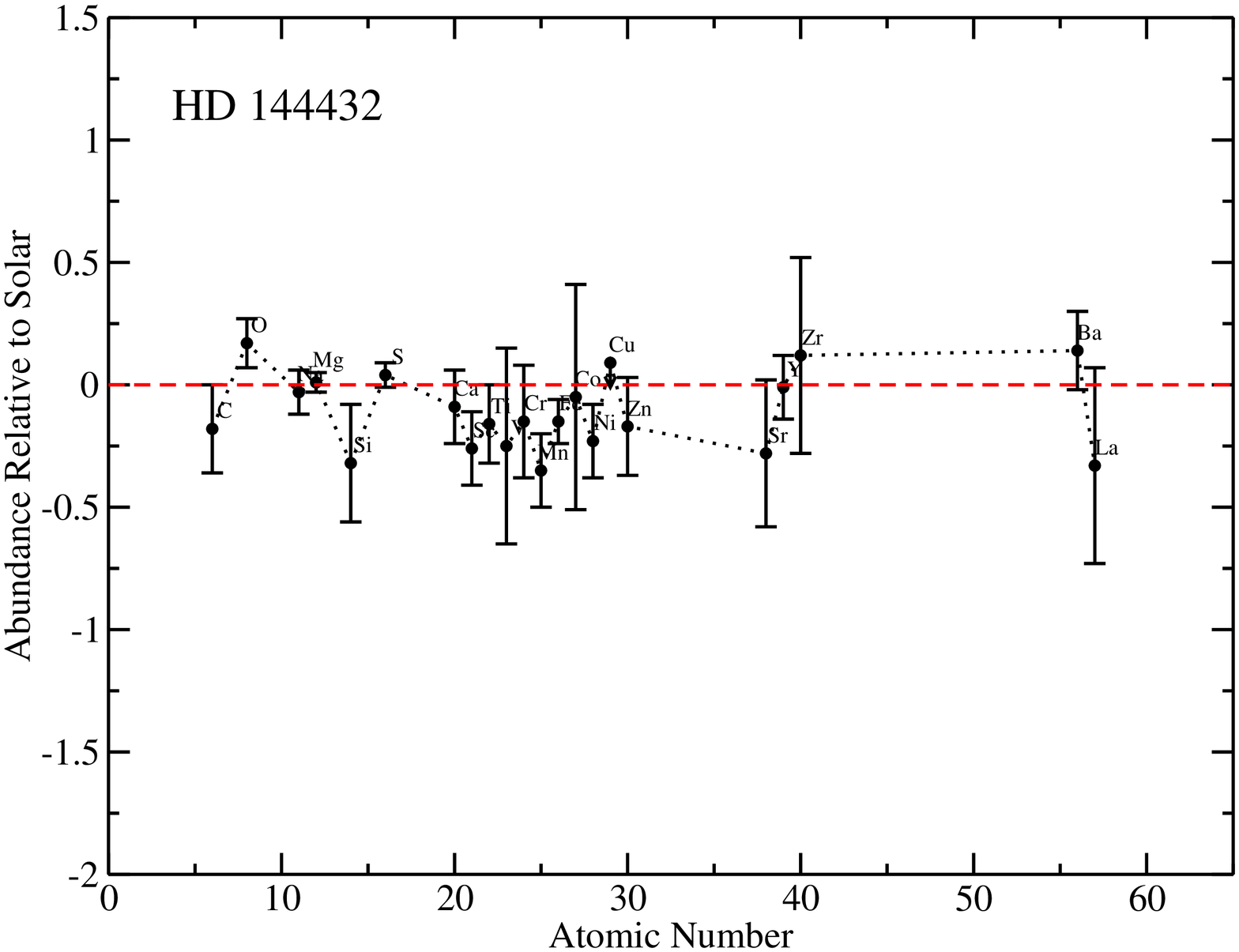}
\includegraphics[width=3.4in]{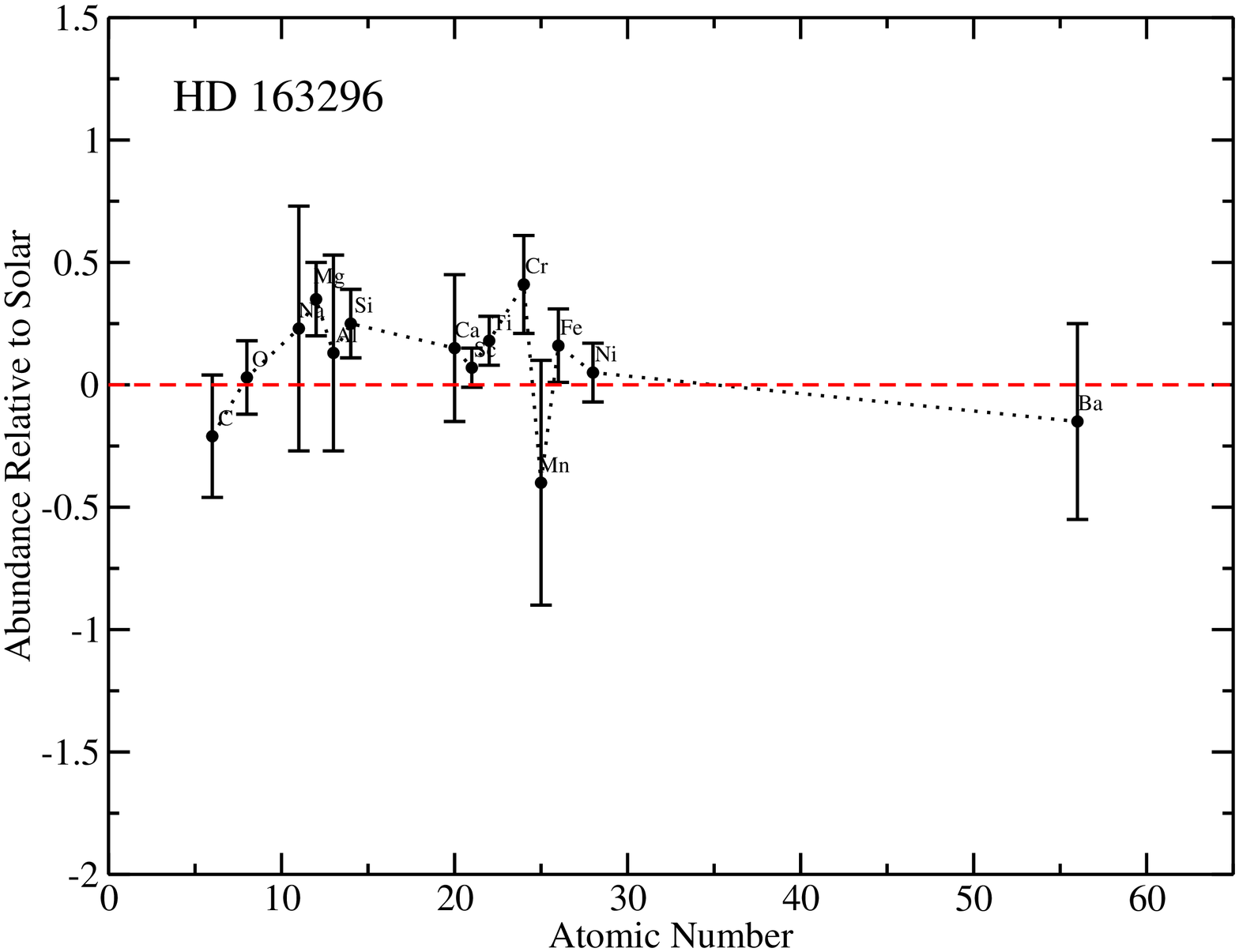}
\includegraphics[width=3.4in]{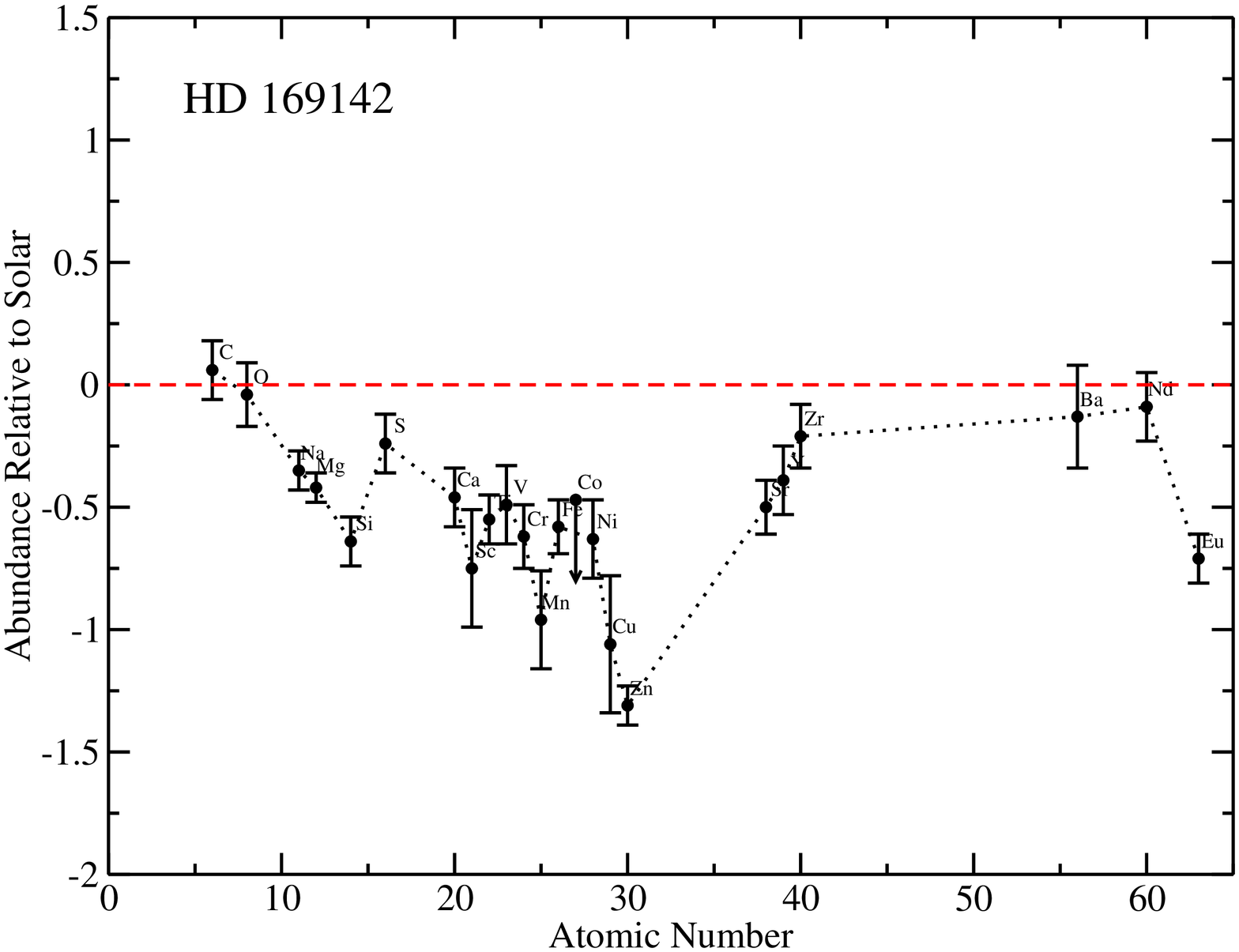}
\includegraphics[width=3.4in]{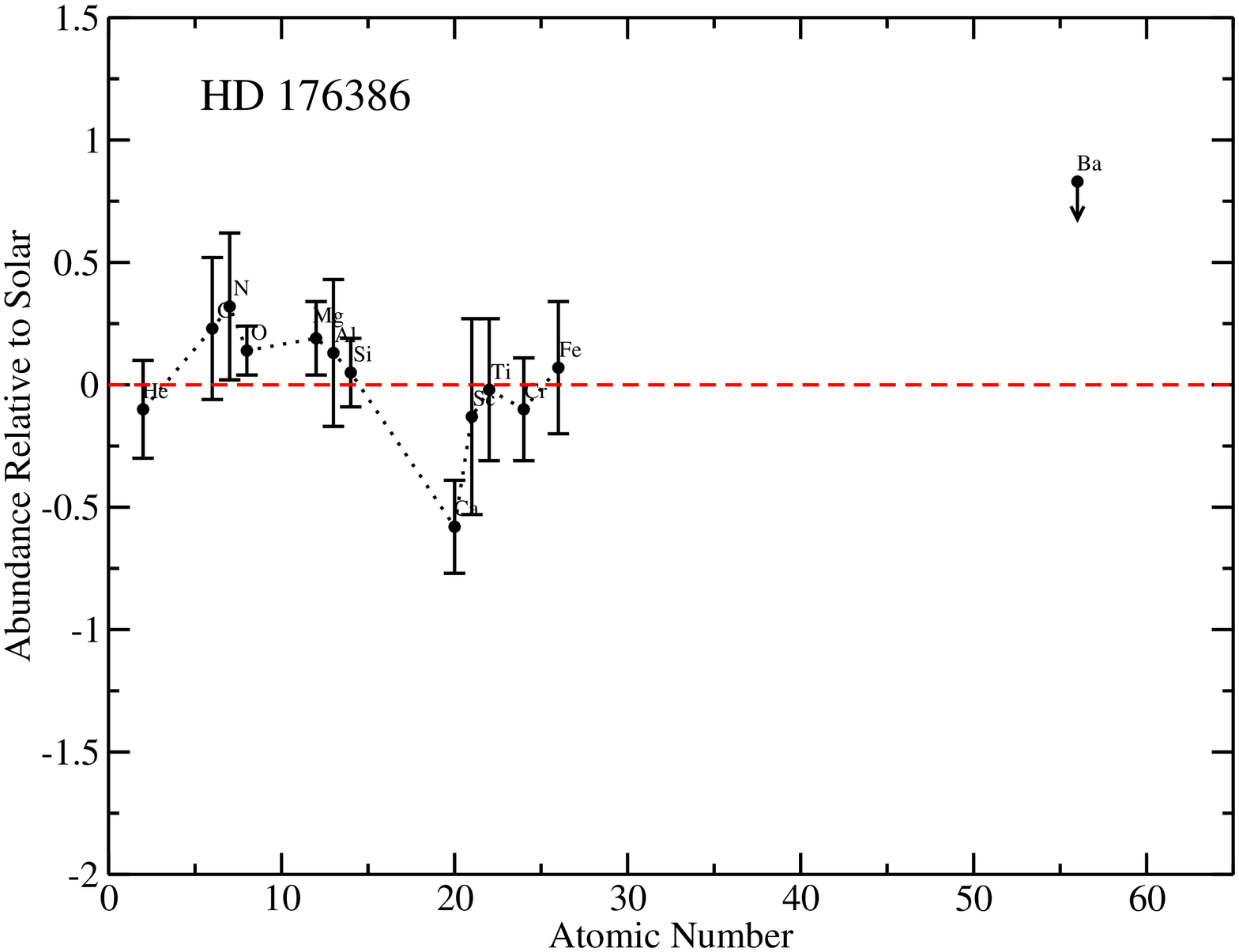}
\caption{Final abundances relative to solar for the stars in this study, 
as in Fig. \ref{abunplots1}.  }
\label{abunplots2}
\end{figure*}

\begin{figure*}
\centering
\includegraphics[width=3.4in]{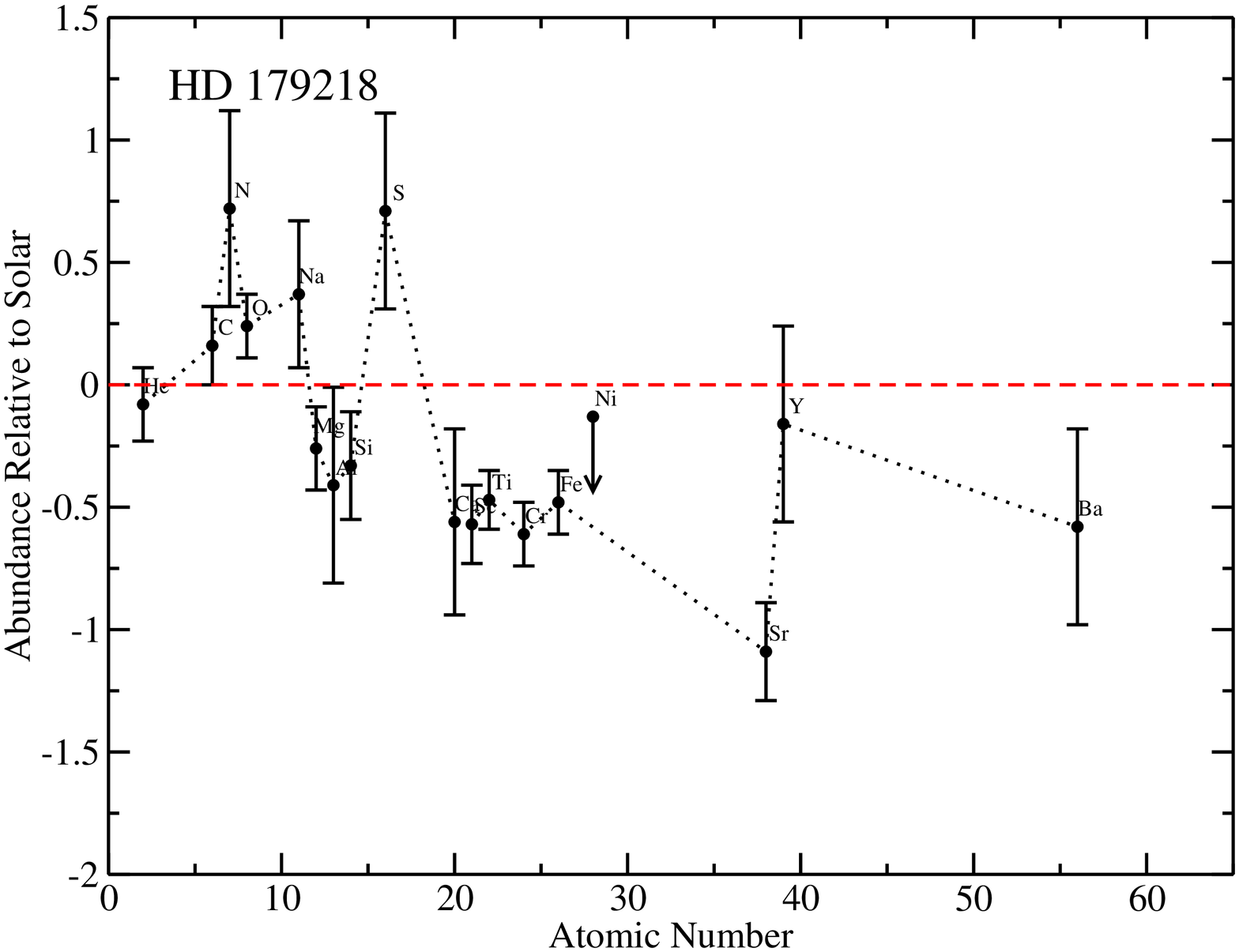}
\includegraphics[width=3.4in]{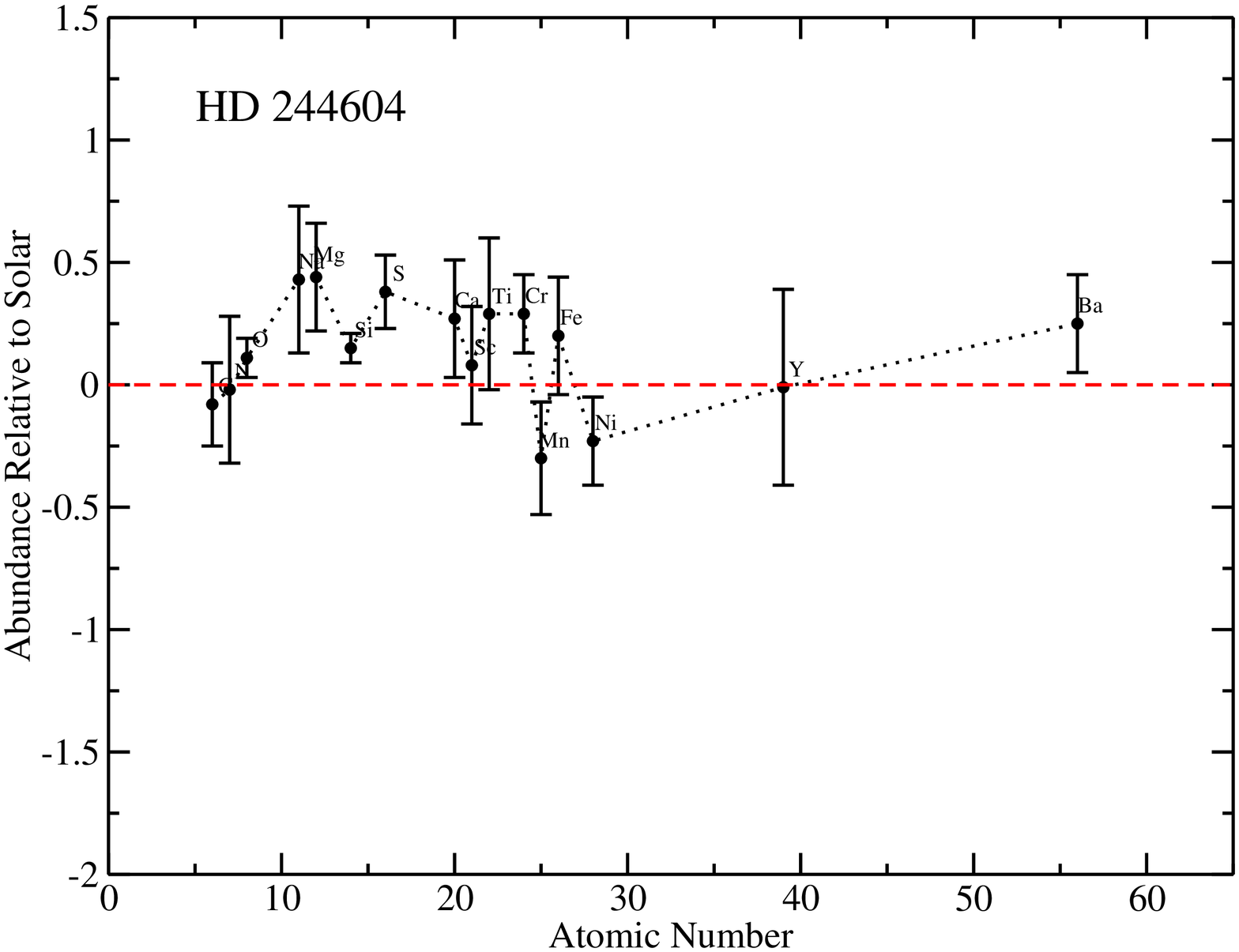}
\includegraphics[width=3.4in]{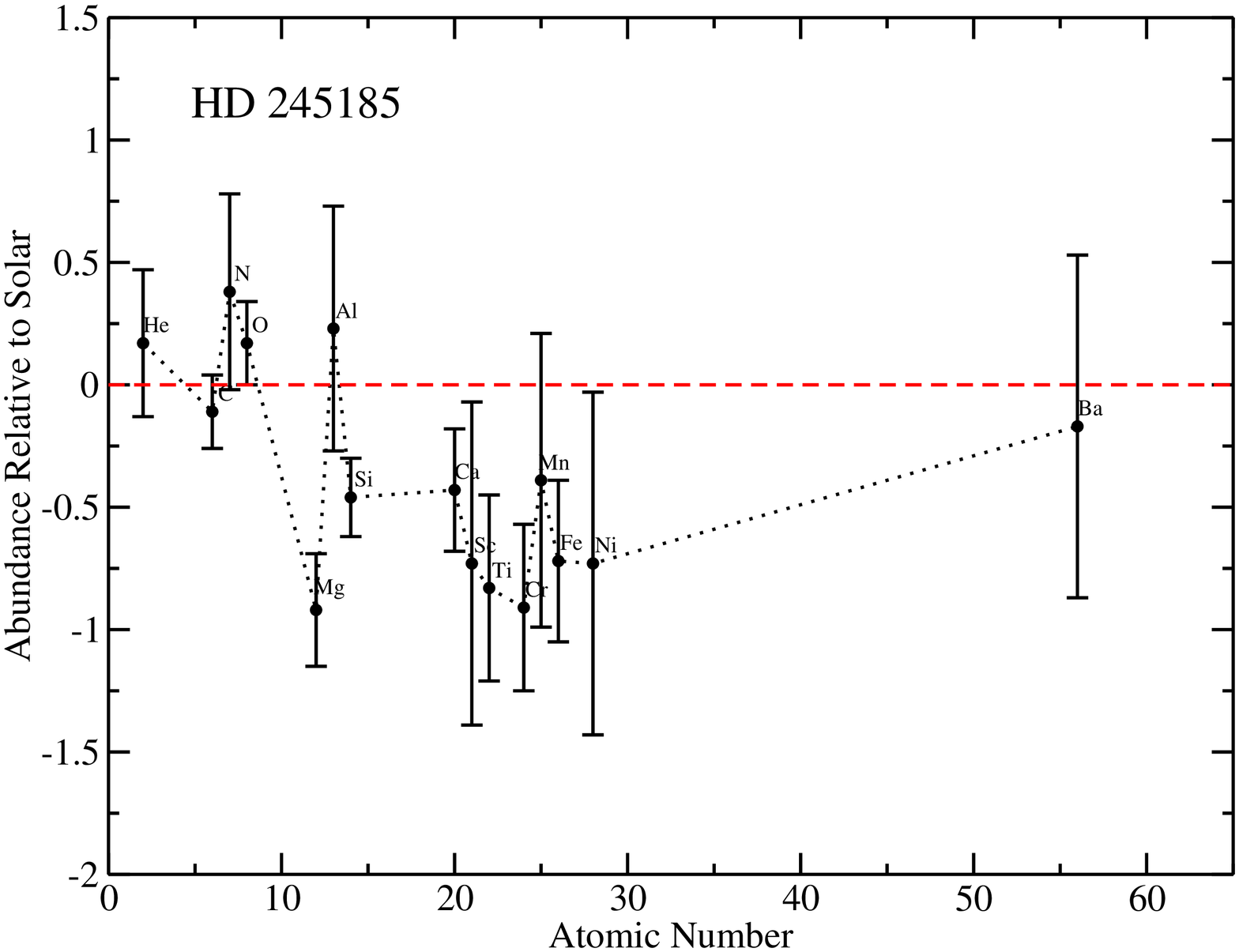}
\includegraphics[width=3.4in]{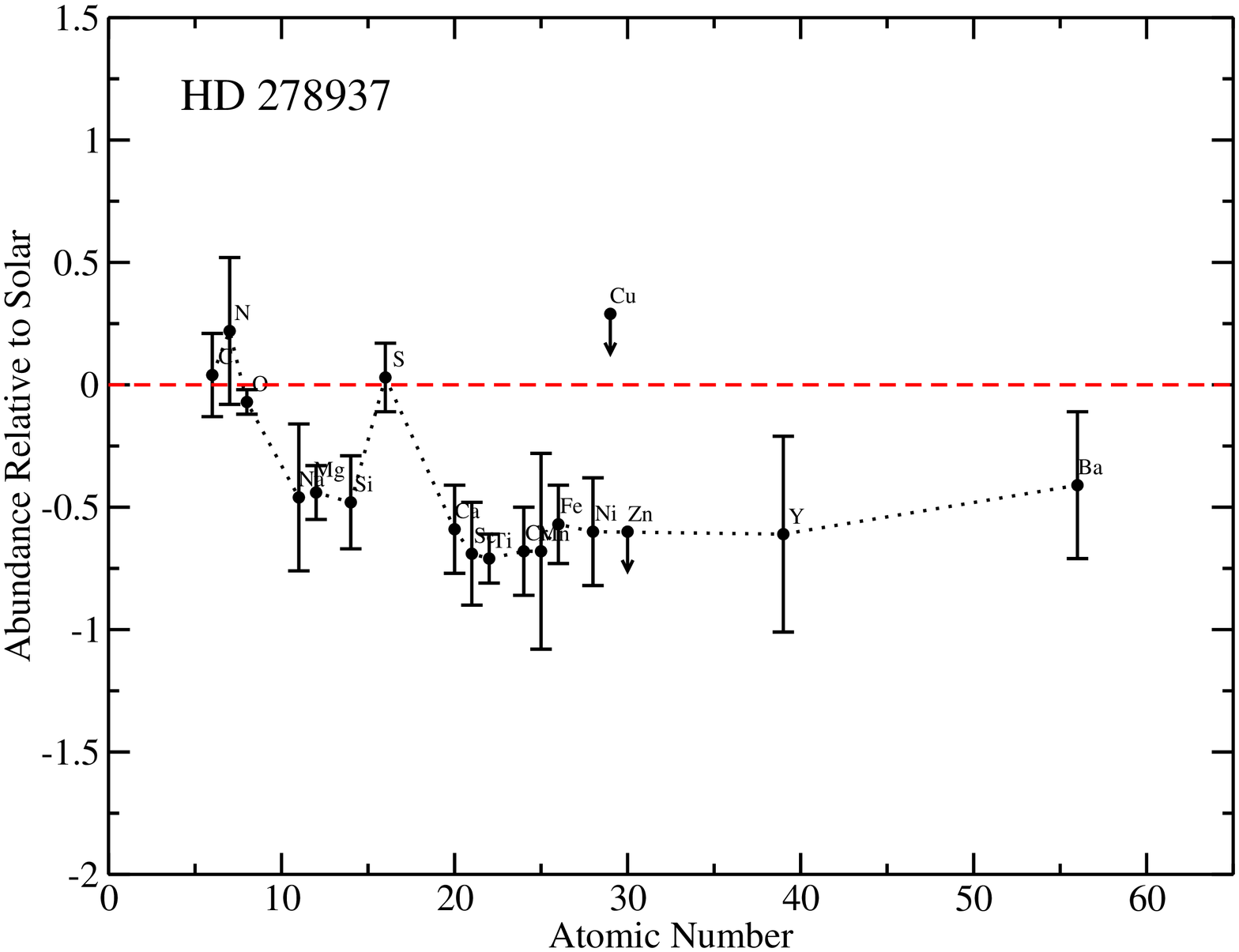}
\includegraphics[width=3.4in]{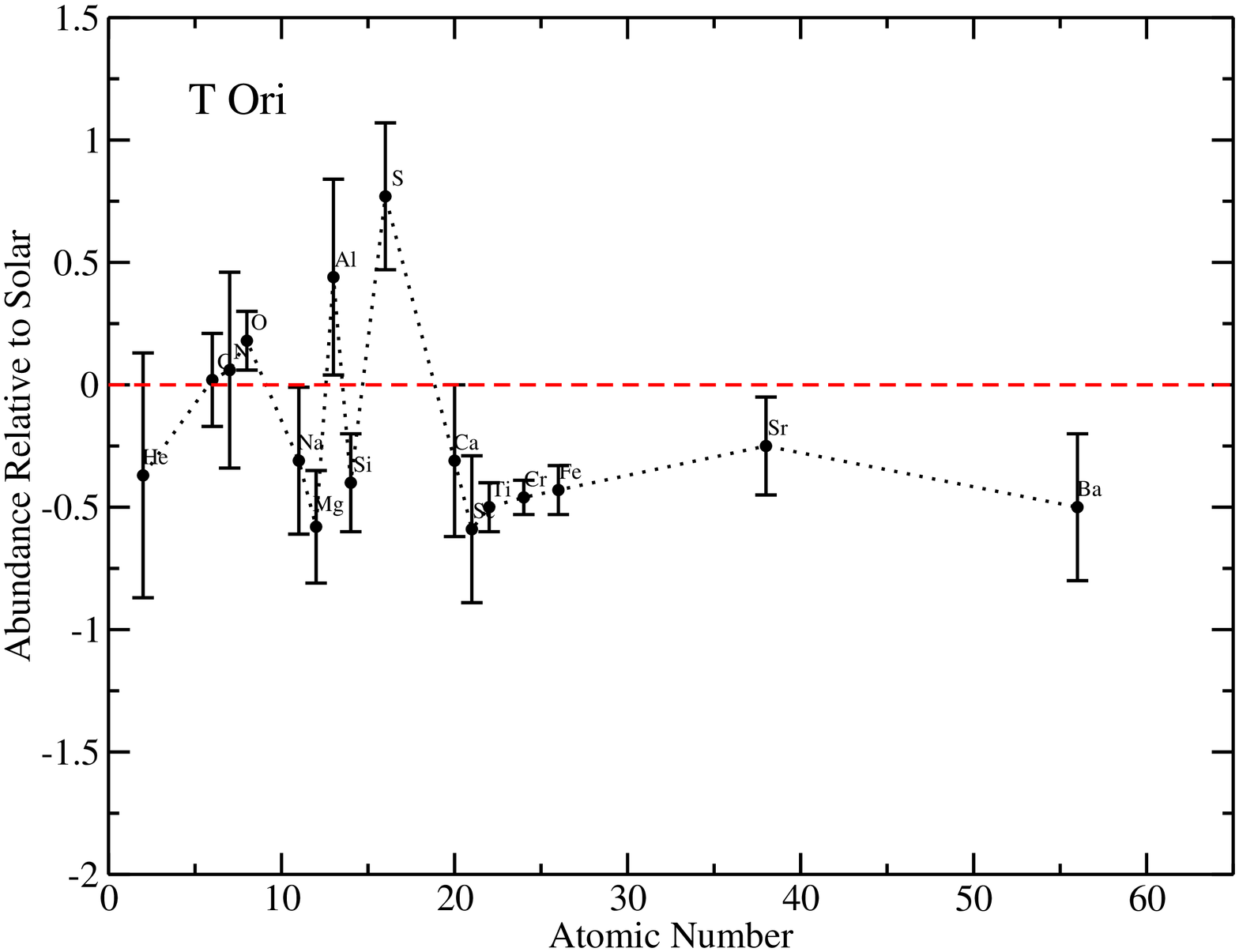}
\includegraphics[width=3.4in]{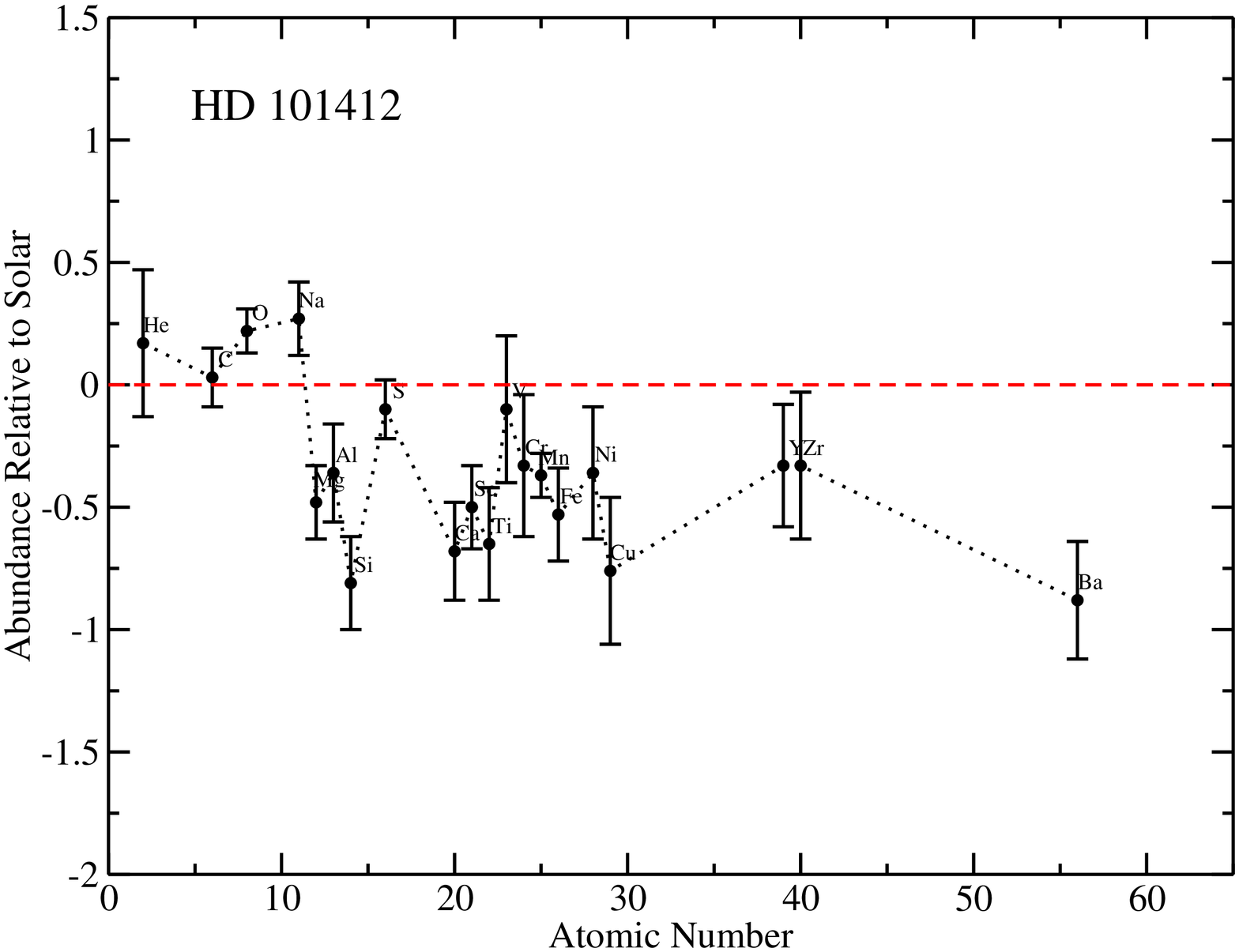}
\caption{Final abundances relative to solar for the stars in this study, 
as in Fig. \ref{abunplots1}.  }
\label{abunplots3}
\end{figure*}

\begin{figure*}
\centering
\includegraphics[width=3.4in]{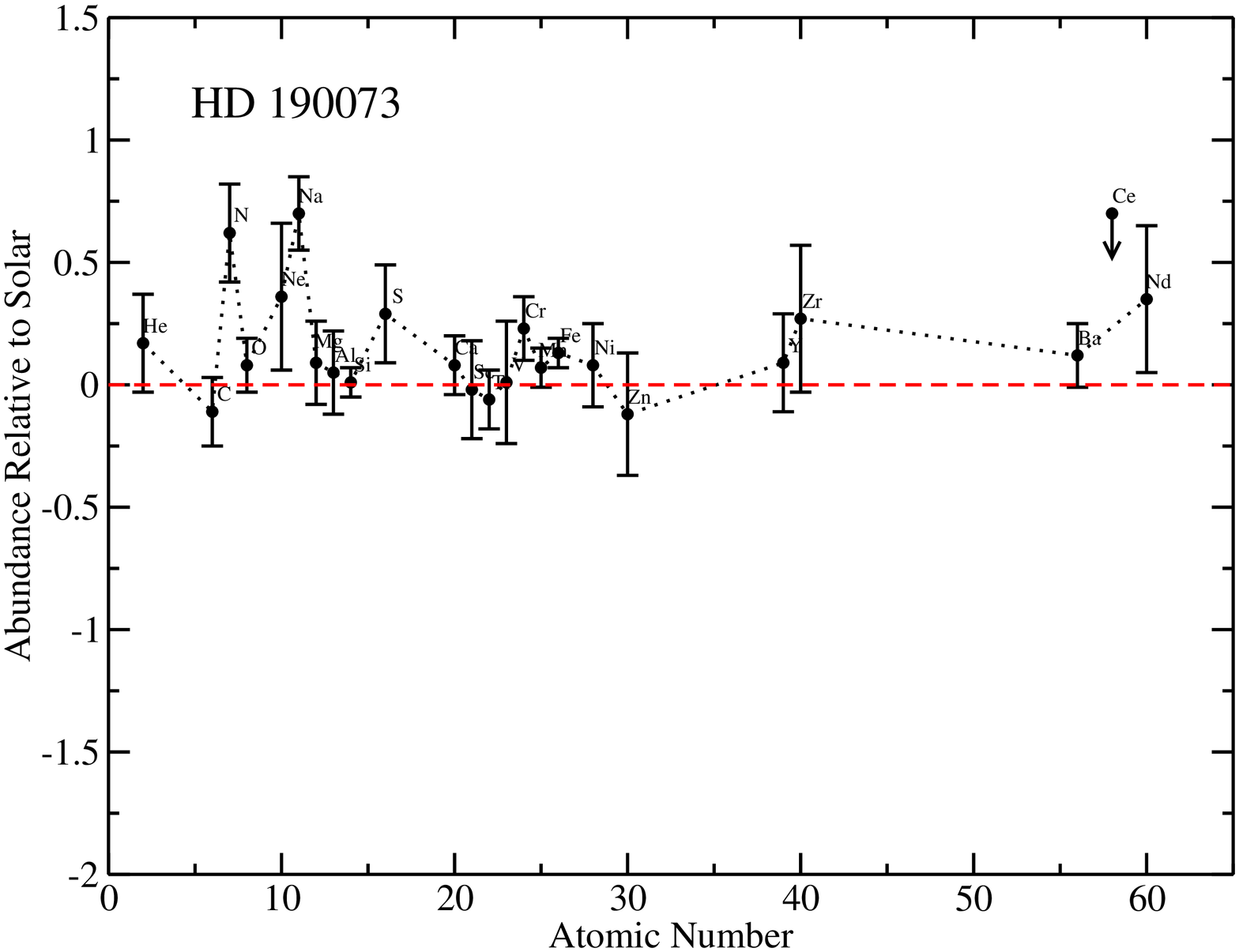}
\includegraphics[width=3.4in]{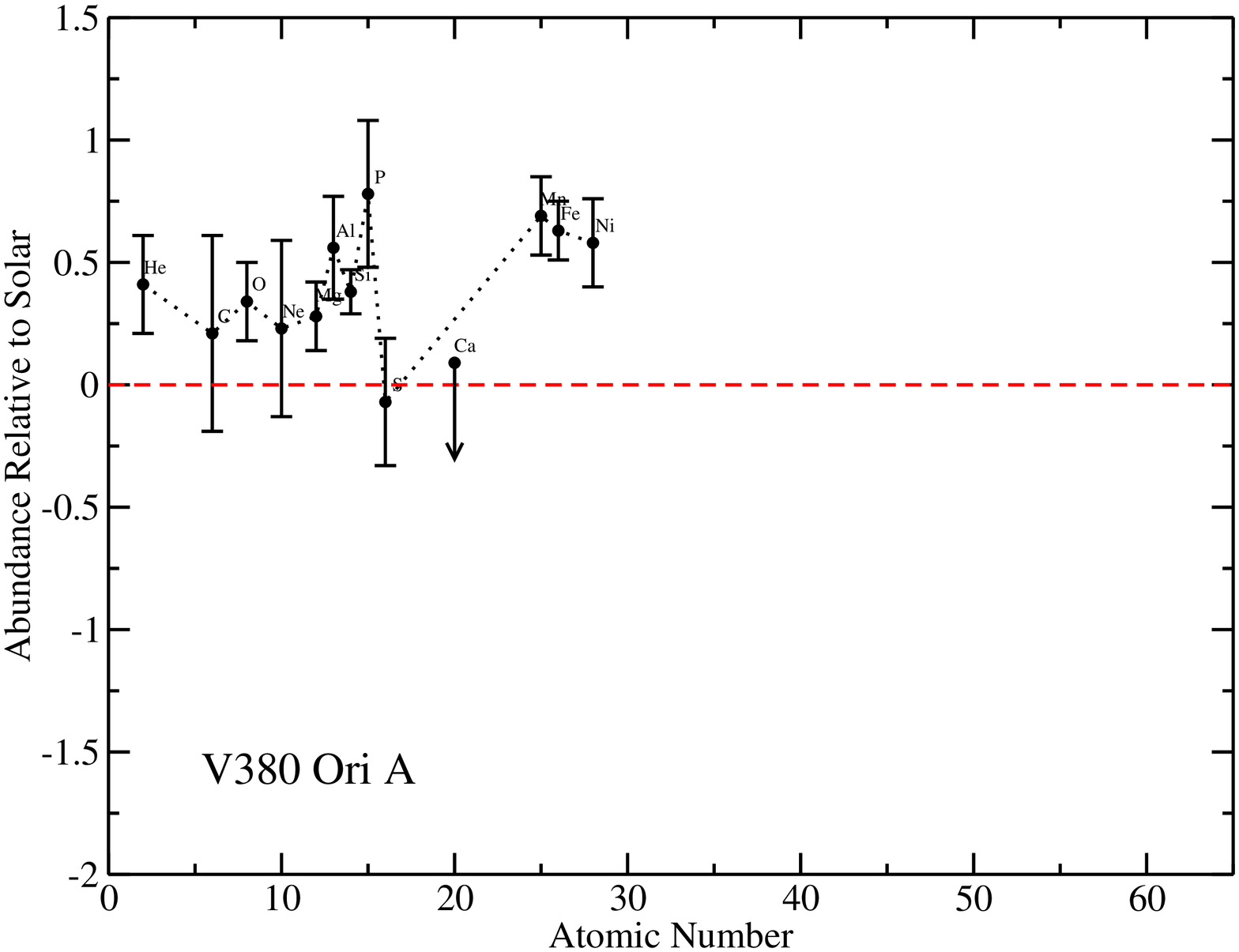}
\includegraphics[width=3.4in]{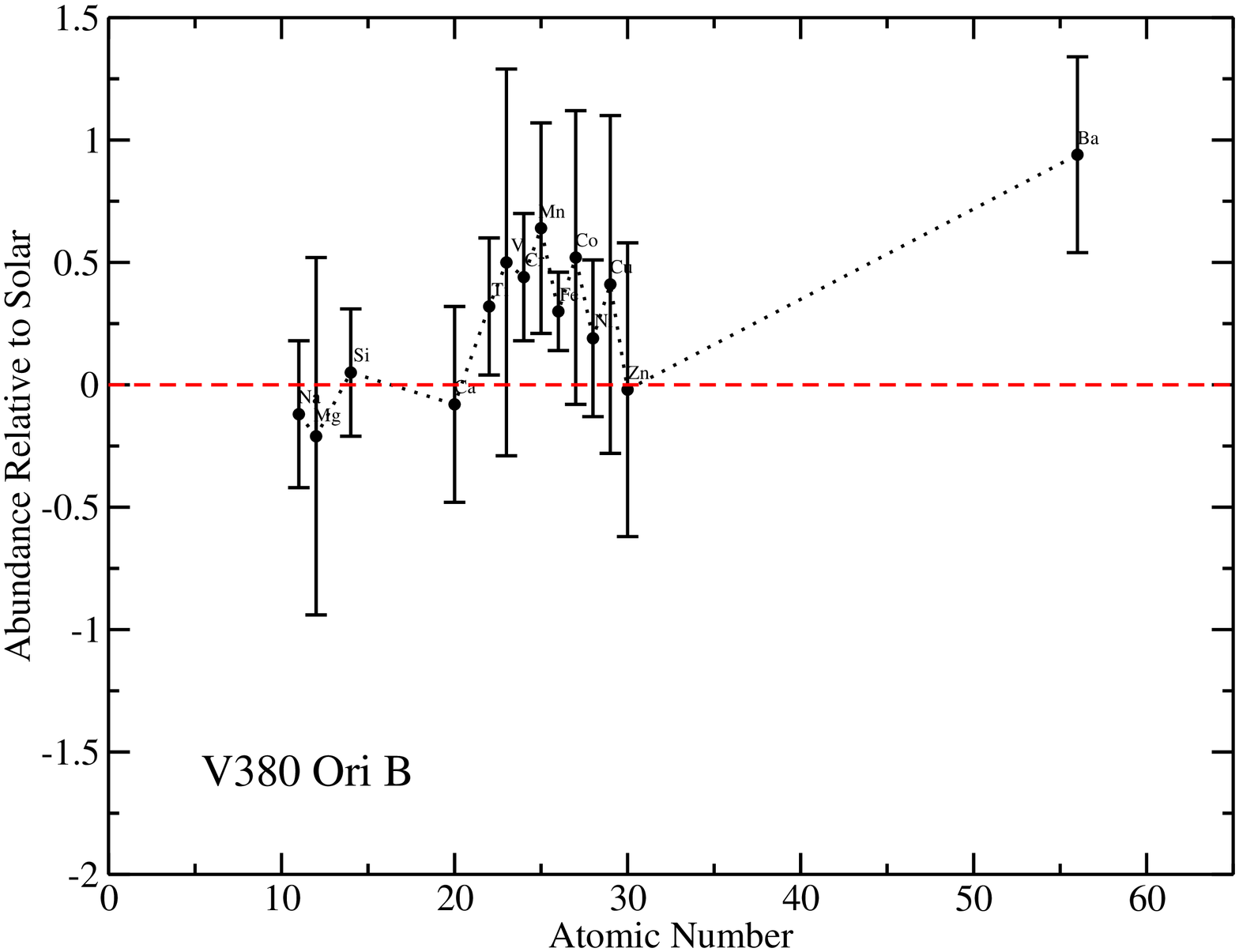}
\caption{Final abundances relative to solar for the stars in this study, 
as in Fig. \ref{abunplots1}.  }
\label{abunplots4}
\end{figure*}


\begin{figure*}
\centering
\includegraphics[width=5.3in]{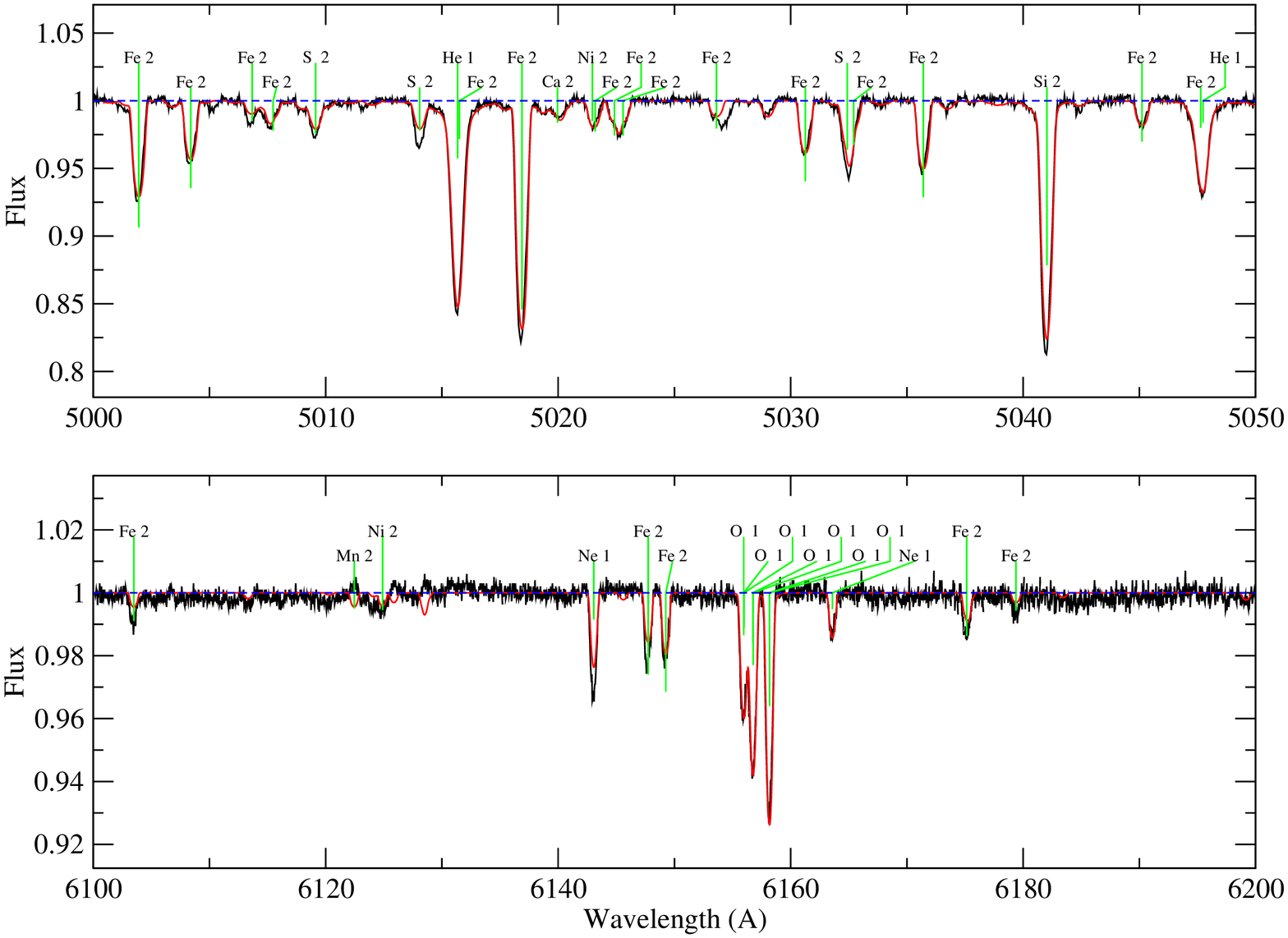}
\caption{Comparison of the observed spectrum (jagged line) to the 
best fit synthetic spectrum (smooth line) for $\pi$ Cet (HD 17081). 
Two independent wavelength regions are presented.  
Lines have been labelled by their major contributing species. 
Gaps in the synthetic spectra indicate regions that were not fit.}
\label{fit-picet}
\label{fit-appendix}
\end{figure*}

\begin{figure*}
\centering
\includegraphics[width=5.3in]{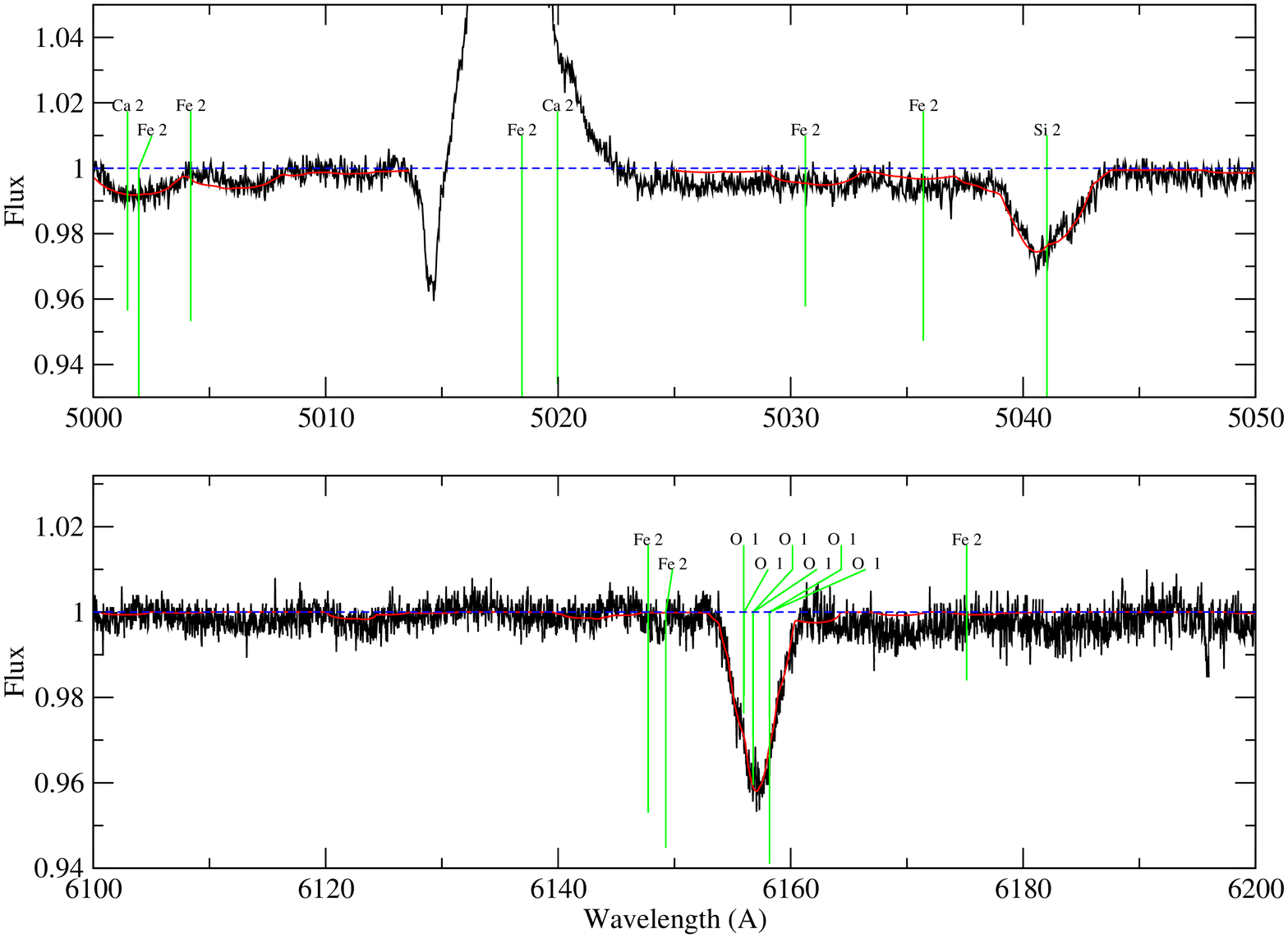}
\caption{Comparison of the observed spectrum to the best fit synthetic 
spectrum for HD 31293, as in Fig. \ref{fit-appendix}}
\label{fit-hd31293}
\end{figure*}

\clearpage

\begin{figure*}
\centering
\includegraphics[width=5.4in]{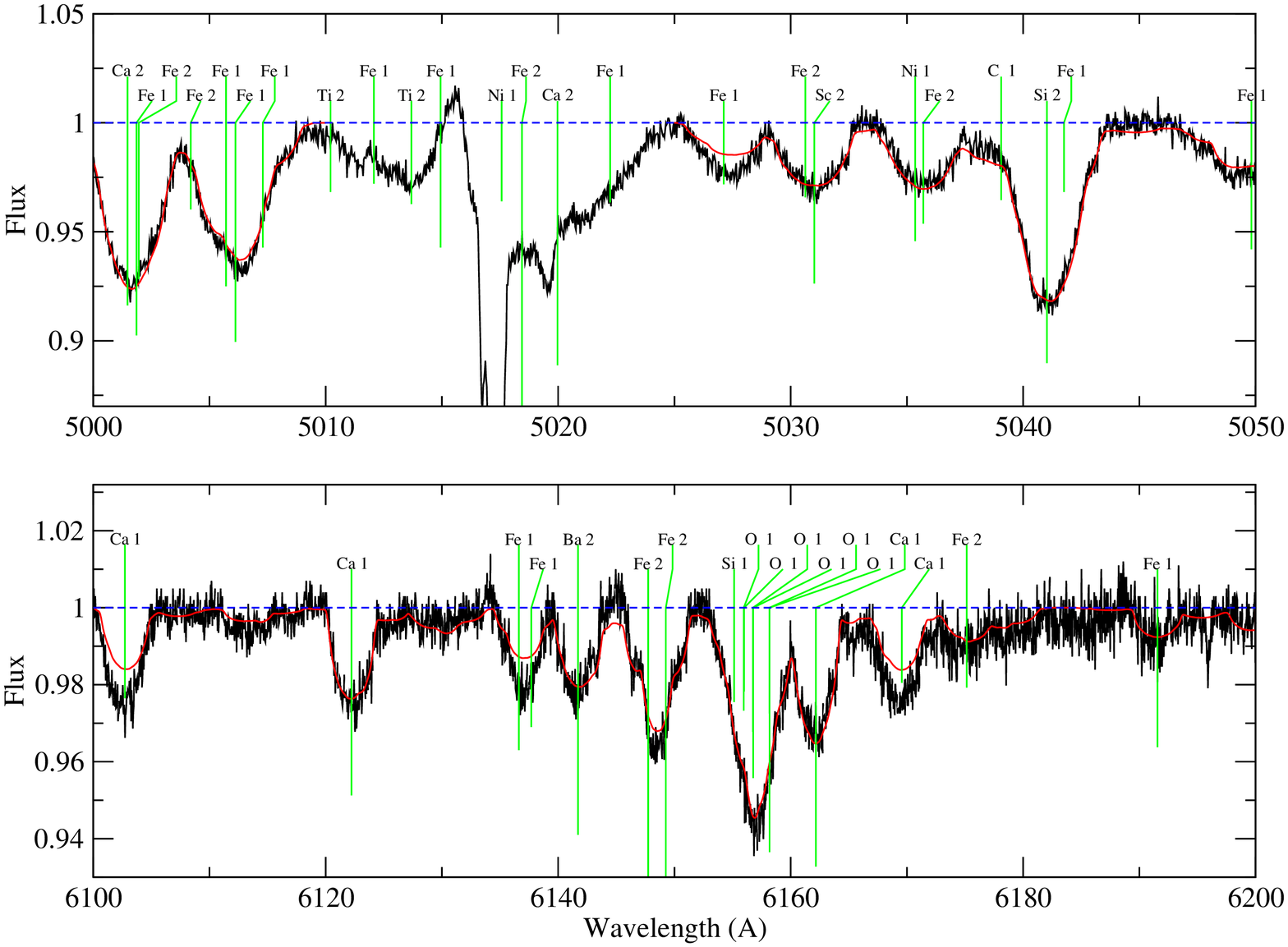}
\caption{Comparison of the observed spectrum to the best fit synthetic 
spectrum for HD 31648, as in Fig. \ref{fit-appendix}}
\label{fit-hd31648}
\end{figure*}

\begin{figure*}
\centering
\includegraphics[width=5.4in]{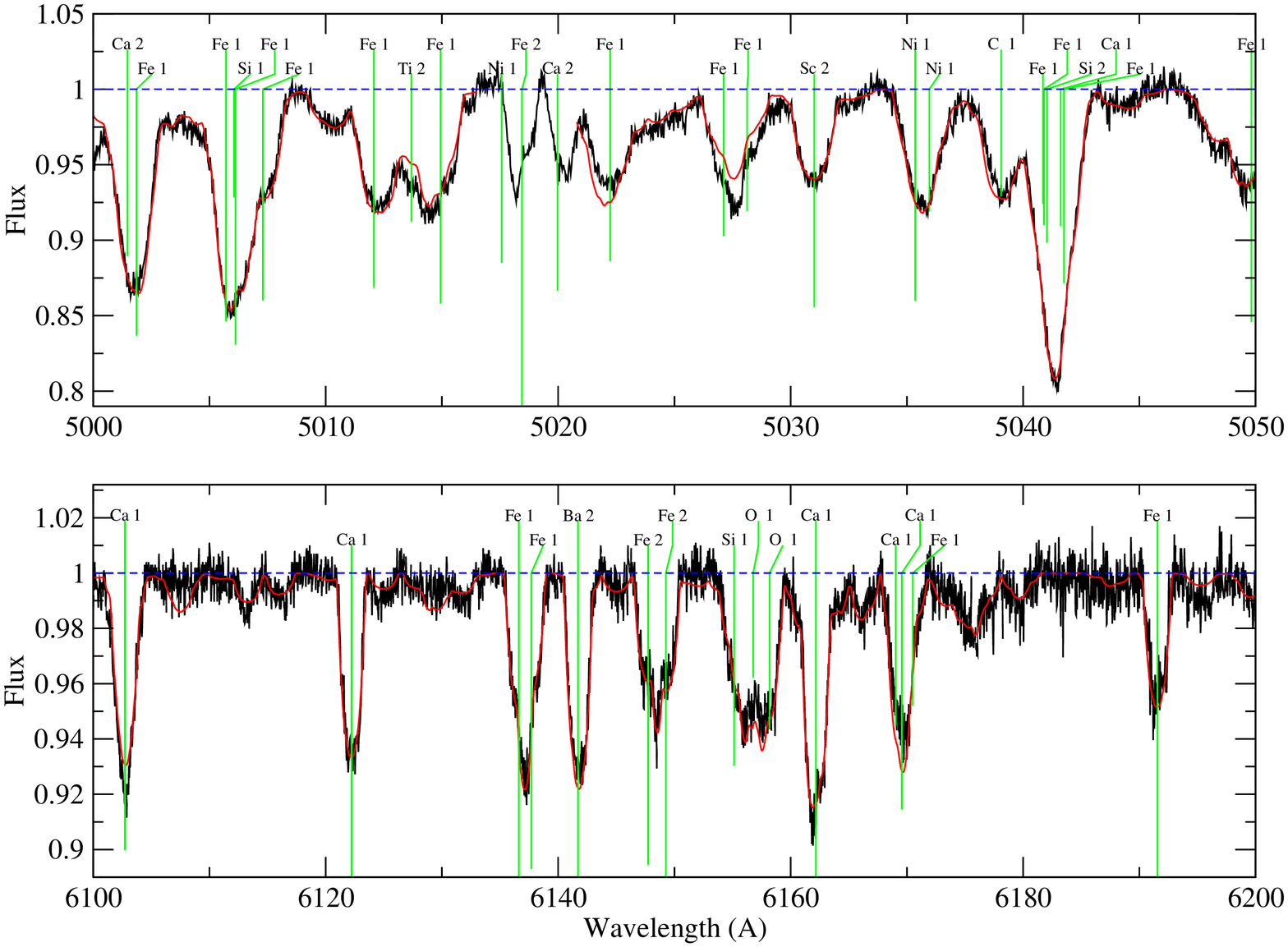}
\caption{Comparison of the observed spectrum to the best fit synthetic 
spectrum for HD 36112, as in Fig. \ref{fit-appendix}}
\label{fit-hd36112}
\end{figure*}

\begin{figure*}
\centering
\includegraphics[width=5.4in]{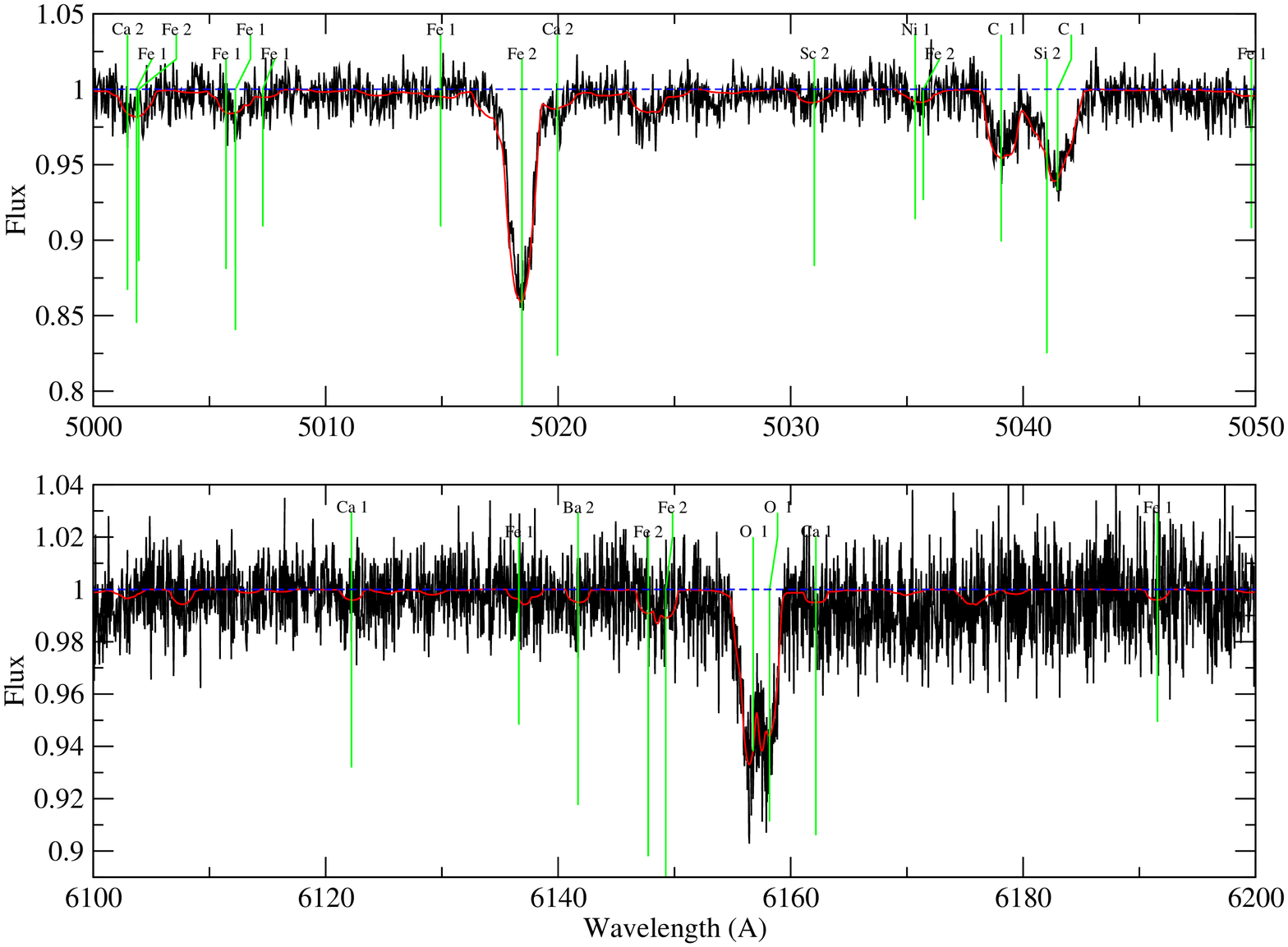}
\caption{Comparison of the observed spectrum to the best fit synthetic 
spectrum for HD 68695, as in Fig. \ref{fit-appendix}}
\label{fit-hd68695}
\end{figure*}

\begin{figure*}
\centering
\includegraphics[width=5.4in]{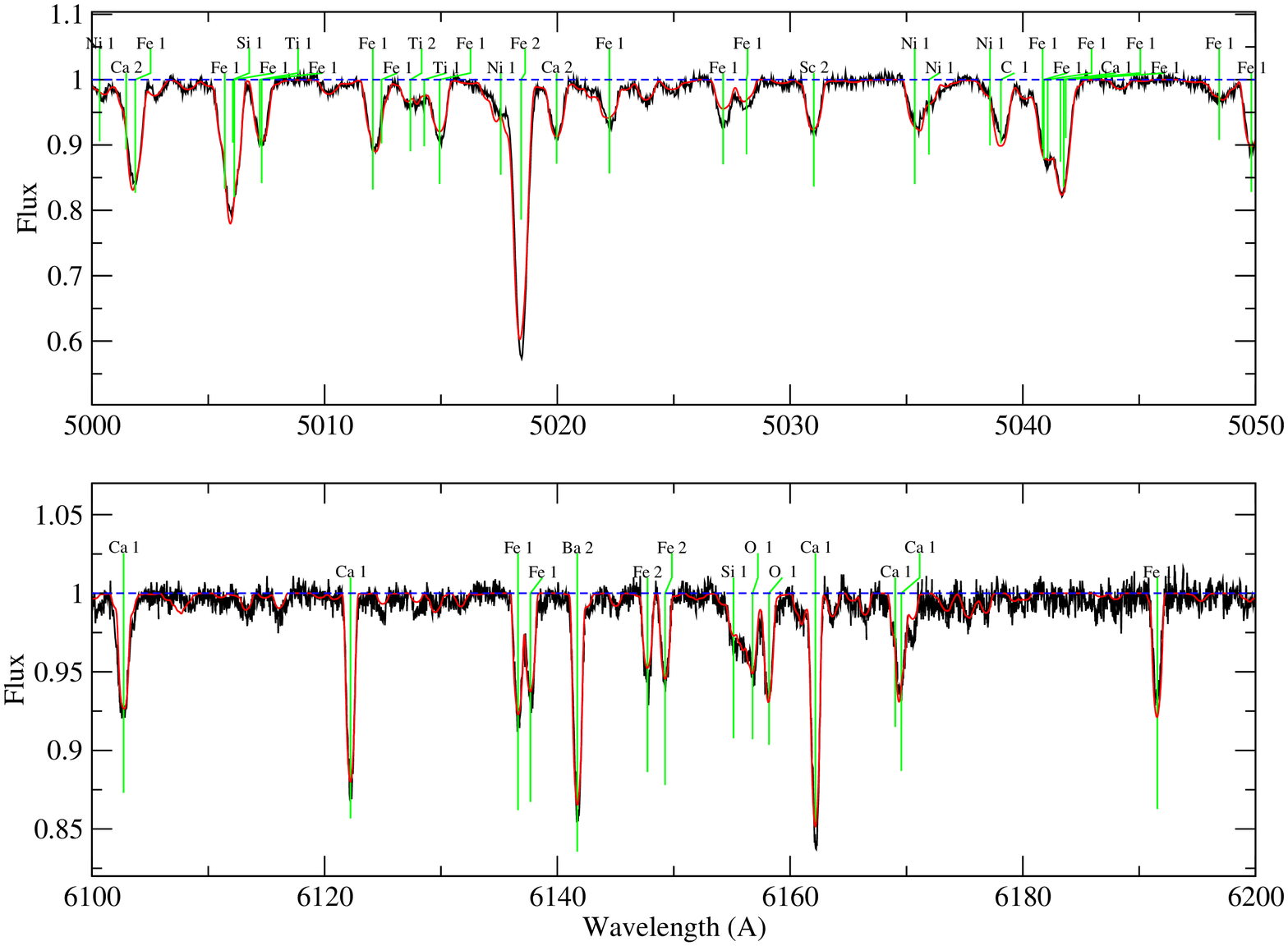}
\caption{Comparison of the observed spectrum to the best fit synthetic 
spectrum for HD 139614, as in Fig. \ref{fit-appendix}}
\label{fit-hd139614}
\end{figure*}

\begin{figure*}
\centering
\includegraphics[width=5.4in]{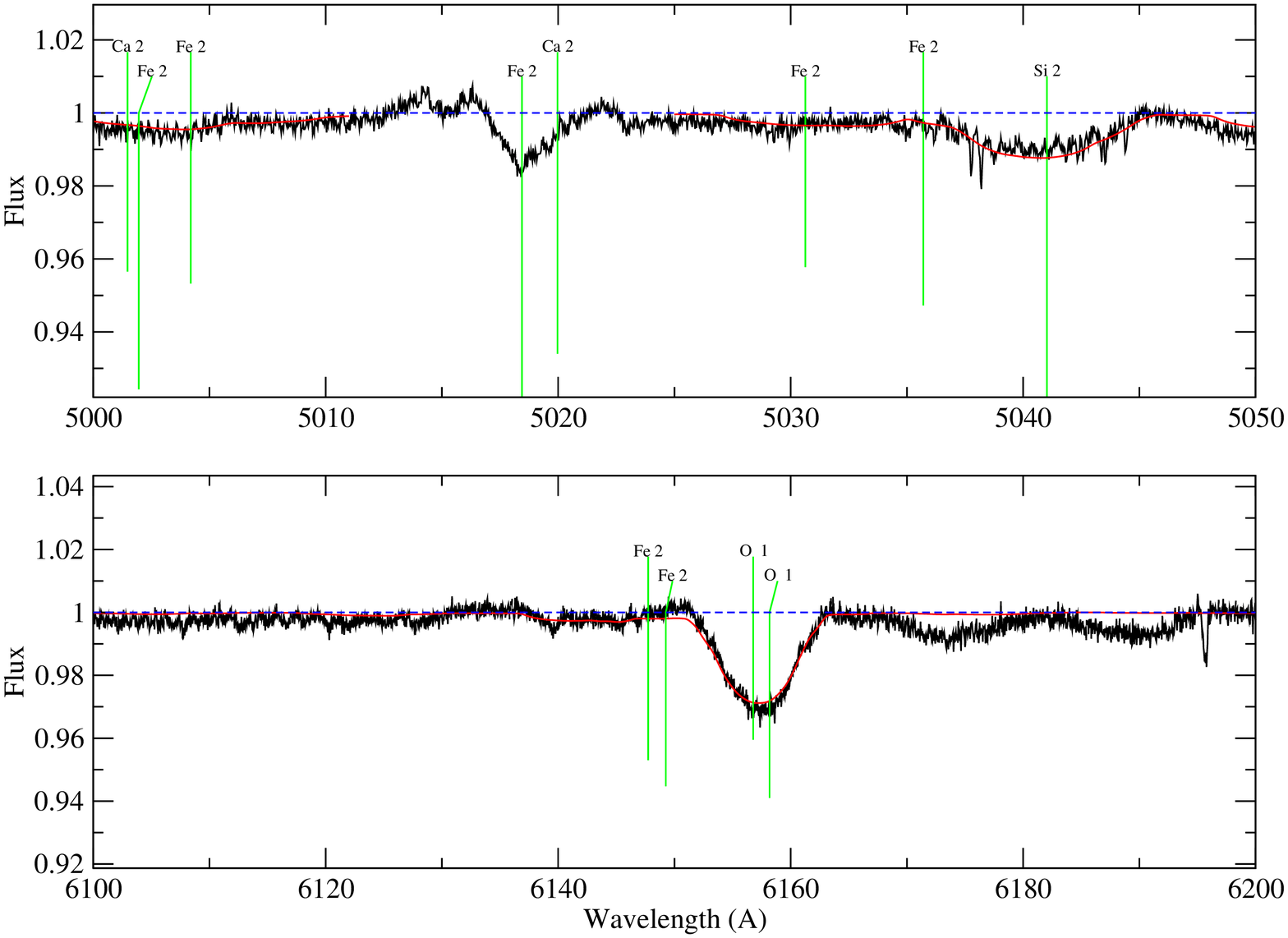}
\caption{Comparison of the observed spectrum to the best fit synthetic 
spectrum for HD 141569, as in Fig. \ref{fit-appendix}}
\label{fit-hd141569}
\end{figure*}

\begin{figure*}
\centering
\includegraphics[width=5.4in]{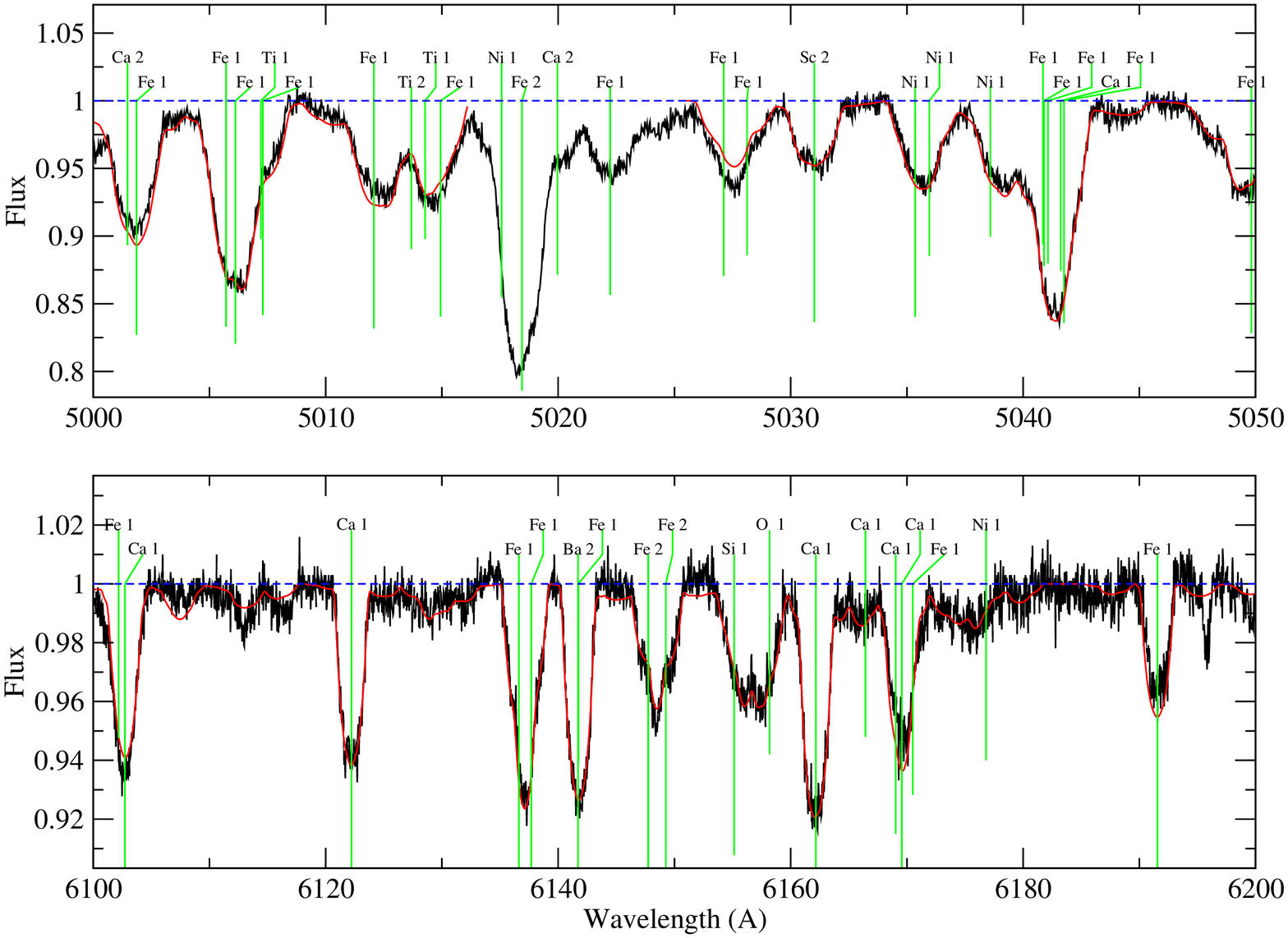}
\caption{Comparison of the observed spectrum to the best fit synthetic 
spectrum for HD 142666, as in Fig. \ref{fit-appendix}}
\label{fit-hd142666}
\end{figure*}

\begin{figure*}
\centering
\includegraphics[width=5.4in]{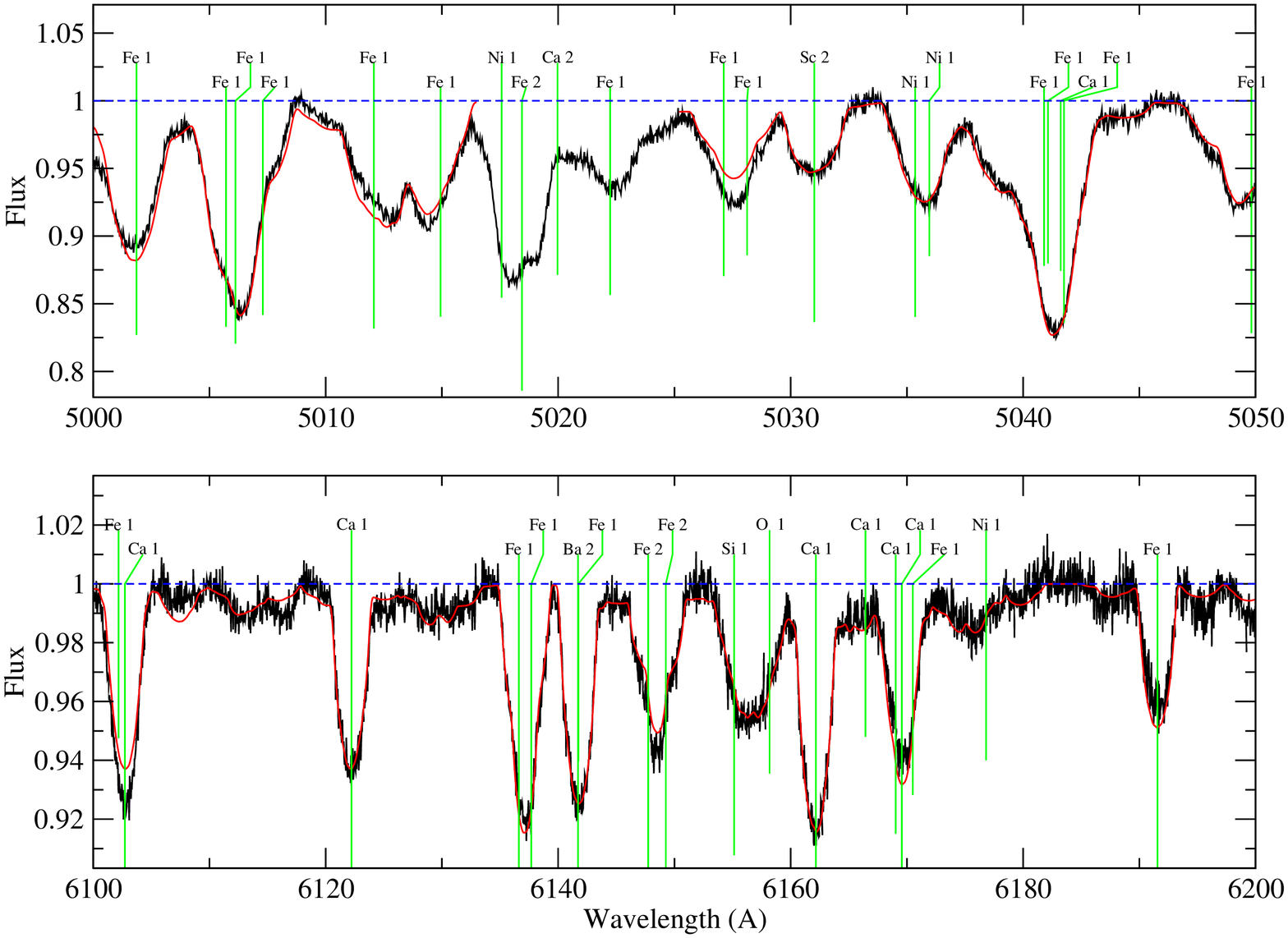}
\caption{Comparison of the observed spectrum to the best fit synthetic 
spectrum for HD 144432, as in Fig. \ref{fit-appendix}}
\label{fit-hd144432}
\end{figure*}

\begin{figure*}
\centering
\includegraphics[width=5.4in]{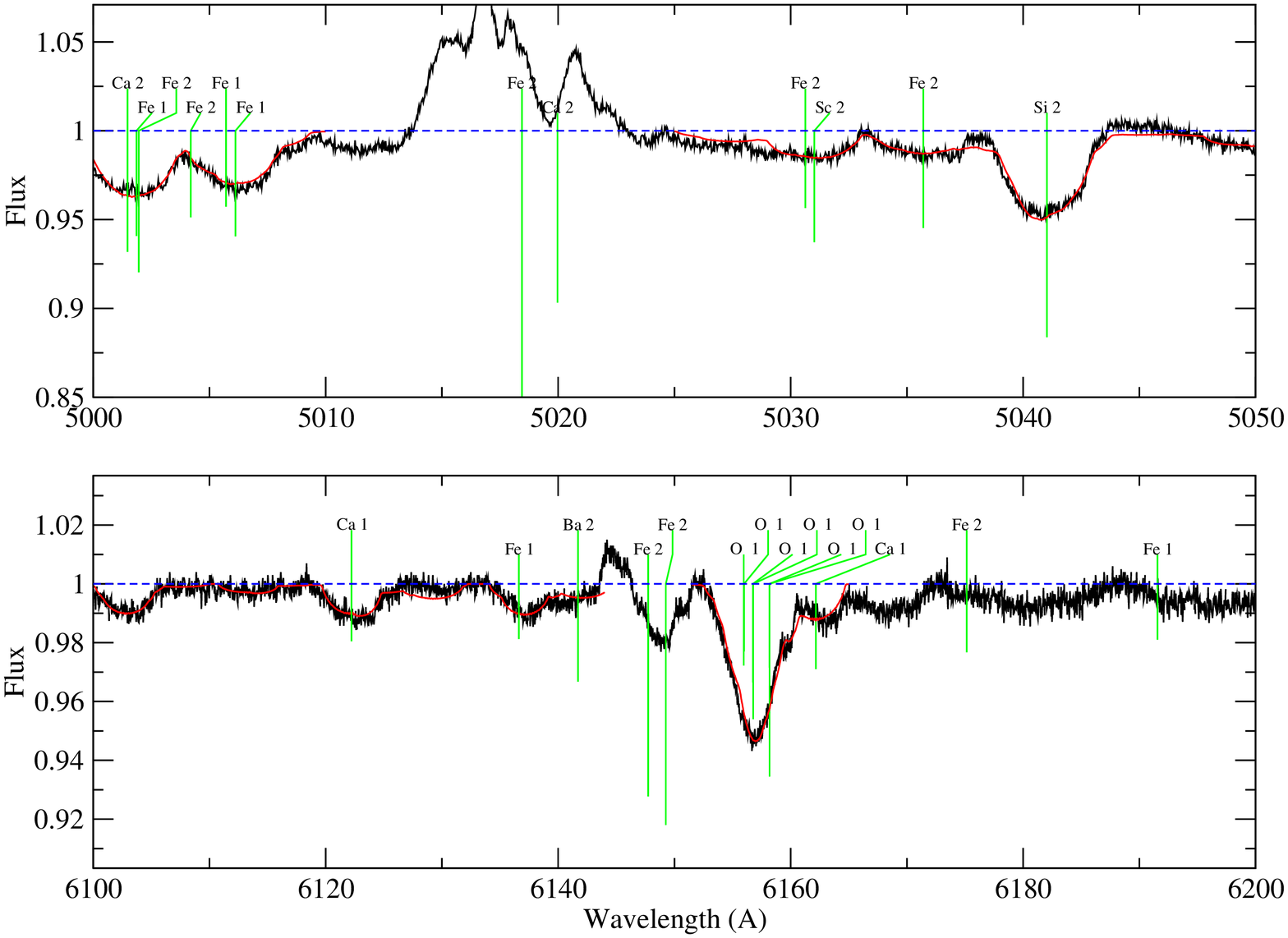}
\caption{Comparison of the observed spectrum to the best fit synthetic 
spectrum for HD 163296, as in Fig. \ref{fit-appendix}}
\label{fit-hd163296}
\end{figure*}

\begin{figure*}
\centering
\includegraphics[width=5.4in]{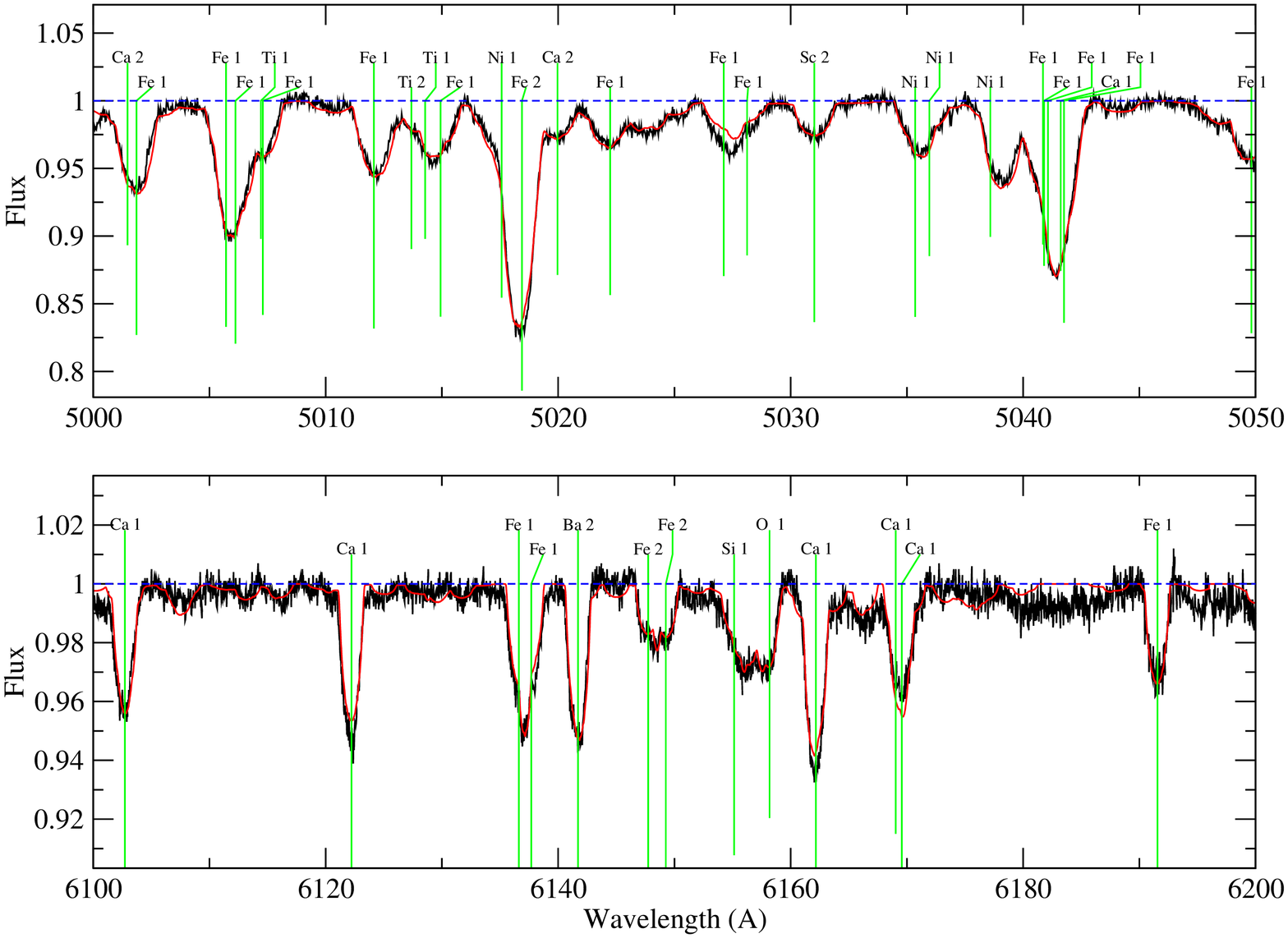}
\caption{Comparison of the observed spectrum to the best fit synthetic 
spectrum for HD 169142, as in Fig. \ref{fit-appendix}}
\label{fit-hd169142}
\end{figure*}

\begin{figure*}
\centering
\includegraphics[width=5.4in]{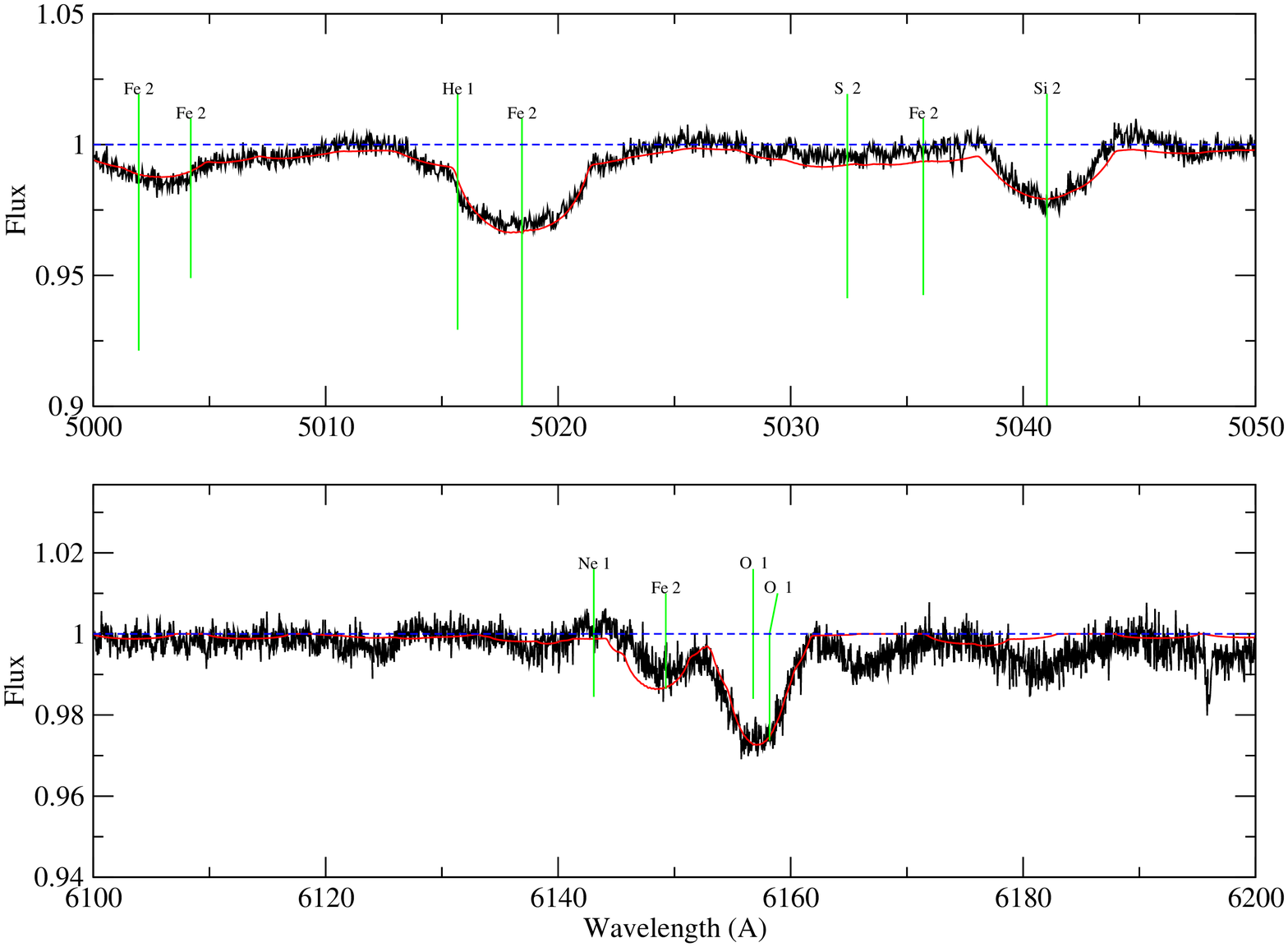}
\caption{Comparison of the observed spectrum to the best fit synthetic 
spectrum for HD 176386, as in Fig. \ref{fit-appendix}}
\label{fit-hd176386}
\end{figure*}

\begin{figure*}
\centering
\includegraphics[width=5.4in]{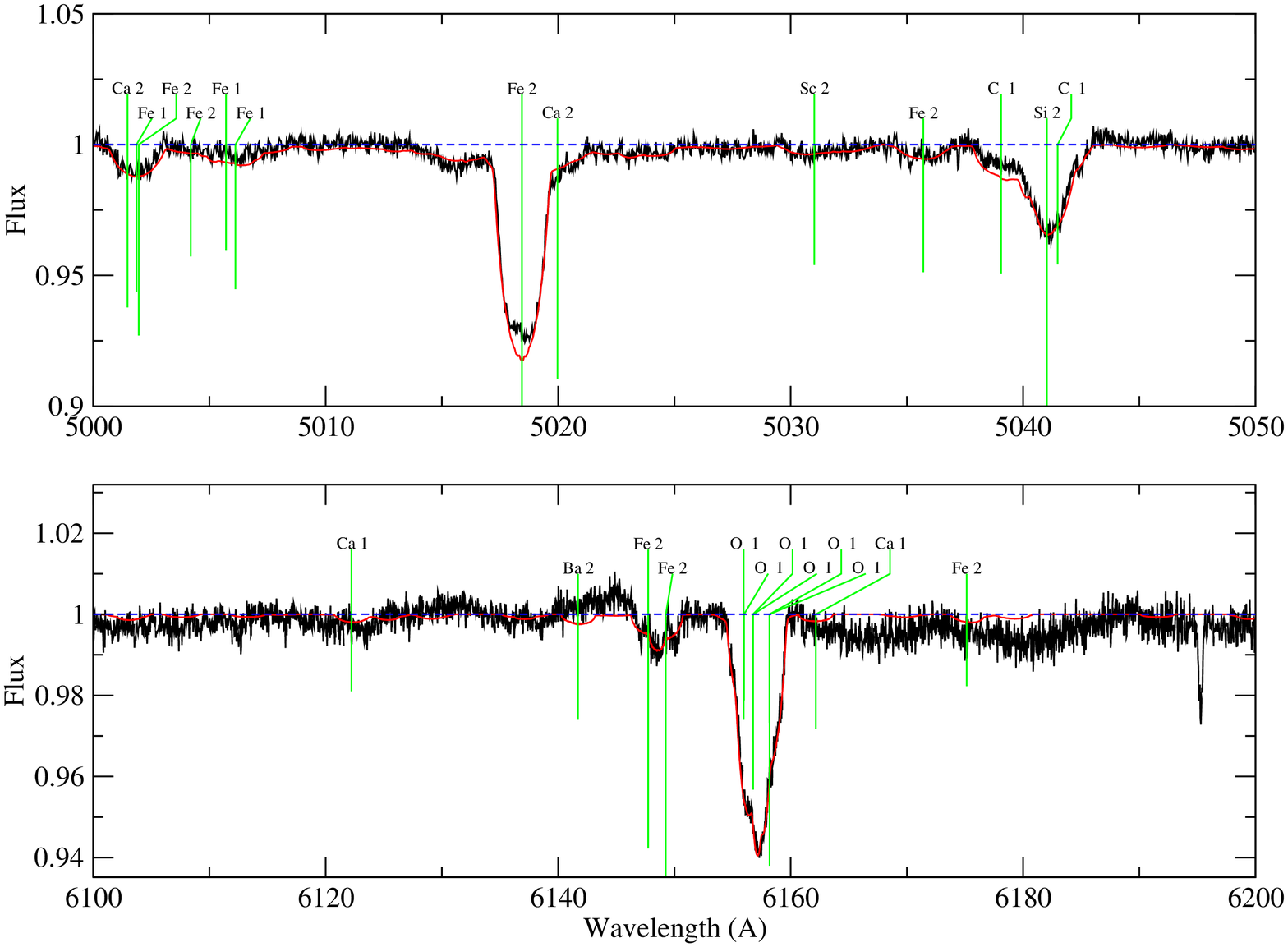}
\caption{Comparison of the observed spectrum to the best fit synthetic 
spectrum for HD 179218, as in Fig. \ref{fit-appendix}}
\label{fit-hd179218}
\end{figure*}

\begin{figure*}
\centering
\includegraphics[width=5.4in]{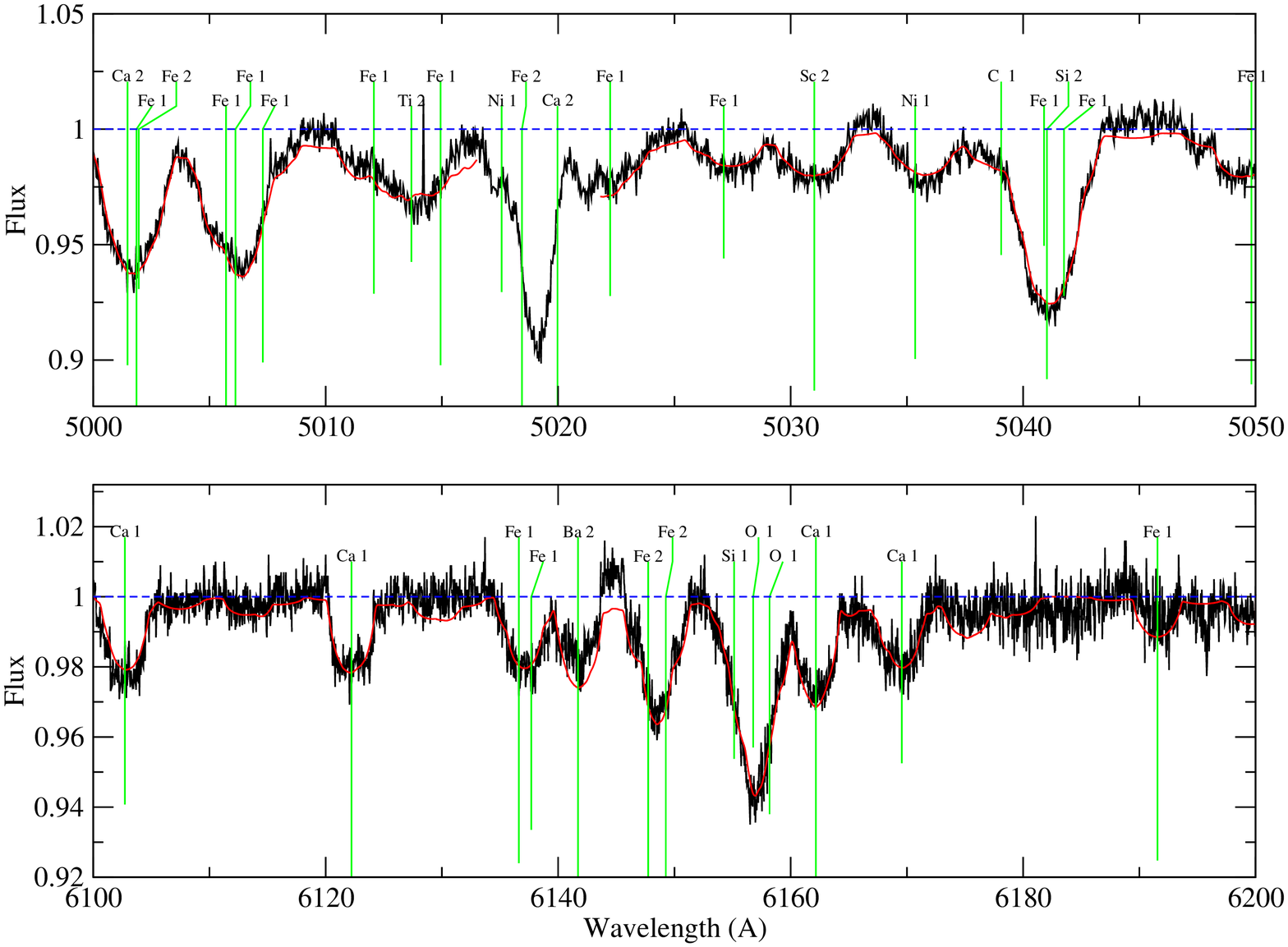}
\caption{Comparison of the observed spectrum to the best fit synthetic 
spectrum for HD 244604, as in Fig. \ref{fit-appendix}}
\label{fit-hd244604}
\end{figure*}

\begin{figure*}
\centering
\includegraphics[width=5.4in]{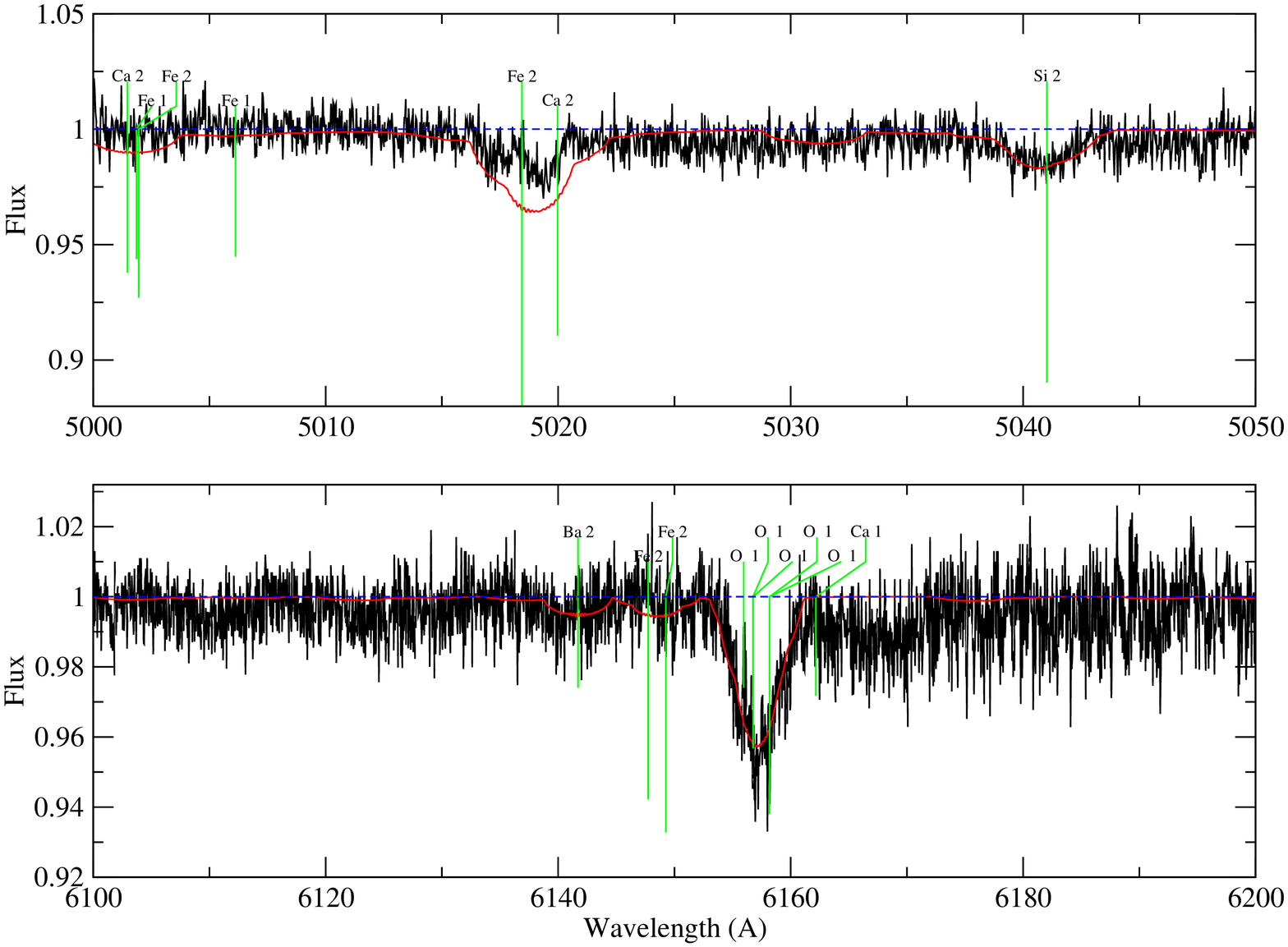}
\caption{Comparison of the observed spectrum to the best fit synthetic 
spectrum for HD 245185, as in Fig. \ref{fit-appendix}}
\label{fit-hd245185}
\end{figure*}

\begin{figure*}
\centering
\includegraphics[width=5.4in]{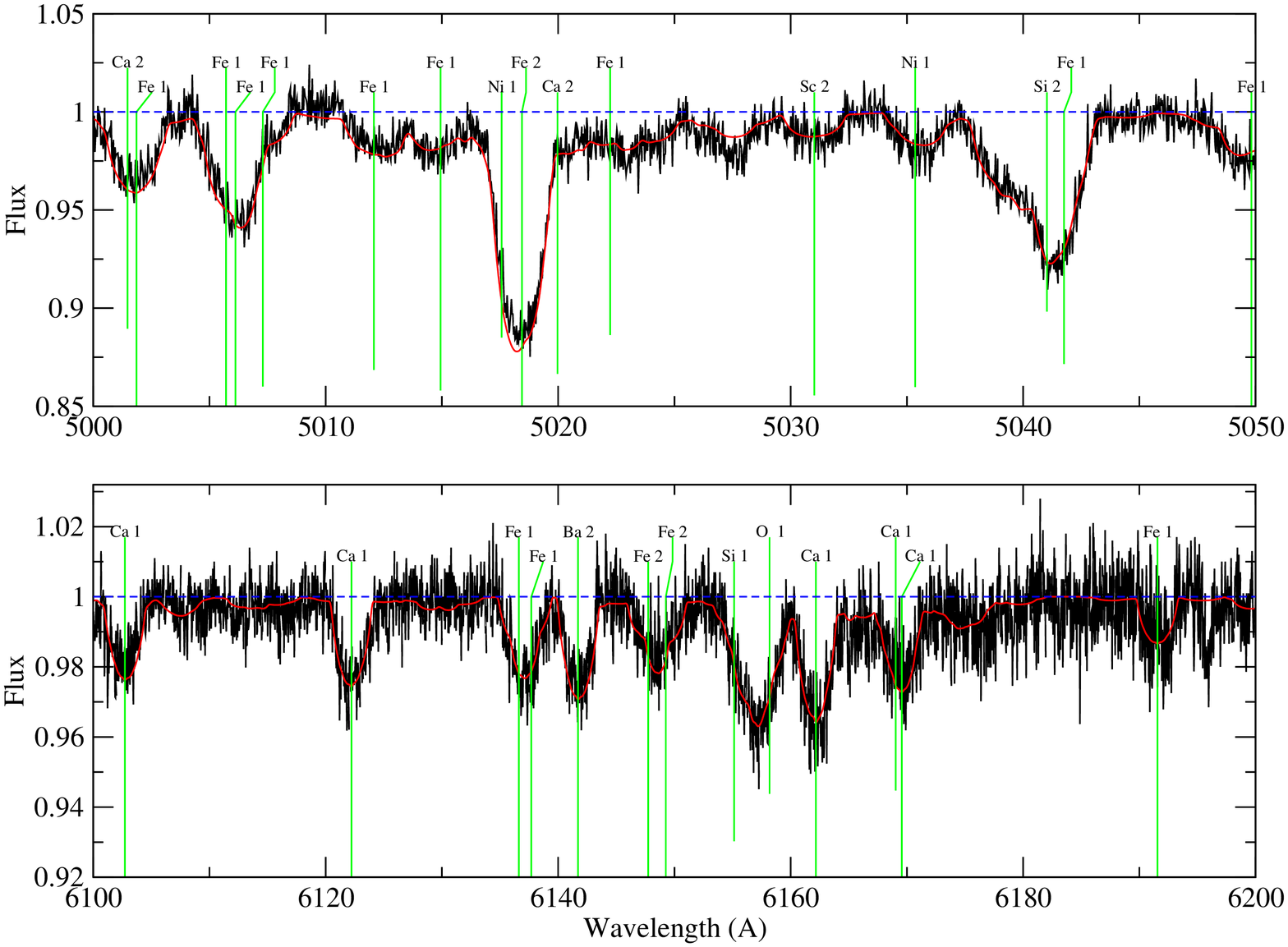}
\caption{Comparison of the observed spectrum to the best fit synthetic 
spectrum for HD 278937, as in Fig. \ref{fit-appendix}}
\label{fit-hd278937}
\end{figure*}

\begin{figure*}
\centering
\includegraphics[width=5.4in]{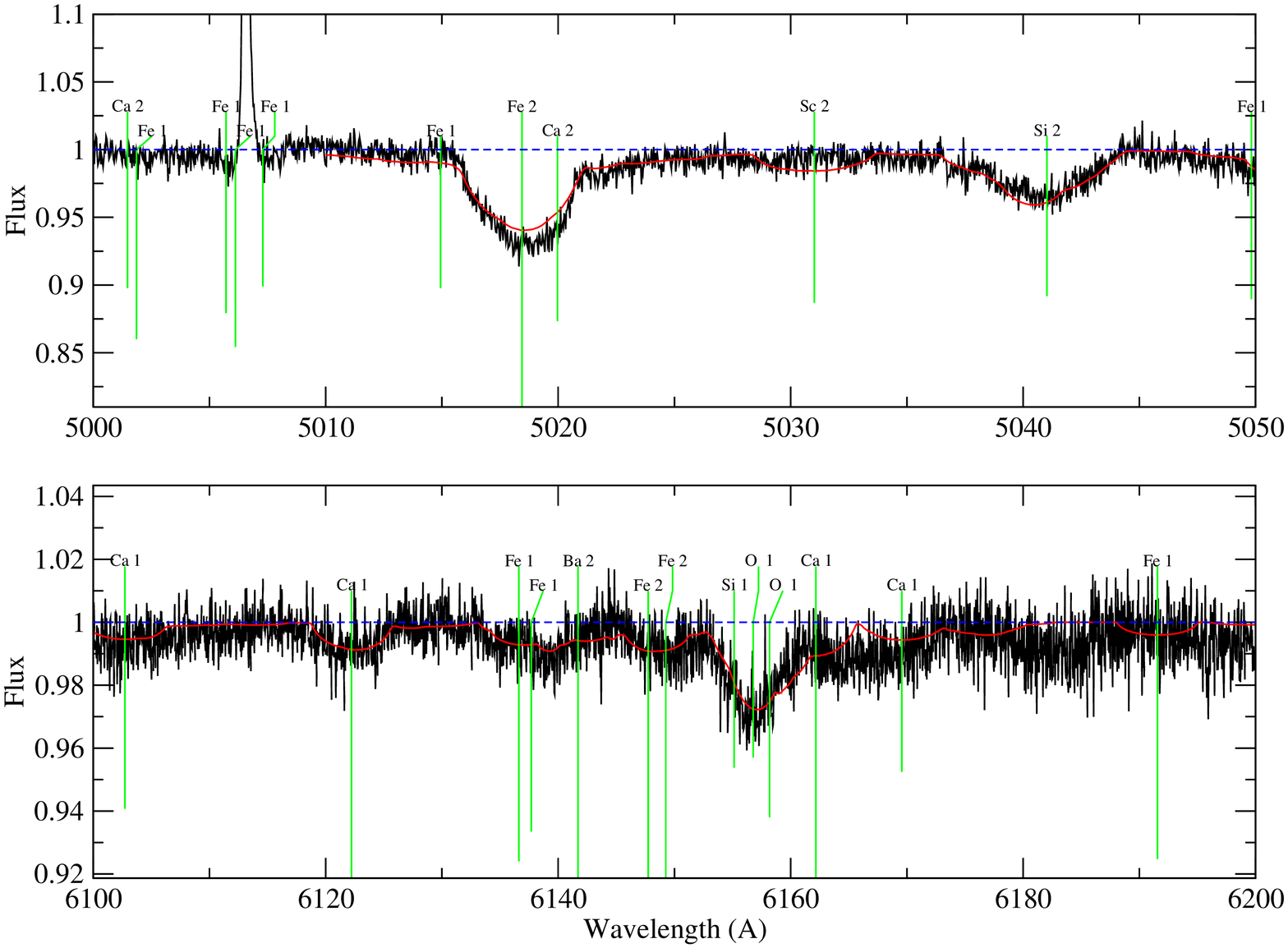}
\caption{Comparison of the observed spectrum to the best fit synthetic 
spectrum for T Ori, as in Fig. \ref{fit-appendix}}
\label{fit-tori}
\end{figure*}

\begin{figure*}
\centering
\includegraphics[width=5.4in]{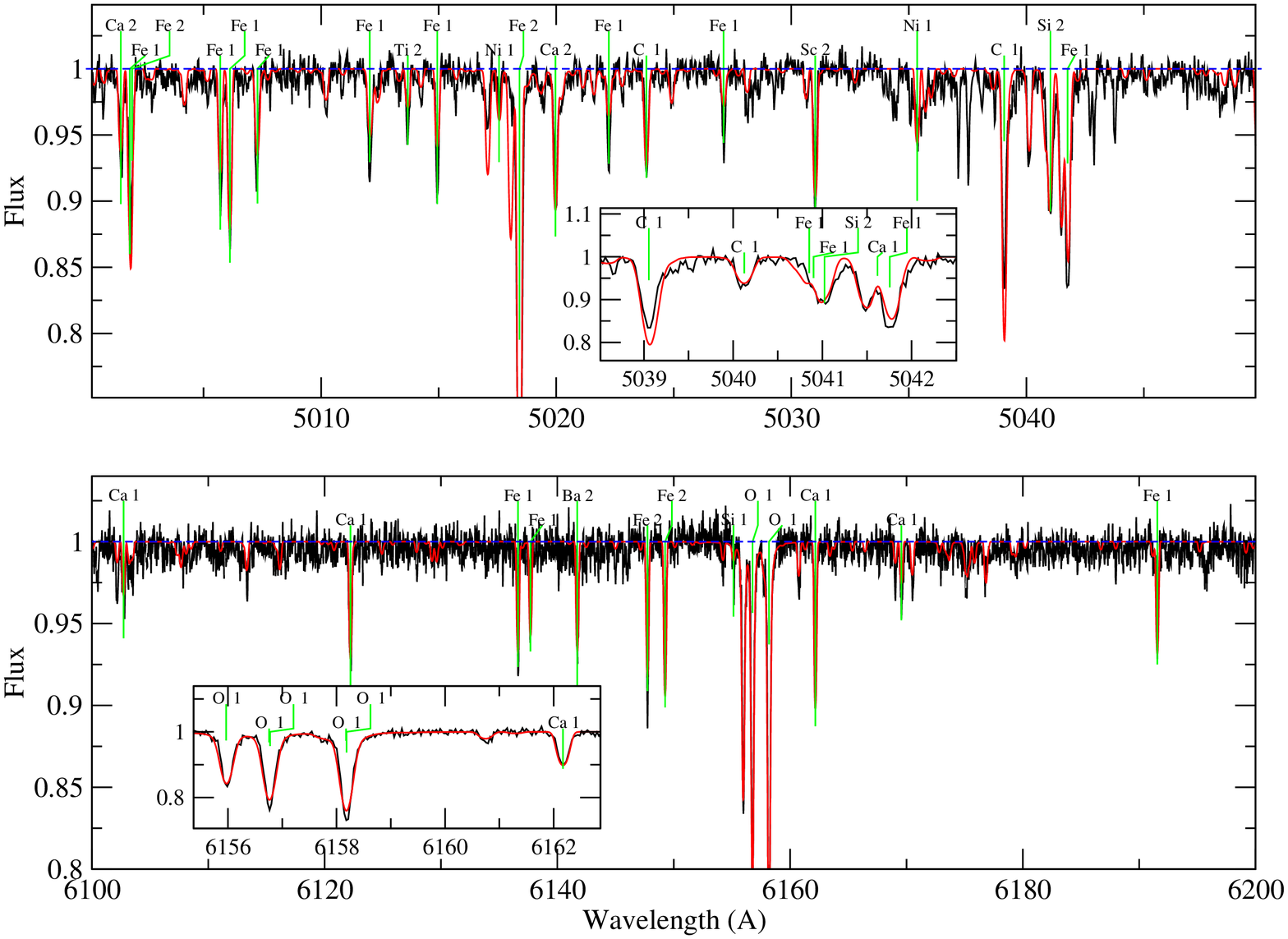}
\caption{Comparison of the observed spectrum to the best fit synthetic 
spectrum for HD 101412, as in Fig. \ref{fit-appendix}}
\label{fit-hd101412}
\end{figure*}

\begin{figure*}
\centering
\includegraphics[width=5.3in]{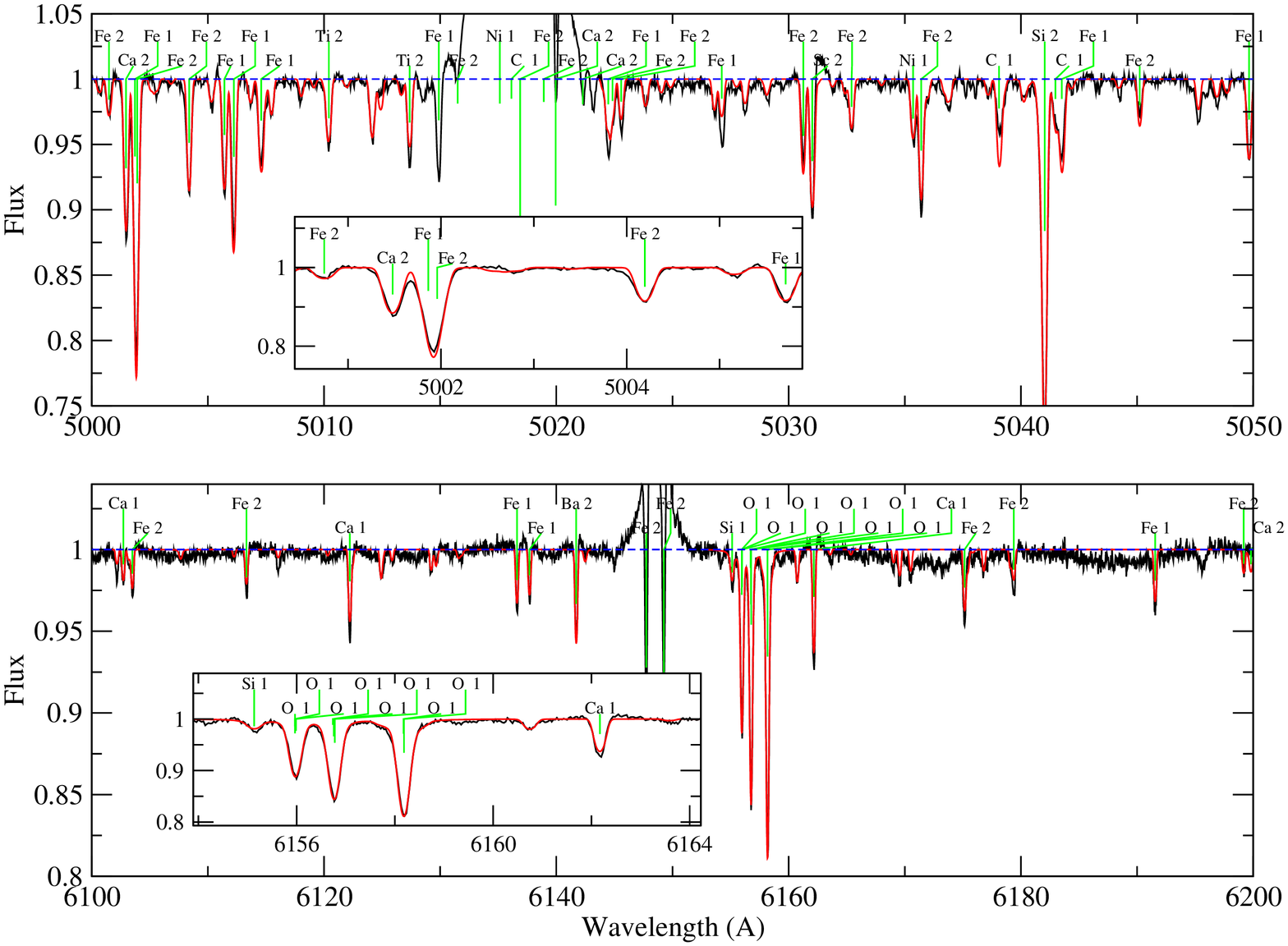}
\caption{Comparison of the observed spectrum to the best fit synthetic 
spectrum for HD 190073, as in Fig. \ref{fit-appendix}}
\label{fit-hd190073}
\end{figure*}

\begin{figure*}
\centering
\includegraphics[width=5.3in]{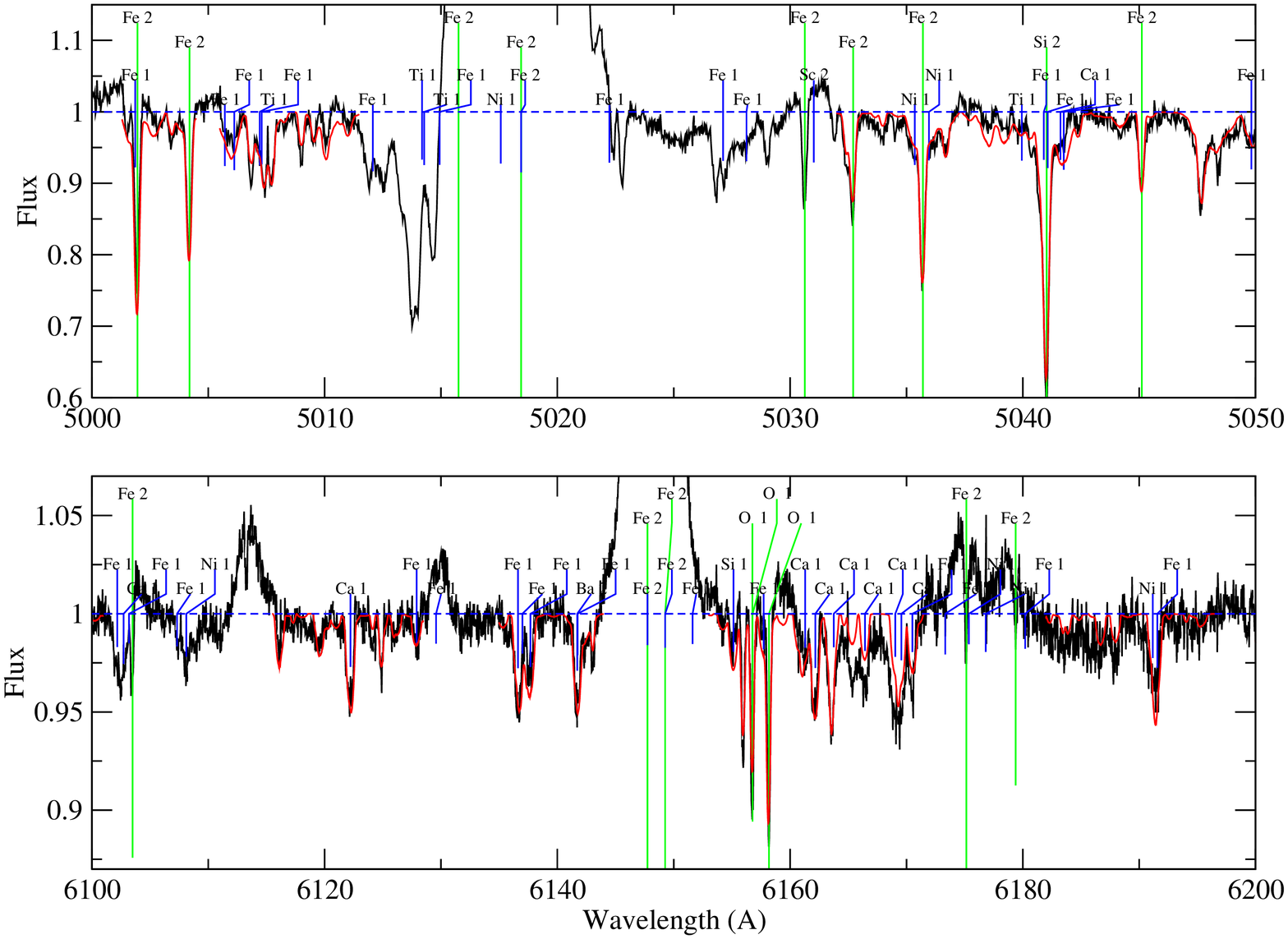}
\caption{Comparison of the observed spectrum (jagged line) to the best fit synthetic 
spectrum of the combined V380 Ori A and B system (smooth line).  
The upper two rows of line labels (green) indicate lines of the primary, 
the lower two rows (blue) indicate lines of the secondary.  
Gaps in the synthetic spectrum indicate regions that were excluded from the fit.  }
\label{fit-v380ori}
\end{figure*}

\clearpage
\bsp 

\label{lastpage}

\end{document}